\documentclass[11pt]{article}
\usepackage{natbib}
\usepackage{amsfonts}
\usepackage{amssymb}
\usepackage{graphicx}
\usepackage{amsmath}
\usepackage{xcolor}
\usepackage{todonotes}
\usepackage[normalem]{ulem} 
\usepackage{tikz} 
\usetikzlibrary{decorations}
\usetikzlibrary{snakes}
\usetikzlibrary{arrows}

\newcommand{\bc}{\begin{center}}
\newcommand{\ec}{\end{center}}
\newcommand{\be}{\begin{equation}}
\newcommand{\ee}{\end{equation}}
\newcommand{\ben}{\begin{equation*}}
\newcommand{\een}{\end{equation*}}
\newcommand{\bea}{\begin{eqnarray}}
\newcommand{\eea}{\end{eqnarray}}
\newcommand{\bean}{\begin{eqnarray*}}
\newcommand{\eean}{\end{eqnarray*}}

\newtheorem{assumption}{Assumption}

\newtheorem{theorem}{Theorem}

\newtheorem{lemma}{Lemma}
\newtheorem{applemma}{Lemma B}

\newcommand{\vect}{\text vec}

\newcounter{pkt}
\newenvironment{tlist}{\begin{list}{(\roman{pkt})}{\usecounter{pkt}\parskip0ex\parsep0ex\itemsep0ex\topsep0ex}}{\end{list}}
\def\EE{\mathord{I\kern-.35em E}}
\def\PP{\mathord{I\kern-.3em P}}
\def\QQ{\mathord{Q\kern-5pt\hbox{\raise1.1pt\hbox{\vrule
        height5pt}}\kern5pt}}
\def\RR{\mathord{I\kern-.3em R}}
\newcounter{saveeqn}

\begin{document}

\title{\textbf{Inference for VARs Identified with Sign Restrictions}}
\author{Eleonora Granziera\thanks{Correspondence: E. Granziera: Bank of Finland, Snellmaninaukio, 00101 Helsinki, Finland.
		E-mail: eleonora.granziera@bof.fi. H.R. Moon:
		Department of Economics, University of Southern California, KAP
		300, Los Angeles, CA 90089. E-mail: moonr@usc.edu. F. Schorfheide: Department of Economics, University
		of Pennsylvania, 3718 Locust Walk, Philadelphia, PA 19104. E-mail:
		schorf@ssc.upenn.edu. 
		We thank Andres Santos (co-editor), Fabio Canova, and Eric Renault, James Stock, several anonymous referees, as well as participants at various conferences 
		and seminars for helpful comments.
		We also thank Mihye Lee for her contributions to the first draft of this paper and Minchul Shin for research assistance.
		Schorfheide gratefully acknowledges financial
		support from the National Science Foundation under Grants SES 1061725 and 1424843.
		The views expressed in this paper are those of the
		authors and do not necessarily reflect those of the Bank of Finland. The Online Technical Appendix as well as
		data and software to replicate
		the empirical analysis are available at {\em https://web.sas.upenn.edu/schorf.}}\\ \emph{Bank of Finland}
	    \and
      	Hyungsik Roger Moon\\
	    \emph{University of Southern California}\\\emph{and Yonsei University}
	    \and Frank Schorfheide \\
     	\emph{University of Pennsylvania,} \\ {\em CEPR, NBER, and PIER}\\[3ex]
}
\date{\today }
\maketitle

\thispagestyle{empty}
\setcounter{page}{0}
\newpage

\begin{abstract}
		
There is a fast growing literature that set-identifies
structural vector autoregressions (SVARs) by imposing
sign restrictions on the responses of a subset of the endogenous
variables to a particular structural shock (sign-restricted SVARs).
Most methods that have been used to construct pointwise coverage bands for impulse responses
of sign-restricted SVARs are justified only from a Bayesian perspective.
This paper demonstrates how to formulate the inference problem
for sign-restricted SVARs within a moment-inequality framework.
In particular, it develops methods of constructing confidence bands for
impulse response functions of sign-restricted SVARs that are valid
from a frequentist perspective. The paper also provides a comparison
of frequentist and Bayesian coverage bands in the context of an empirical
application - the former can be substantially wider than the latter. (JEL: C1, C32)
\end{abstract}


\noindent KEY\ WORDS: Bayesian Inference, Frequentist Inference, Set-Identified Models,
Sign Restrictions, Structural VARs.

\thispagestyle{empty}
\setcounter{page}{0}

\newpage

\section{Introduction}
\label{sec_intro}

During the three decades following \cite{Sims1980} ``Macroeconomics and Reality,''
structural vector autoregressions (SVARs) have become an important tool
in empirical macroeconomics. They have been used for macroeconomic forecasting
and policy analysis, as well as to investigate the sources of business cycle fluctuations
and to provide a benchmark against which modern dynamic macroeconomic theories
can be evaluated. The most controversial step in the specification of a structural
VAR is the mapping between-reduced form one-step-ahead forecast errors
and orthogonalized, economically interpretable structural innovations. Traditionally, SVARs
have been constructed by imposing sufficiently many restrictions such that
the relationship between structural innovations and forecast errors is one-to-one.

Over the past decade, starting with \cite{Faust1998}, \cite{CanovaDeNicolo2002}, and \cite{Uhlig2005},
empirical researchers have used more agnostic approaches that generate bounds on structural impulse
response functions by restricting the sign of certain responses. We
refer to this class of models as sign-restricted SVARs. They have been employed, for instance,
to measure the effects of monetary policy shocks (e.g., \cite{Faust1998}, \cite{CanovaDeNicolo2002}, \cite{Uhlig2005}),
technology shocks (e.g., \cite{DedolaNeri2007}, \cite{PeersmanStraub2009}), government spending shocks
(\cite{MountfordUhlig2009}, \cite{Pappa2009}), oil price shocks (e.g., \cite{BaumeisterPeersman2013}, \cite{KilianMurphy2012}), and financial shocks (e.g., \cite{HristovHuelsewigWollmershaeuser2012}, \cite{GambettiMusso2017}).

Because impulse responses in sign-restricted SVARs can only be restricted to a bounded set, they belong
to the class of set-identified econometric models, using the terminology
of \cite{Manski2003}.\footnote{The microeconometrics literature uses the terms {\em set} and {\em partially}
identified model interchangeably. In the VAR literature a {\em partially} identified structural
VAR is one in which the researcher tries to identify only a subset of the structural shocks. To avoid confusion
we shall use the term {\em set} identified because we are focusing on models in which impulse responses
can only be bounded.} With the exception of \cite{FaustRogersSwansonWright2003} and
\cite{FaustSwansonWright2004} (since the two papers are methodologically equivalent, we are using the
abbreviation FRSW to refer to both of them), researchers have exclusively reported Bayesian
credible bands for sign-restricted VARs, and a general method for constructing uniformly asymptotically valid
frequentist confidence intervals was absent from the literature when the first draft of this paper was written; see \cite{MSGL2011}. As shown in detail in \cite{MoonSchorfheide2012}, the
large-sample numerical equivalence of frequentist confidence sets and Bayesian credible sets
breaks down in set-identified models, which means that Bayesian credible bands
may not be interpreted as approximate frequentist confidence bands.\footnote{ \color{black} Treatments of Bayesian inference in sign-restricted SVARs can be found, for instance, in \cite{Uhlig2005}, \cite{RubioRamirezWaggonerZha2010}, \cite{BaumeisterHamilton2015},  \cite{KilianLuetkepohl2017}, and the references cited therein.}

The goal of this paper is to provide researchers with an easy-to-use tool to construct valid frequentist
confidence bands for impulse responses and other measures of the dynamic effects of structural shocks (e.g., variance decompositions) of sign-restricted SVARs.
The specific contributions are the following:
First, we formulate the problem of analyzing set-identified sign-restricted SVAR models in a moment-inequality-based minimum distance
framework. Second, we find an easily interpretable sufficient condition for
the non-emptiness of the identified set of sign-restricted structural impulse
responses and we propose a consistent
estimator of the identified set that is straightforward to compute.
Third, using our minimum distance framework, we formally analyze Bonferroni confidence sets. Fourth, we provide step-by step recipes for
practitioners on how to compute these confidence sets.\footnote{The contribution of this
paper is meant to be positive. We do not criticize the use of Bayesian inference methods
as long as it is understood that their output needs to be interpreted from a Bayesian
perspective. We provide applied researchers who are interested in impulse response confidence
bands that are valid from a frequentist perspective with econometric tools to compute
such bands.} 

At an abstract level, our inference problem
is characterized by a vector of point-identified reduced-form parameters $\phi$,
a vector of structural parameters (impulse responses or variance decompositions) $\theta$, and a vector of nuisance parameters $q$.
The sign restrictions generate an identified set for $q$, $F^{q}(\phi)$. 
Conditional on $q$ and $\phi$ the vector $\theta$ is point identified, but because $q$ is set-identified, so is $\theta$ and we denote its identified set by $F^\theta(\phi)$. 
To obtain a confidence set for $\theta$ we pursue a Bonferroni  
approach: we construct a confidence interval for the set-identified nuisance parameter $q$ and then take the union of standard Wald confidence sets for $\theta$ that are generated conditional on all $q$ in the first-stage confidence set. The Bonferroni inequality is used
to ensure the desired coverage probability of the resulting confidence
set for $\theta$. We also show that the plug-in estimator
$F^\theta(\hat{\phi})$ delivers a consistent estimate of the identified
set for $\theta$, denoted by $F^\theta(\phi)$.

In the first draft of this paper, we used a projection approach instead; see \cite{MSGL2011}.\footnote{ \color{black}Inference procedures for subvectors  have been further developed by \cite{ChaudhriZivot2011}, \cite{KaidoMolinariStoye2016}, \cite{Andrews2017}, and \cite{BugniCanayShi2017}.}. We constructed a joint confidence interval for the set-identified pair $(q,\theta)$ and projected it onto the $\theta$ ordinate. Here $\theta$ is the response of a particular variable at a particular horizon to a particular shock. In order to generate point-wise confidence bands for impulse response functions we repeated the computations for different definitions of $\theta$. In subsequent research we compared the projection approach and the Bonferroni approach and found that there is no clear ranking of the two types of confidence sets.
However, the Bonferroni approach has a clear computational advantage: the irregular confidence set for the nuisance parameter $q$ only has to be computed once. Conditional on $F^q(\phi)$ one can easily generate standard confidence sets for impulse responses of different variables at different horizons, of vectors of responses, and of variance decompositions and then take unions over $q$. Thus, we decided to focus on the Bonferroni set in the current version of the paper.

The Bonferroni approach has a long history in the time series literature. For instance,
\cite{CavanaghElliottStock1995} and \cite{CampbellYogo2006} use it to
eliminate nuisance parameters that characterize the persistence of error terms
or regressors. In the context of structural VARs, the Bonferroni
approach has been used by FRSW. However, FRSW's setup is quite different from ours.
In their framework the set identification of $q$ arises from a rank deficiency
in equality restrictions, which depend on estimated parameters. While FRSW restrict
$q$ further by imposing inequality conditions, these inequality conditions do not
depend on estimated parameters. Our analysis, on the other hand, focuses on inequality restrictions
for $q$ that may or may not be binding and do depend on estimated parameters.
This generalization is essential to cover the wide range of empirical applications referenced above. In the Monte Carlo analysis we explore ideas by \cite{CampbellYogo2006} and \cite{McCloskey2017} to tighten the Bonferroni sets.

Building upon recent advances in the moment-inequality literature in microeconometrics,
in particular, \cite{ChernozhukovHongTamer2007},
\cite{Rosen2008}, \cite{AndrewsGuggenberger2009}, and \cite{AndrewsSoares2010a},
we provide an asymptotic analysis of the Bonferroni approach to constructing confidence sets for the dynamic effects of shocks in sign-restricted SVARs.
Because the number of linearly independent
moment conditions is a function of the nuisance parameter $q$, we need to modify some of the existing microeconometric theory. As is common in the literature, we use a point-wise testing procedure to obtain a confidence set for $q$.
We use Andrews and Soares' (2010a) moment selection procedure to tighten the critical values for the point-wise testing procedures.\footnote{A recent survey of the moment-inequality literature is provided by \cite{CanayShaik2017}. Alternative procedures include \cite{AndrewsBarwick2012} and \cite{RomanoShaikhWolf2014}. Because of a potentially large number of inequality restrictions, the refinements to the moment selection proposed in \cite{AndrewsBarwick2012} did not seem practical.} We adapt their theory to account for the $q$-dependent rank of the set of moment inequalities in our model 
and prove that the proposed confidence sets are asymptotically valid
in a uniform sense. Our results on the non-emptiness of identified sets for
impulse responses complement the equality-restriction-based VAR identification results reported
in \cite{RubioRamirezWaggonerZha2010}.


Since the working paper versions of this paper have been written, a series of alternative approaches for the construction of frequentist confidence bands of sign-restricted SVARs have been proposed. \cite{GafarovMeierMontielOlea2016b} developed an alternative projection-based approach that starts from a Wald confidence ellipsoid for the reduced-form VAR parameters 
and takes unions of the identified sets $F^\theta(\phi)$. Because it relies on the regular behavior of the estimator of the reduced-form coefficients, this method also has the interpretation of delivering Bayesian credible sets for the identified set $F^\theta(\phi)$. The downside is that it is quite conservative.
\cite{GafarovMeierMontielOlea2016a} propose a $\delta$-method confidence interval for sign-restricted SVARs which relies on a closed-form characterization of the endpoints of the identified set. While the resulting intervals are less conservative, their drawbacks are that they are only pointwise, but not uniformly consistent and that they can only be applied to scalar $\theta$'s. 
Finally, \cite{GiacominiKitagawa2015} construct robust Bayesian credible sets for impulse response functions in set-identified SVARs that have good frequentist properties.

The remainder of the paper is organized as follows. Section~\ref{sec_idea} develops the notation used in this paper and 
provides a simple example of a sign-restricted SVAR. We describe how set-identification arises in this model and sketch the Bonferroni approach to the construction of confidence intervals for the dynamic effects of structural shocks. 
Section~\ref{sec_implementation} is geared toward practitioners and discusses the
step-by-step implementation and computational aspects of the proposed inference method in the context of a general SVAR. Technical assumptions and large sample results are presented in Section~\ref{sec_largesample}. Some extensions are discussed in Section~\ref{sec_extensions}.
To illustrate the proposed methods,
we conduct a Monte Carlo study in Section~\ref{sec_mc}
and generate confidence bands for output, inflation, interest rate, and money responses to a monetary policy shock in an empirical application in Section~\ref{sec_empirical}.
Finally, Section~\ref{sec_conclusion} concludes. Proofs and detailed derivations as well as further information about the Monte Carlo experiments and the empirical analysis are relegated to a supplemental Online Appendix.

We use the following notation throughout the remainder of the paper:
${\cal I}\{ x \ge a \}$ is the indicator function that is one if $x \ge a$ and zero otherwise.
$0_{n \times m}$ is an $n \times m$ matrix of zeros and $I_n$ is the
$n \times n$ identity matrix. We use $[AB]_{(i.)}$ to denote the $i$'th row of the matrix $(A \cdot B)$
and $[AC]_{i}$ to denote the $i$'th element of the vector $(A\cdot C)$.
$\otimes$ is the Kronecker product, $vec(\cdot)$ stacks the columns of a matrix,
and $vech(\cdot)$ vectorizes the lower triangular part of a square matrix.
We use $diag(A_1,\ldots,A_k)$ to denote a quasi-diagonal
matrix with submatrices $A_1$, $\ldots$, $A_k$ on its diagonal and zeros elsewhere.
If $A$ is an $n \times m$ matrix, then $\| A \|_W = \sqrt{ tr[WA'A] }$. In the special case of a vector, our
definition implies that $\|A\|_W = \sqrt{A'WA}$. If the weight matrix is the identity
matrix, we omit the subscript. We write $x \gg 0$ to mean that all elements of the vector $x$
are strictly greater than zero; we write $x > 0$ to mean that all elements of $x$ are greater than
or equal to zero but not all equal to zero, that is, $x \not= 0$; finally, we write $x \ge 0$ to mean that
all elements of $x$ are greater than or equal to zero.
We use $\propto$ to indicate proportionality
and ``$\stackrel{p}{\longrightarrow}$'' and ``$\Longrightarrow$'' to indicate convergence in probability and convergence in distribution, respectively, as $T\longrightarrow \infty$.
A multivariate normal distribution is denoted by $N(\mu,\Sigma)$.
We use $\chi^2_m$ to denote a $\chi^2$ distribution
with $m$ degrees of freedom.

\section{General Setup and Illustrative Example}
\label{sec_idea}

Throughout this paper we consider an $n$-dimensional VAR with $p$ lags, which takes the form
\be
y_t = A_1 y_{t-1} + \ldots + A_p y_{t-p} + u_t, \quad \mathbb{E}[u_t | {\cal F}_{t-1}] = 0, \quad
\mathbb{E}[u_t u_t'|{\cal F}_{t-1}] = \Sigma_u.
\label{eq_idea_varrf}
\ee
Here $y_t$ is an $n \times 1$ vector and the information set ${\cal F}_{t-1} = \{ y_{t-1}, y_{t-2},\ldots \}$ is composed
of the lags of $y_t$'s. Constants and deterministic trend terms are omitted because they are irrelevant for the
subsequent discussion. The one-step-ahead forecast errors (reduced-form shocks) $u_t$
are linear functions of a vector of fundamental innovations (structural shocks) $\epsilon_t$:
\be
u_t = A_\epsilon \epsilon_t = \Sigma_{tr} \Omega_\epsilon \epsilon_t, \quad \mathbb{E}[\epsilon_t | {\cal F}_{t-1}] = 0, \quad
\mathbb{E}[\epsilon_t \epsilon_t'|{\cal F}_{t-1}] = I_n,
\label{eq_idea_ueps}
\ee
where $\Sigma_{tr}$ is the lower triangular Cholesky factor
of $\Sigma_u$ and $\Omega_\epsilon$ is an arbitrary orthogonal matrix.
Assuming that the lag polynomial associated with the VAR in~(\ref{eq_idea_varrf}) is invertible,
one can express $y_t$ as the following infinite-order vector moving average (VMA) process:
\be
y_t = \sum_{h=0}^\infty C_h(A_1,\ldots,A_p) \Sigma_{tr} \Omega_\epsilon \epsilon_{t-h}.
\ee
We assume that the object of interest
is the propagation of the first shock, $\epsilon_{1,t}$, and denote the first column of the
matrix $\Omega_\epsilon$ by $q$, where $q$ is a unit-length vector.  The domain of $q$ is the $n$-dimensional unit sphere $\mathbb{S}^n = \big\{ q \in \mathbb{R}^n \, \big| \, \|q\|=1 \big\}$.
In Section~\ref{subsec_idea_setid} we discuss how imposing sign restrictions on some impulse responses
generates set identification of the dynamic effects of $\epsilon_{1,t}$. Section~\ref{subsec_idea_ex} provides
an illustration in the context of a bivariate VAR(0). Section~\ref{subsec_idea_rankreductions} introduces some 
important notation. We present the construction of a Bonferroni confidence set for the dynamic effects of $\epsilon_{1,t}$ in Section~\ref{subsec_idea_bonferroni}. The Bonferroni set is based on a confidence set for $q$, which is described in Section~\ref{subsec_idea_csq}.

\subsection{Sign Restrictions and Set Identification}
\label{subsec_idea_setid}

The SVAR identification problem arises because the one-step-ahead forecast error covariance matrix $\Sigma_u$
is invariant to the orthogonal matrix $\Omega_\epsilon$, which implies that $\Omega_\epsilon$ and its first column $q$
are not identifiable from the data.  Point identification could be 
achieved by selecting a particular $q$ as a function of $(A_1,\ldots,A_p,\Sigma_{tr})$.
The recent SVAR literature has pursued
a more agnostic approach and restricted the set of admissible $q$ by a collection
of sign restrictions on impulse responses.\footnote{
	We assume that these sign restrictions do not encode equality restrictions (e.g., by
	representing $a=0$ as $a \le 0$ and $a \ge 0$.)
	The extension to models that combine sign-restrictions and equality restrictions
	is deferred to Section~\ref{sec_extensions}.} The impulse response of variable $y_{i,t}$ to $\epsilon_{1,t}$ at horizon $h$ is given by 
\be
IRF(i,h|\epsilon_{1,t}=1) = [C_h(A_1,\ldots,A_p) \Sigma_{tr}]_{(i.)} q.
\label{eq_idea_irf}
\ee
We define the $n \times 1$ vector
\[
    {\color{black}\phi_j} = [C_h(A_1,\ldots,A_p) \Sigma_{tr}]_{(i.)}'
\]
as the responses of a variable $i$ at horizon $h$ to the vector of reduced-form innovations $u_t$ and summarize the sign-restricted impulse responses as
\be
 \Phi_q'q \ge 0, \quad \mbox{where} \quad \Phi_q = [\phi_1,\ldots,\phi_r],   
\label{eq_idea_ineqcanonical}
\ee  
{\color{black}$r$ is the number of restrictions}, and $\Phi_q$ is a $n \times r$ matrix.
The vectors $\phi_j$ are 
possibly multiplied by $-1$ to restrict the response to be weakly negative rather than weakly
positive. Moreover, the notation in~(\ref{eq_idea_ineqcanonical}) is general 
enough to accommodate sign restrictions on cumulative impulse responses over $\bar{h}$ periods,
obtained from $\sum_{h=0}^{\bar{h}} C_h(A_1,\ldots,A_p) \Sigma_{tr} q$.

The object of inference is a $k$-dimensional parameter defined as
\be
\theta = f \big( \Phi_\theta(A_1,\ldots,A_p,\Sigma_{tr}),q \big) \in \Theta \subset \mathbb{R}^k.
\label{eq_idea_thdef}
\ee
The parameter set $\Theta$ is chosen to be consistent with potential sign restrictions for elements of $\theta$ implied by (\ref{eq_idea_ineqcanonical}). Our leading example of $f(\cdot)$ is a vector of impulse responses, which can be expressed as a linear function of the reduced-form impulse responses
\be
f \big( \Phi_\theta(A_1,\ldots,A_p,\Sigma_{tr}),q \big) 
= \Phi'_\theta q,
\label{eq_idea_thleadcase}
\ee
where the definition of $\Phi_\theta$ is similar to the definition of $\Phi_q$ in (\ref{eq_idea_ineqcanonical}). 
In addition to impulse responses, researchers often report variance decompositions. For instance, the fraction of the one-step-ahead forecast error variance of variable $y_{1,t}$ explained by shock $\epsilon_{1,t}$ is given by
\[
   f \big( \Phi_\theta(A_1,\ldots,A_p,\Sigma_{tr}),q \big) = \frac{\iota_1' \Sigma_{tr}qq'\Sigma_{tr}' \iota_1}{\iota_1' \Sigma_{tr}\Sigma_{tr}' \iota_1},
\]
where $\iota_1 = [1,0,\ldots,0]'$ is a $n\times 1$ vector. 

While $\Phi_q$ and $\Phi_\theta$ can be consistently estimated, the vector $q$ as well 
as the object of interest $\theta$ are only set identified. 
We use $F^q(\Phi_q)$ and $F^\theta(\Phi_q,\Phi_\theta)$ to denote the identified sets of $q$ and $\theta$, respectively.
Formally, they are defined as
\begin{eqnarray}
F^q(\Phi_q) &=& \big\{ q \in \mathbb{S}^n \, \big| \, \Phi_q' q \ge 0 \big\}
\label{eq_idea_fqphi} \\
F^\theta(\Phi_q,\Phi_\theta) &=& \big\{ \theta \in \Theta \; \big| \; \exists \; q \in F^q(\Phi_q) \; \mbox{s.t.} \;
\theta = f(\Phi_\theta,q) \big\}.
\label{eq_idea_fthphi} 
\end{eqnarray}
The goal is to construct a confidence set for $\theta$. As an intermediate step in a Bonferroni approach we will also construct a confidence set for $q$.

\subsection{Identified Sets $F^q(\cdot)$ and $F^\theta(\cdot)$ in a Bivariate VAR(0)}
\label{subsec_idea_ex}

For concreteness consider the following example. Suppose the vector $y_t$ is composed
of inflation and output growth and the one-step-ahead forecast errors are linear functions
of structural demand and supply shocks, stacked
in the vector $\epsilon_t = [\epsilon_{D,t}, \epsilon_{S,t}]'$.
In order to obtain bounds for the effects of
a demand shock, we impose the sign restriction that, contemporaneously, a demand shock moves prices and output in the same direction
and the normalization restriction that a positive demand shock increases prices:
\be
\phi_1'q= \big[ \Sigma_{11}^{tr} \quad 0 \big] \left[ \begin{array}{c} q_1 \\q_2 \end{array} \right] \ge 0, \quad  
\phi_2'q= \big[ \Sigma_{21}^{tr} \quad \Sigma_{22}^{tr} \big] \left[ \begin{array}{c} q_1 \\q_2 \end{array} \right]  \ge 0, \quad \Phi_q = [\phi_1,\phi_2].
\label{eq_ex_ineq}
\ee
Suppose that the object of interest, $\theta$, is the contemporaneous inflation response to a demand shock $\epsilon_{D,t}$: 
\be
\theta = \phi_1'q, \quad \Phi_\theta = \phi_1.
\label{eq_ex_th}
\ee

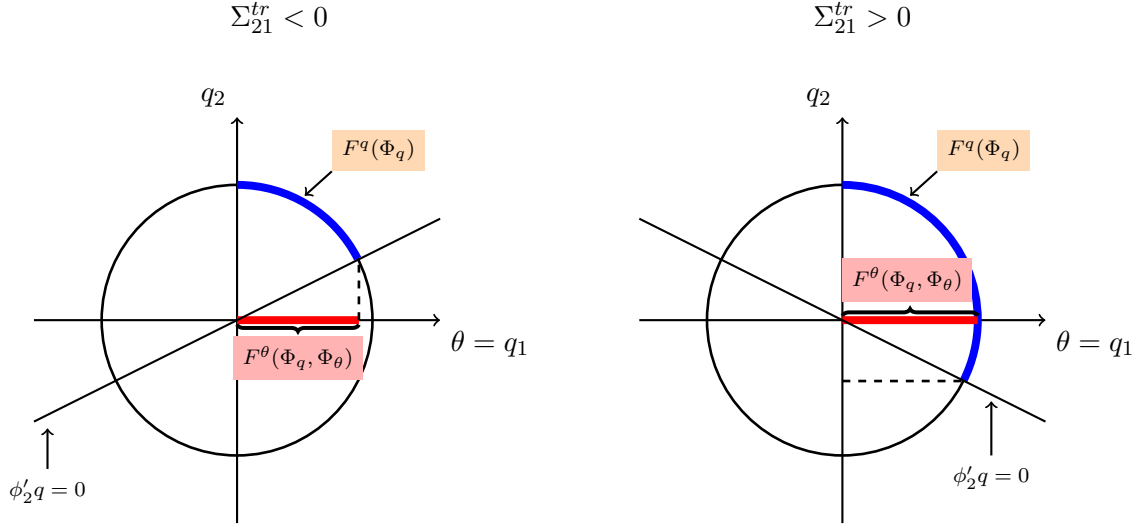
\begin{figure}[t!]
	\caption{Identified Sets for VAR(0)}
	\label{f_VAR0IDset}
		
	\begin{center}
		\begin{tabular}{cc}
			\vspace{0.2in}
			
			$\Sigma_{21}^{tr} < 0 \quad \quad $ & $\Sigma^{tr}_{21} > 0 \quad \quad$ \\
			\begin{tikzpicture}[scale=1.8]
			\draw [->, line width = 0.8] (-1.5,0) -- (1.5,0) node [below right] {$\theta=q_{1}$};
			\draw [->, line width = 0.8] (0, -1.5) -- (0,1.5) node [above left] {$q_{2}$};
			\draw [black, line width = 1] (0,0) circle (1cm);
			
			\draw [blue, line width = 3] (27:1) arc (27:90:1cm);
			\draw [<-, black, line width = 0.8] (0.5, 0.92) -- (0.7,1.1) node [fill=orange!30, above right, line width = 0.5] {\scriptsize{$F^{q}(\Phi_q)$}};
			
			
			\draw [red, line width = 3] (0,0) -- (0.9,0);
			\draw[decorate,decoration={brace,mirror, raise=1.7},red, black, line width = 1.5] (0,0) -- (0.9,0);
			\node at (0.45, -0.28) [fill=red!30, line width = 0.5] {\scriptsize{$F^{\theta}(\Phi_q,\Phi_\theta)$}};
			
			\draw [black, line width = 0.8] (-1.5,-0.75) -- (1.5,0.75);
			\draw [<-, black, line width = 0.8] (-1.4, -0.8) -- (-1.4,-1.1) node [below] {\scriptsize{$\phi_2'q = 0$}};
			
			\draw [black, dashed, line width = 1] (0.9,0) -- (0.9, 0.45);
			\end{tikzpicture}
			
			\quad \quad \quad &
			
			\begin{tikzpicture}[scale=1.8]
			
			\draw [->, line width = 0.8] (-1.5,0) -- (1.5,0) node [below right] {$\theta=q_{1}$};
			\draw [->, line width = 0.8] (0, -1.5) -- (0,1.5) node [above left] {$q_{2}$};
			\draw [black, line width = 1] (0,0) circle (1cm);
			
			\draw [blue, line width = 3] (90:1) arc (90:-26.5:1cm);
			\draw [<-, black, line width = 0.8] (0.5, 0.92) -- (0.7,1.1) node [fill=orange!30, above right, line width = 0.5] {\scriptsize{$F^{q}(\Phi_q)$}};
			
			
			\draw [red, line width = 3] (0,0) -- (1,0);
			\draw[decorate,decoration={brace, raise=1.7},red, black, line width = 1.5] (0,0) -- (1,0);
			\node at (0.48, 0.30) [fill=red!30, line width = 0.5] {\scriptsize{$F^{\theta}(\Phi_q,\Phi_\theta)$}};
			
			\draw [black, line width = 0.8] (-1.5,0.75) -- (1.5,-0.75);
			\draw [<-, black, line width = 0.8] (1.1, -0.6) -- (1.1,-1) node [below] {\scriptsize{$\phi_2'q = 0$}};
			
			\draw [black, dashed, line width = 1] (0,-0.45) -- (0.9, -0.45);
			\end{tikzpicture}
			
		\end{tabular}
	\end{center}
\end{figure}

Figure~\ref{f_VAR0IDset} provides an illustration of the two inequality constraints
and the resulting identified sets. To simplify the graphical illustration, we assume
that $\Sigma_{11}^{tr}=1$ which implies $\theta = q_1$. Because the Cholesky factorization of
the covariance matrix $\Sigma_u$ is normalized such that $\Sigma_{11}^{tr}$ and
$\Sigma_{22}^{tr}$ are nonnegative, (\ref{eq_ex_ineq}) implies that
\be
q_1 \ge 0 \quad \mbox{and} \quad q_2 \ge -\left(\frac{\Sigma_{21}^{tr}}{\Sigma_{22}^{tr}} \right) q_1.
\label{eq_ex_ineq2}
\ee
The $x$-axis of both
panels corresponds to $\theta=q_1$ and the $y$-axis represents $q_2$.
Each panel depicts a unit circle as well
as the locus $\phi_2'q= 0$.
In the left panel $\Sigma_{21}^{tr} <0$, whereas in the right panel $\Sigma_{21}^{tr} > 0$.
The identified set $F^{q}(\Phi_q)$ is given by the arc that ranges from the intersection
of the unit circle with the $y$-axis to the intersection with $\phi_2'q =0$.
The identified set $F^\theta(\Phi_q,\Phi_\theta)$ is given by the projection of the arc onto
the $x$-axis.  The identified set is a singleton only if $\Sigma_{22}^{tr}=0$ and {\color{black} $\Sigma_{21}^{tr}<0$},
which means that the one-step-ahead forecast error covariance matrix is singular. We will rule out this case
because in practice it is not empirically relevant.


\subsection{Rank Reductions and Notation}
\label{subsec_idea_rankreductions}

In order to develop a notation for the general inference problem we need to accommodate two types of rank reductions.
First, consider the matrix $\Phi_q$ for the bivariate VAR(0):
\[
    \Phi_q = \left[ \begin{array}{cc} \Sigma_{11}^{tr} & \Sigma_{21}^{tr} \\
                                       0               & \Sigma_{22}^{tr}  \end{array} \right].                                  
\]
To construct a confidence set for $q$, we will replace the unknown $\Phi_q$ by a sample estimate $\hat{\Phi}_q$ and start from the 
high-level assumption that $\hat{\Phi}_q$ has a normal limit distribution. Because of the zeros in the Cholesky factorization of $\Sigma_u$ (and possibly other restrictions imposed on the reduced-form parameters), the covariance matrix 
of $\hat{\Phi}_q$ has a rank-reduction. We circumvent this issues as follows. Let 
\[
\bar{S}' = \left[ \begin{array}{cccc} 1 & 0 & 0 & 0 \\ 0 & 0 & 1 & 0 \\ 0 & 0 & 0 & 1 \end{array} \right]
\]
be the selection matrix that deletes the zero elements in $\vect{(\Phi_q)}$ and such that
\ben
\phi_q = \bar{S}' \vect{(\Phi_q)} \quad \mbox{and} \quad \bar{S} \phi_q = \vect{(\Phi_q)}.
\een
Then, we can express the sign restrictions as 
\be
  \tilde{S}(q)\phi_q = \left( I \otimes q'\right) \bar{S} \phi_q \geq  0. \label{eq_idea_ineqcanonical.2}
\ee

The second type of rank reduction arises as follows. After eliminating $\Sigma_{12}^{tr}=0$, we obtain $\phi_q = [ \Sigma_{11}^{tr}, \Sigma_{21}^{tr}, \Sigma_{22}^{tr} ]'$ in the bivariate VAR(0). In turn, the matrix $\tilde{S}(q)$ takes the form:
\[
\tilde{S}(q)  =
\left[ \begin{array}{ccc} q_1 &  0  & 0 \\ 0 &  q_1 & q_2 \end{array} \right].
\]
The first row of $\tilde{S}(q)$ becomes zero if $q=[q_1,q_2]'=[0,1]$. 
As a consequence the covariance matrix of $\tilde{S}(q) \hat{\phi}_q$ is singular at $q=[0,1]'$ and cannot be inverted to form a weight matrix for the estimation of $q$. 
In order to eliminate the rows of zeros in $\tilde{S}(q)$, we introduce the selection matrix $V(q)$ and define
\be
S(q) = V(q) \tilde{S}(q).
\label{eq_sqvq}
\ee
The row dimension of $S(q)$ is  $r(q)$. By construction,
the matrix $S(q)$ has the full row rank for all $q$.

Inference about $q$ will be based on the objective function
\be
G(q;\phi_{q},W(\cdot))
= \min_{\mu \ge 0} \; \left\| S(q) \phi_{q} - V(q) \mu  \right\|^2_{W(q)}.
\label{eq_idea_popobjfcn}
\ee
The vector $\mu$ captures the slackness in the inequalities generated by the sign restrictions in (\ref{eq_idea_ineqcanonical.2}). $W(q)$ is a symmetric and positive-definite weight matrix with a dimension that adjusts to the dimension of $V(q)$:
\[
  W(q) = V(q)\tilde{W}(q)V(q)'.
\]
The matrix $\tilde{W}(q)$ is a weight matrix that conforms with $\tilde{S}(q) \phi_q$. An important example of a $\tilde{W}(q)$ is the inverse of the asymptotic covariance matrix of $\sqrt{T} \tilde{S}(q)( \hat{\phi}_q-\phi_q)$. It can be verified that 
\be
     q \in F^q(\phi_q) \quad \mbox{if and only if} \quad 
     G(q;\phi_{q},W(\cdot))=0,
\ee
where we now write $F^q(\phi_q)$ instead of $F(\Phi_q)$.

\subsection{A Bonferroni Confidence Interval for $\theta$}
\label{subsec_idea_bonferroni}

Let $S_{\theta}'$ be the selection matrix that deletes the zeros other pre-determined elements in $\vect{(\Phi_\theta)}$ and 
define $\phi_\theta = S_{\theta}' \vect{(\Phi_\theta)}$. Moreover, let $\phi = [\phi_q',\phi_\theta]'$. We will 
often write $F^\theta(\phi)$ to abbreviate $F^\theta(\Phi_q,\Phi_\theta)$.
The goal is to obtain a confidence interval $CS^\theta(\hat{\phi})$ that satisfies the condition
\be
\liminf_{T \longrightarrow \infty} \; \inf_{\rho \in {\cal R}} \;  \inf_{\theta \in F^\theta(\phi(\rho))} \;  P_\rho \big\{ \theta \in CS^\theta(\hat{\phi}) \} \ge 1 - \alpha.
\label{eq_csthetacond}
\ee
The vector $\phi$ may be a subvector of
a larger reduced-form parameter vector $\rho$ with domain ${\cal R}$, that characterizes the distribution of the data $y_1,\ldots,y_T$,
which is why we write $\phi(\rho)$. For instance, in a Gaussian VAR $\rho$ comprises the elements of $A_1$, $\ldots$, $A_p$, and $\Sigma_u$; see (\ref{eq_idea_varrf}).
The parameter $\theta$ does not appear as an index of the probability distribution
$P$, because conditional on $\phi$ the parameter $\theta$ does not affect the distribution of the estimator
of the reduced-form parameters $\hat{\phi}$. The confidence interval is indexed by $\hat{\phi}$ because
it is a sufficient statistic in our setup. 

As mentioned in the Introduction, in our application it is quite natural to use a Bonferroni approach to compute an asymptotic $1-\alpha$ confidence set for $\theta$. 
Under the Bonferroni approach, one first constructs a confidence set for $q$. This is a ``non-standard'' object because $q$ is a set-identified parameter. Conditional on $q$, however, inference for $\theta$ becomes ``standard,'' because $\theta$ is point-identified. 
Let $\alpha = \alpha_1+\alpha_2$. Bonferroni confidence steps can be obtained in three steps:

\begin{enumerate}
	\item Construct a $1-\alpha_1$ confidence interval $CS^q(\hat{\phi}_q)$
	for $q$ with the property that
	\be
	\liminf_{T \longrightarrow \infty} \; \inf_{\rho \in {\cal R}} \; \inf_{q \in F^q(\phi_q)} \; P_\rho \{ q \in CS^q(\hat{\phi}_q) \} \ge 1-\alpha_1.
	\label{eq_idea_csqcond}
	\ee
	
	\item Generate a confidence set for $\theta$ conditional
	on $q$. This is a ``regular'' problem because conditional on $q$, the vector $\theta$ is point identified. For instance, if $\theta$ is scalar and defined as $\theta=\phi_\theta'q$, then
	one can use a Wald confidence interval of the form:
	\be
	CS_{q}^\theta(\hat{\phi}_\theta) = \Theta \bigcap \bigg[ \hat{\phi}_\theta'q - z_{\alpha_2/2} \hat\sigma_{\hat{\theta}}, \; \hat{\phi}_\theta'q + z_{\alpha_2/2} \hat\sigma_{\hat{\theta}} \bigg].
	\label{eq_idea_csthq}
	\ee
	Here $z_{\alpha_2/2}$ is the two-sided $\alpha_2$ critical value associated with the $N(0,1)$ distribution and $\hat{\sigma}_{\hat{\theta}}$ is a standard error estimate for $\hat{\theta} = \hat{\phi}_\theta' q$. The intersection of the symmetric confidence interval for $\theta$ can be used, for instance, to truncate the symmetric interval at zero, if $\theta$ is a response that is assumed to be non-negative.
	
	\item Construct the confidence set for $\theta$ by taking the following union of $CS_{q}^\theta(\hat{\phi}_\theta)$ sets:
	\be
	CS^\theta(\hat{\phi}) = \bigcup_{q \in CS^q(\hat{\phi}_q)} \; CS^q_\theta(\hat{\phi}_\theta).
	\label{eq_excsBFtheta}
	\ee
\end{enumerate}

\subsection{A Confidence Set for $q$}
\label{subsec_idea_csq}

The main contribution of this paper is to adapt an inference procedure from the moment inequality literature to obtain a  confidence set for $q$ in the first step of the Bonferroni procedure. The confidence set is generated as a level set based on the sample analogue  
of the objective function in (\ref{eq_idea_popobjfcn}):
\be
CS^q(\hat{\phi}_q) =  \bigg\{ q \in \mathbb{S}^n \, \bigg| \, G\big(q;\hat{\phi}_q,W(\cdot)\big) \le c^{\alpha_1}(q) \bigg\}. 
\label{eq_idea_csq}
\ee 
Here $c^{\alpha_1}(q)$ is a critical value that guarantees that 
the confidence set satisfies (\ref{eq_idea_csqcond}).
In the remainder of this subsection, we outline the derivation of
the critical value $c^{\alpha_1}(q)$ for the bivariate VAR(0) example.

For illustrative purposes, suppose that the estimates of the reduced-form parameters have an exact standard normal distribution: 
\[
   \sqrt{T} ( \hat{\phi}_q - \phi_q ) \sim N \big(0,I_3\big).
\]
We parameterize the slackness in the inequality restrictions as
\[
  q_1 \phi_{q,1} = \tilde{\mu}_1, \quad q_1 \phi_{q,1} + q_2 \phi_{q,2} = \tilde{\mu}_2
\] 
and use the weight matrix that standardizes the distribution 
of $\tilde{S}(q) \hat{\phi}_q$:
\[
   W(q) = T \left[ \begin{array}{cc} \frac{1}{q_1^2} & 0 \\ 0 & 1 \end{array} \right]. 
\]
In turn, we can express the sample analogue of the objective function in (\ref{eq_idea_popobjfcn}) as
\begin{eqnarray*}
  G\big(q;\hat{\phi}_q,W(\cdot)\big)
  &=& \min_{\mu \ge 0} \quad \bigg[ \sqrt{T}(\hat{\phi}_{q,1}-\phi_{q,1}) + \sqrt{T} (\mu_1-\tilde\mu_1)/|q_1|  \bigg]^2 {\cal I} \big\{ q \not= [0,1]'   \big\} \\
  &&+ \bigg[ \sqrt{T}(\hat{\phi}_{q,2} - \phi_{q,2})q_1 + \sqrt{T}(\hat{\phi}_{q,3} - \phi_{q,3})q_2  + \sqrt{T}(\mu_2 - \tilde\mu_2) \bigg]^2 \\
  &=& \min_{\nu \ge -\sqrt{T} \tilde{\mu}} \quad \big[ Z_1 - \nu_1/|q_1| \big]^2 {\cal I}\big\{ q \not= [0,1]' \big\}
  + \big[ Z_2 - \nu_2 \big]^2,
\end{eqnarray*}  
where $\mu = [\mu_1,\mu_2]'$, $\nu = [ \nu_1,\nu_2]'$, and $Z_1$ and $Z_2$ are two independent $N(0,1)$ random variables. 

A conservative upper bound on the sample objective function can be 
obtained by assuming that both inequalities are binding, that is, $\tilde{\mu}_1 = \tilde{\mu}_2 = 0$:
\[
  G\big(q;\hat{\phi}_q,W(\cdot)\big) \le Z_1^2 {\cal I}\{ Z_1 \le 0 \} + Z_2^2 {\cal I} \{ Z_2 \le 0\}.
\]
Critical values for the distribution of the bound can be easily obtained by simulation. A sharper bound and a smaller critical value that leads to a smaller confidence set can be obtained by realizing that at most one inequality is binding. Thus, we will use the moment selection approach of \cite{AndrewsSoares2010a} to eliminate non-binding moment conditions constructing critical values for $G\big(q;\hat{\phi}_q,W(\cdot)\big)$. This means that in our illustrative example, the critical value can be essentially reduced to the 100$(1-\alpha_1)$ quantile of the distribution of $Z_1^2{\cal I}\{Z_1 \le 0\}$.

%
%

\section{Implementation}
\label{sec_implementation}

This section focuses on the implementation of the proposed inference methods. A formal statement of assumptions and a rigorous analysis of the large sample properties of the confidence set will follow in Section~\ref{sec_largesample}. The remainder of this section is organized as follows. Section~\ref{subsec_implementation_phihat}
briefly discusses the estimation of $\phi$. Section~\ref{subsec_implementation_csq} describes how we construct the confidence set for $q$. The calculation of confidence sets for $\theta$ given $q$ is reviewed in Section~\ref{subsec_implementation_csthetaq}. Finally, we provide some additional details for the computation of confidence bands for impulse responses in Section~\ref{subsec_implementation_irfbands}. 
Throughout this section we assume that the impulse responses are not restricted
through equality conditions (e.g., the restriction that certain responses have to be zero). Extensions of our approach to a setting
in which some identifying information is extracted from equality conditions are straightforward but notationally cumbersome and discussed in Section~\ref{sec_extensions}.

\subsection{Estimating the Reduced-Form Coefficients $\phi$}
\label{subsec_implementation_phihat}

We start from the assumption that 
$\hat{\phi}_q$ and $\hat{\phi}_\theta$ have  Gaussian limit distributions and that the asymptotic
covariance matrices can be estimated consistently:
\begin{eqnarray}
  \sqrt{T}(\hat{\phi}_q-\phi_q) \Longrightarrow N \big(0,\Lambda_{qq} \big)
  \quad \mbox{and} \quad \hat{\Lambda}_{qq} \stackrel{p}{\longrightarrow} \Lambda_{qq} > 0 
  \label{eq_phihatcovergence} \\
  \sqrt{T}(\hat{\phi}_\theta-\phi_\theta) \Longrightarrow N \big(0,\Lambda_{\theta \theta} \big)
  \quad \mbox{and} \quad \hat{\Lambda}_{\theta \theta} \stackrel{p}{\longrightarrow} \Lambda_{\theta \theta} > 0. \nonumber
\end{eqnarray}
This assumption
requires that all roots of the characteristic polynomial associated with the difference
equation~(\ref{eq_idea_varrf}) lie outside of the unit circle. Throughout the paper, we are ruling out the presence of unit roots and are assuming that $y_t$ is trend stationary.

{\color{black} We will also assume that $\Lambda_{qq}$  and $\Lambda_{\theta \theta}$ are full rank. Because most impulse response function confidence bands are pointwise, the dimension of $\theta$ is typically one, which immediately leads to $\Lambda_{\theta \theta} > 0$.
Whether or not $\Lambda_{qq} > 0$ is satisfied depends on the number of imposed sign restrictions.
In a first-order approximation, the reduced-form responses
stacked in $\phi_q$ are linear functions of the $n^2p+n(n+1)/2$ coefficients in $(A_1,\ldots,A_p,\Sigma_{tr})$. In order to restrict $r$ structural responses, we need 
at least $nr - n(n-1)/2$ reduced form responses (recall that $n(n-1)/2$ responses upon impact are zero because $\Sigma_{tr}$ is lower triangular). Thus, for $r> n(p+1)$ the matrix $\Lambda_{qq}$ cannot be of full rank. In practice, most applications will satisfy the rank condition because we previously eliminated the $n(n-1)/2$ zero elements of the lower triangular matrix $\Sigma_{tr}$ from the vector $\phi_q$ and the number of sign restrictions is small relative to the number of reduced-form VAR parameters.\footnote{ \color{black} 
Consider a 4-variable VAR(4) and suppose that the responses
of 3 of the 4 variables are restricted upon impact and for the subsequent 3 periods.
The number of estimated reduced-form coefficients is $4 \cdot 16 + 10 = 74$. In order
to construct the sign-restricted responses, the number of elements in the vector $\phi_q$ is bounded by $3 \cdot 4 \cdot4-3=45$ (because the impact effect of the structural shocks depends on a minimum of 3 reduced-form responses that are zero).}
}

The VAR
coefficient matrices $A_1,\ldots,A_p$ and $\Sigma_u$ can be estimated by OLS.
An estimate of $\Sigma_{tr}$ is obtained by applying the Cholesky decomposition
to $\hat{\Sigma}_u$. We then evaluate the functions $\Phi_q(\cdot)$ and $\Phi_\theta(\cdot)$ at  $\hat{A}_1,\ldots,\hat{A}_p,\hat{\Sigma}_{tr}$ to obtain $\hat{\phi}_q$ and $\hat{\phi}_\theta$.
We obtain $\hat{\Lambda}_{qq}$ and $\hat{\Lambda}_{\theta \theta}$ by using a parametric bootstrap procedure: conditional on $\hat{A}_1,\ldots,\hat{A}_p,\hat{\Sigma}_u$ we simulate $n_\Lambda$ bootstrap samples $Y^*_{1:T}$ from the VAR in~(\ref{eq_idea_varrf}). {\color{black}  Innovations
$u_t^*$ can either be drawn by resampling the residuals $\hat{u}_t$ or by $iid$ sampling from a $N(0,\hat{\Sigma}_u)$ distribution. In Sections~\ref{sec_mc} and~\ref{sec_empirical} we do the latter.} From each bootstrap
sample we compute $\hat{\phi}_q^*$. Finally we compute the bootstrap sample covariance matrix
of $\hat{\phi}_q^*$ and scale it appropriately to obtain $\hat{\Lambda}_{qq}$. The same approach is used to compute $\hat{\Lambda}_{\theta \theta}$.

%
%

\subsection{Confidence Set for $q$}
\label{subsec_implementation_csq}

The confidence interval for $q$ is obtained by verifying whether 
\[
   G\big(q;\hat{\phi}_q,W(\cdot)\big) \le c^{\alpha_1}(q)
\]
for $q \in {\cal Q}$. This requires the selection of a grid ${\cal Q}$ and the evaluation of the critical-value function $c^{\alpha_1}(q)$.

\noindent {\bf Generating a Grid for $q$.} 
We generate $n_{\cal Q}$ grid points for $q \in \mathbb{S}^n$ from a distribution that is uniform under rotations using a well-known result
by James (1954). Let $Z^{(j)}$, $j = 1,\ldots,n_{\cal Q}$, be a sequence of $n\times 1$
vectors of $iid N(0,I_n)$ random vectors and define $q^{(j)} = Z^{(j)} / \|Z^{(j)} \|$. Then,
$q^{(j)}$ is uniformly distributed on the unit sphere $\mathbb{S}^n$.
We define the grid as ${\cal Q} = \{q^{(1)},\ldots,q^{(n_{\cal Q})} \}$.
{\color{black}  For the confidence intervals to be asymptotically valid, the number of grid points has to expand faster than the sample size. Our theoretical analysis in Section~\ref{sec_largesample} abstracts from the discretization of $\mathbb{S}^n$.}

\noindent {\bf Weight Matrix for Sample Objective Function.} 
The weight matrix $\hat{W}_q(q)$ is obtained as follows.
We denote the asymptotic covariance matrix of $\sqrt{T}S(q)(\hat{\phi}_q-\phi_q)$ as $\Sigma(q)  = S(q)\Lambda_{qq} S(q)'$. 
A consistent estimator is given by
\[
   \hat{\Sigma}(q) = S(q) \hat{\Lambda}_{qq} S'(q) = \hat{D}^{1/2}(q) \hat{\Omega}(q) \hat{D}^{1/2}(q),
\]
where $\hat{\Omega}(q)$ is the correlation matrix associated with $\hat{\Sigma}(q)$
and $\hat{D}^{1/2}(q)$ is a diagonal matrix of standard deviations.
We then let
\be
    \hat{W}(q) = T \hat{D}^{-1/2}(q)\hat{B}(q) \hat{D}^{-1/2}(q)
\ee
and focus on two particular choices of $\hat{B}(q)$: $\hat{B}(q) = \hat{\Omega}^{-1}(q)$ and $\hat{B}(q)=I$. {\color{black} The choice of $\hat{B}(q) = \hat{\Omega}^{-1}(q)$ clearly requires our assumption in (\ref{eq_phihatcovergence}) that $\Lambda_{qq} > 0$. In the case of $\hat{B}(q)=I$, one could in principle allow for a singular covariance matrix $\Lambda_{qq} > 0$. In the formal analysis in Section~\ref{sec_largesample} one would have to replace $\Lambda_{qq}$ by its singular value decomposition. We did not pursue this extension below because it would make the notation and exposition more cumbersome.}

Overall, this leads to the sample objective function
\begin{eqnarray}
   G\big(q;\hat{\phi}_q,\hat{W}(\cdot)\big)
     &=& \min_{\mu \ge 0} \;
      T \left\| \hat{D}^{-1/2}(q) S(q) \hat{\phi}_q - \hat{D}^{-1/2}(q)V(q) \mu  \right\|_{\hat{B}(q)}^2 \label{eq_gobjdef} .
\nonumber
\end{eqnarray}
The  function
$G\big(q;\hat{\phi}_q,\hat{W}(\cdot)\big)$ has the same structure as the objective functions considered in the literature on moment inequality models, e.g.,
\cite{ChernozhukovHongTamer2007}, \cite{Rosen2008}, \cite{AndrewsGuggenberger2009}, and \cite{AndrewsSoares2010a}.
The main difference of our set up for the VAR application is 
the dimension of $S(q)\hat{\phi}_q$, $\hat{D}(q)$ and $\hat{B}(q)$ varies with $q$
and the limit $\hat{D}(q)$ as a function of $q$ can be singular.

\noindent {\bf Critical Values.}
In order to obtain the critical value function $c^{\alpha_1}(q)$ we apply the moment selection
approach of \cite{AndrewsSoares2010a}. The moment selection tries to eliminate clearly non-binding inequality conditions in the weak limit of the objective function $G\big(q;\hat{\phi}_q,\hat{W}(\cdot)\big)$ and compute the required critical value.\footnote{One can show that in population at most $n-1$ inequality conditions (recall that $n$ is the dimension of $y_t$) can be binding; see also the example in Section~\ref{subsec_idea_bonferroni}. Thus, we experimented with an algorithm that orders the inequalities based on the strength of their violation to select a subset of $n-1$ binding conditions in case the Andrews-Soares procedure classifies more than $n-1$ inequalities as binding. In finite samples, the selection of no more than $n-1$ restrictions could potentially sharpen the confidence set for $q$. However, in our experiments the gains (if any) were so small that they were essentially not noticeable when we constructed the confidence bands for $\theta$. Thus, we did not explore this idea further in this paper.}
An estimate of the slackness in inequality condition $j=1,\ldots,r(q)$ is provided by
\be
   \hat{\xi}_{j,T}(q) = \hat{D}_{jj}^{-1/2}(q)  [S(q)]_{(j.)}  \sqrt{T} \hat{\phi}_q.
   \label{eq_xijT}
\ee
Inequality condition $j$ is deemed non-binding if
\be
    \hat{\xi}_{j,T}(q) \ge \kappa_T,
    \label{eq_xijTcond}
\ee
where $\kappa_T$ is a sequence that diverges \emph{slowly} to infinity, e.g., $\kappa_T = 1.96 \ln (\ln T)$.
Thus, estimates of the number of non-binding and binding moment inequality constraints are given by
\be
   \hat{r}_2(q) = \sum_{j=1}^{r(q)} {\cal I} \{ \hat{\xi}_{j,T}(q) \ge \kappa_T \}
   \quad \mbox{and} \quad \hat{r}_{1}(q) = r(q) - \hat{r}_{2}(q),
\ee
respectively.

Define the $(\hat{r}_{1}(q) \times r(q))$ selection
matrix $M_{\hat{\xi}}(q)$ that deletes rows of $\hat{D}^{-1/2}(q)S(q)\hat{\phi}_q$
that correspond to non-binding inequality conditions in the sense of~(\ref{eq_xijTcond}).
Moreover, let $m$ be the dimension of the vector $\phi_q$,
\[
    Z_m \sim N(0,I_m), \quad \mbox{and} \quad \hat{A}'(q) = \hat{D}^{-1/2}(q) S(q) \hat{L}_{qq}, \quad \mbox{where} \quad \hat{\Lambda}_{qq} = \hat{L}_{qq} \hat{L}_{qq}'.
\]
{\color{black} Conditional on $\hat{B}(q)$ and $M_{\hat{\xi}}(q)$, define the random function
	(the randomness is induced through $Z_m$)}
\be
   \bar{\cal G}_{ \color{black} Z_m} \big(q;\hat{B}(q),M_{\hat{\xi}}(q) \big)
      = \min_{\nu \ge 0} \; \left\| M_{\hat{\xi}}(q) \hat{A}'(q) Z_m
              - \nu \right\|^2_{M_{\hat{\xi}}(q) \hat{B}(q) M_{\hat{\xi}}'(q) },
              \label{eq_calgbar}
\ee
where $\nu$ is a $\hat{r}_{1}(q) \times 1$ vector. We adopt the convention that
$\bar{\cal G}_{Z_m} \big(q;\hat{B}(q),M_{\hat{\xi}}(q) \big) = 0$ if $\hat{r}_{1}(q)=0$.
The critical value $c^{\alpha_1}(q)$ is defined as
\be
    c^{\alpha_1}(q) =  1-\alpha_1 \; \mbox{quantile of } \bar{\cal G}_{Z_m} \big(q;\hat{B}(q),M_{\hat{\xi}}(q) \big)
    \label{eq_cqalpha}
\ee
and can be obtained from a simulation approximation of the limit objective function.\footnote{
For $j=1,\ldots,n_Z$ generate random vectors $Z_m^{(j)}$ and compute $\bar{\cal G}_{Z_m}^{(j)} \big(q;\hat{B}(q),M_{\hat{\xi}}(q) \big)$. Then compute the $1-\alpha_1$ percentile of the empirical distribution of the simulated limit objective functions.} If $\hat{B}(q) = I$, then the evaluation of $\bar{\cal G}_{Z_m}(\cdot)$ is fast because the quadratic programming problem has the following closed-form solution: 
\begin{eqnarray}
\bar{\cal G}_{Z_m} \big(q;\hat{B}(q),M_{\hat{\xi}}(q) \big)
&=&  \sum_{j=1}^{\hat{r}_{1}(q)} [ M_{\hat{\xi}}(q) \hat{A}'(q) Z_m ]_j^2 {\cal I} \big\{ [ M_{\hat{\xi}}(q) \hat{A}'(q) Z_m ]_j < 0 \big\}. \nonumber
\end{eqnarray}

\subsection{Confidence Set for $\theta$ Conditional on $q$} 
\label{subsec_implementation_csthetaq}

Conditional on $q$, the dynamic effects of the shock $\epsilon_{1,t}$ on $y_t$ are point-identified and the inference about impulse responses and variance decompositions becomes regular. Methods on how to construct confidence intervals for these objects date back to \cite{Runkle1987}, who proposed to use either the $\delta$-method in combination with numerical derivatives of the mapping from reduced-form VAR coefficients into IRFs and variance decompositions or to use a residual-based bootstrap. \cite{Luetkepohl1990} derived asymptotic distributions based on analytical derivatives for the $\delta$-method and \cite{MittnikZadrzny1993} provided extensions to VARMA models. A recent survey of the literature on frequentist inference for IRFs and variance decompositions in point-identified settings is provided by \cite{KilianLuetkepohl2017}. Any of these methods can be embedded into our Bonferroni approach.

%
%

\subsection{Special Case: Confidence Bands for IRFs}
\label{subsec_implementation_irfbands}

Confidence bands for impulse responses in the VAR literature predominantly depict pointwise confidence intervals, which means that
we can express the scalar parameter $\theta$ as $\tilde{S}_\theta(q) \phi_\theta$, where $\phi_\theta$ summarizes the reduced-form impulse responses that are necessary to generate the structural response $\theta$ and $\tilde{S}_\theta(q)$ is defined similarly as $\tilde{S}(q)$ in Section~\ref{subsec_idea_rankreductions}. For this important special case one can show that $F^\theta(\phi_q,\phi_\theta)$ is convex and bounded, which simplifies computations and reporting of results.

\begin{lemma}
	\label{l_boundedsets}
	Suppose that $F^q(\phi_q)$ is non-empty and not a singleton. Moreover $\theta = S_\theta(q)\phi_\theta$ and $k=\mbox{dim}(\theta)=1$. Then $F^\theta(\phi_q,\phi_\theta)$ is convex and bounded.
\end{lemma}

\noindent {\bf Approximating $F^q(\hat{\phi}_q)$ and $F^\theta(\hat{\phi})$.} An estimate of the identified set for $q$
can be obtained from
\[
F^q(\hat{\phi}_q) \approx \hat{F}^q(\hat{\phi}_q)= \big\{ q \in {\cal Q} \; | \; \tilde{S}(q) \hat{\phi}_q \ge 0 \big\}.
\]
Thus, for every $q \in {\cal Q}$ one checks whether $\tilde{S}(q) \hat{\phi}_q \ge 0$ and retains the $q$'s for which the condition is satisfied. Denote the elements of $\hat{F}^q(\hat{\phi}_q)$
by $q^{(j)}, \; j=1,\ldots,n_q$, where $n_q \le n_{\cal Q}$. Compute $\theta^{(j)} = \hat{\phi}_\theta'q^{(j)}$.
We show in the Online Appendix that $F^\theta(\hat{\phi})$ is a bounded interval.
Thus, we define the interval
\[
\hat{F}^\theta(\hat{\phi}) = \big[ ( \min_{j=1,\ldots,n_q} \theta^{(j)}) ,\; ( \max_{j=1,\ldots,n_q} \theta^{(j)}) \big].
\]

\noindent {\bf Computing $CS^q(\hat{\phi}_q)$ and $CS^\theta(\hat{\phi})$}. The computation of $CS^q(\hat{\phi}_q)$ follows the steps outlined in Section~\ref{subsec_implementation_csq}.
Note that by construction $\hat{F}^q(\hat{\phi}_q) \subseteq CS^q(\hat{\phi}_q)$. Denote the elements of $CS^q(\hat{\phi}_q)$
by $q^{(j)}, \; j=1,\ldots,n_q$. Then, for each $q^{(j)}$, compute 
the Wald interval with bounds
\[
   \theta^{(j)}_l = \hat{\phi}_\theta'q^{(j)} - z_{\alpha_2/2} \sqrt{q^{(j)'} \hat{\Lambda}_{\theta \theta} q^{(j)}/T} \quad \mbox{and} \quad 
   \theta^{(j)}_u = \hat{\phi}_\theta'q^{(j)} + z_{\alpha_2/2} \sqrt{q^{(j)'} \hat{\Lambda}_{\theta \theta} q^{(j)}/T},
\]
and let
\[
   CS^\theta(\hat{\phi}) = \Theta \bigcap \big[ ( \min_{j=1,\ldots,n_q} \theta^{(j)}_l) ,\; ( \max_{j=1,\ldots,n_q} \theta_u^{(j)}) \big].
\]
The intersection with $\Theta$ can be used to restrict the confidence interval to values of $\theta$ that are consistent with the assumed sign restriction.
While $CS^q(\hat{\phi}_q)$ has to be calculated only once, the computations for $CS^\theta(\hat{\phi})$ 
have to be repeated
for every response $\theta=\partial y_{i,t+h}/\partial \epsilon_{1,t}$ of interest. Here $i$ potentially
ranges from $i=1,\ldots,n$ and $h=0,1,\ldots, h_{max}$.

\section{Large Sample Analysis}
\label{sec_largesample}

This section formally establishes the consistency of the plug-in estimators $F^q(\hat{\phi}_q)$ and $F^\theta(\hat{\phi})$ and the asymptotic validity of the confidence sets $CS^q(\hat{\phi}_q)$ and $CS^\theta(\hat{\phi})$. As mentioned in Section~\ref{subsec_idea_bonferroni}, the vectors $\phi_q$ and $\phi$ may generally not be sufficient to characterize the sampling distribution of data and estimators. Thus, we again will use $\rho$  to characterize the distribution of the data under the reduced-form VAR model~(\ref{eq_idea_varrf}). {\color{black}  We denote this distribution by $P_\rho$.}
The statements about uniform asymptotic coverage probabilities will be made for $\rho \in {\cal R}$. Some of the regularity conditions will be required to hold for a slightly larger, $\delta$-inflated open set
\be
{\cal R}^\delta = \big\{ \tilde{\rho} \in \bar{\cal R} \; \big| \; \exists \rho \in {\cal R} \; \mbox{s.t.} \| \tilde{\rho} - \rho \| < \delta \big\},
\label{eq_calRdelta}
\ee
where $\bar{\cal R} \supset {\cal R}$ and $\delta>0$.\footnote{For instance, suppose $\rho$ is an autocorrelation parameter for an AR(1) model. We could define ${\cal R} = [0,\,0.999]$, ${\cal R}^\delta = [0,1)$ for $\delta=0.001$, and $\bar{\cal R} = [0,1]$.}
Asymptotic inference for $q$ is discussed in Section~\ref{subsec_largesample_q} and Section~\ref{subsec_largesample_theta} considers inference for $\theta$. 

\subsection{Asymptotic Inference for $q$} 
\label{subsec_largesample_q}

We begin by stating some high-level assumptions.

\begin{assumption} \label{a_all}
There exists a compact reduced-form parameter set ${\cal R}$ and a $\delta$-inflated superset
${\cal R}^\delta$ defined in~(\ref{eq_calRdelta})
{\color{black} such that $ {\cal R} \subset {\cal R}^\delta \subset \bar{\cal R}$ and}:
\begin{tlist}
\item For every $\rho \in {\cal R}^\delta$, there does not exist an $r \times 1$ vector $\lambda > 0$ such that
      \[
          \Phi_q \lambda = 0.
      \]
\item $\phi_q(\rho)$ is continuously differentiable for all $\rho \in {\cal R}^\delta$.
\item There exists an estimator $\hat{\phi}_{q}$ of $\phi_q(\rho_T)$ and a matrix $\Lambda_{qq}^{-1/2}(\rho_T)$
      such that for each sequence $\{\rho_T\} \in {\cal R}$
      (a) $ \hat{\phi}_{q} - \phi_q(\rho_T)  \stackrel{p}{\longrightarrow} 0$; (b)
      $\sqrt{T} \Lambda_{qq}^{-1/2}(\rho_T) (\hat{\phi}_{q} - \phi_q(\rho_T)) \Longrightarrow N(0,I)$.
\item For each $\rho \in {\cal R}$ the matrix $\Lambda_{qq}(\rho)$ is continuous, positive definite, and there
      exists a  full-rank positive-definite matrix $\Lambda_{min}$ such that $\Lambda_{qq}(\rho) - \Lambda_{min} \ge 0$ for
      all $\rho \in {\cal R}$.
\item There exists an estimator $\hat{\Lambda}_{qq}$ of $\Lambda_{qq}(\rho_T)$ such that
      $\|\hat{\Lambda}_{qq} - \Lambda_{qq}(\rho_T)  \| \stackrel{p}{\longrightarrow} 0$ for any converging sequence $\{ \rho_T\} {\color{black}\in {\cal R}}$.
\end{tlist}
\end{assumption}

Condition~(i) of Assumption~\ref{a_all} states that the convex cone generated
by the columns of the reduced-form impulse response matrix $\Phi_q$
does not contain the zero vector.  This assumption is sufficient to ensure that
the identified set $F^q(\phi_q(\rho))$ is non-empty and that the plug-in estimator $F^q(\hat{\phi}_{q})$ is consistent
whenever $\hat{\phi}_{q} \stackrel{p}{\longrightarrow} \phi_q$ (see Theorem~\ref{t_setidentified} below).
Assumption~\ref{a_all}(i) rules out, for instance, that equality conditions
are coded as pairs of inequalities, and, more generally, that
linear combinations of inequalities constrain impulse responses to be equal to zero.
We discuss in Section~\ref{sec_extensions}
how our framework can be extended to allow for a mixture of inequality and equality restrictions
on impulse responses.

Condition~(i) is typically not satisfied for all values of the
reduced-form parameter $\rho \in \bar{\cal R}$, which is why we only require it to hold on
the set ${\cal R}^\delta \subset \bar{\cal R}$. For instance, consider a VAR(1) generalization of the bivariate VAR(0)
in Section~\ref{subsec_idea_ex} with autoregressive coefficient matrix $A_1$. As before, suppose $y_t$
is composed of inflation and output growth and the investigator imposes the sign restriction
that in response to a (positive) demand shock inflation and output responses are both non-negative
upon impact and one period after impact. In this case
\[
   \Phi_q' = \left[ \begin{array}{c} \Sigma_{tr} \\ A_1 \Sigma_{tr} \end{array} \right].
\]
If $A_1 = diag(\rho_1,\rho_2)$ and $\rho_1,\rho_2 < 0$, then
Condition~(i) is violated. Conditional on these reduced-form parameters, the identified set is empty.
Assumption~\ref{a_all} excludes these values of $\rho$ from ${\cal R}^\delta$.
From a practitioner's perspective, an empty confidence set $CS^q(\hat{\phi}_{q})$ provides
evidence that the imposed sign restrictions are inconsistent with the estimated
reduced-form parameters.

The continuity in Condition (ii) is with respect to the Euclidean norm. While $\rho$ could in principle be infinite-dimensional if the distribution of the error terms is treated nonparametrically, the function $\phi_q(\cdot)$ only depends on the finite-dimensional subvector of $\rho$ that contains the reduced-form parameters $A_1,\ldots,A_p,\Sigma_{u}$; see Equation (\ref{eq_idea_irf}). {\color{black} In combination with the compactness of ${\cal R}$,  Condition~(ii) implies that the domain of $\phi_q$, which is given by $\{\phi_q(\rho) : \rho \in \mathcal{R} \}$,  is compact.} Conditions~(iii) and~(v) require that $\hat{\phi}_q$
and $\hat{\Lambda}_{qq}$ converge uniformly for $\rho \in {\cal R}$.
{\color{black} Note that the stated convergences in probability and in distribution are assumed to hold 
	under the sequence of distributions $P_{\rho_T}$.}\footnote{ \color{black} E.g., 
	$\| \hat{\phi}_{q} - \phi_q(\rho_T) \| \stackrel{p}{\longrightarrow} 0$ is shorthand for 
	$P_{\rho_T} \{ \| \hat{\phi}_{q} - \phi_q(\rho_T) \| > \epsilon \} \longrightarrow 0$ as $T \longrightarrow \infty$ for any $\epsilon > 0$.}
The uniform convergence of $\hat{\phi}_{q}$ to a Gaussian limit distribution
also requires a restriction of the domain of $\rho$
because it breaks down at the boundary of the stationary region
in the VAR parameter space. For instance, in the context of an AR(1) model $y_t = \rho_T y_{t-1} + u_t$ with autoregressive coefficient
$\rho_T = 1-c/T$, an estimator of an impulse response at horizon $h=1$, that is, $\phi_q(\rho)=\rho$,
behaves according to
\[
   \sqrt{T} \big(1-\rho_T^2\big)^{-1/2} (\hat{\phi}_T - \rho_T)
   = \frac{ \frac{1}{T} \sum y_{t-1} u_t}{ \sqrt{c(2-c/T)}\frac{1}{T^2} \sum y_{t-1}^2} \not\Longrightarrow N(0,1).
\]
Uniform convergence to a Gaussian limit distribution can be achieved
if ${\cal R}$ is restricted to the interval $[-1+\epsilon, \; 1 - \epsilon]$ for some $\epsilon > 0$.\footnote{See
\cite{GiraitisPhillips2004} for a more general discussion.}
From a practitioner's perspective we are essentially assuming that
the researcher has applied some stationarity-inducing transformations, e.g.,
transformed prices into inflation rates. Because some authors, e.g.,
\cite{Uhlig2005}, prefer to specify  VARs in terms of variables that exhibit (near) non-stationary dynamics,
our Monte Carlo experiments in Section~\ref{sec_mc} include designs in which the roots of the
vector autoregressive lag polynomial are close to the unit circle.\footnote{An
extension of our analysis to VARs with unit
roots or cointegration restrictions
is beyond the scope of this paper. The construction
of uniformly valid confidence intervals for reduced-form parameters
in itself is a very challenging task; see \cite{Mikusheva2007}.}

Our first theorem establishes that the identified set $F^q(\phi_q)$ is non-empty and not a singleton, that is, the dynamic effects of $\epsilon_{1,t}$ are set-identified instead of point-identified.
This result can be deduced from Assumption~\ref{a_all}(i) using Gordan's Alternative Theorem (see, for instance, Border (2007)).

\begin{theorem}
\label{t_setidentified}
Suppose Assumption~\ref{a_all}(i) is satisfied. Then, the identified set
$F^q(\phi_q(\rho))$ is non-empty and is not a singleton for all $\rho \in \cal{R}^\delta$.
\end{theorem}

The second theorem focuses on asymptotic inference. The first part establishes the consistency of the plug-in
estimator $F^q(\hat{\phi}_q)$.
The consistency is stated in terms
of the Haussdorf distance. We denote the Hausdorff distance
between two sets $A$ and $B$ by $d_{H}\left(A,B \right)$.\footnote{Formally,
	the Hausdorff distance is defined as
	$d( A,B) = \max \; \left\{ d( A|B), \; d(B|A) \right\}$, where
	$d(A|B) = \sup_{a\in A} \; d( a,B)$ and $d(a,B)=\inf_{b\in B}\; \left\Vert a-b\right\Vert$.
	We set $d( A,B) = \infty$ if either $A$ or $B$ is empty.}
The consistency relies on the compactness of $F^q(\phi_q)$ and the continuity
of the correspondences with respect to $\phi_q$.
Unlike in some of
the models studied by \cite{ChernozhukovHongTamer2007}, it is not necessary to inflate the set $F^q(\hat{\phi}_q)$ by $\epsilon_T \downarrow 0$
to achieve consistency.\footnote{A result similar to ours in a general GMM setting is provided by \cite{Yildiz2012}. We prove the result directly based on Assumption~\ref{a_all}.} The second part of Theorem~\ref{t_asymptotics_q} establishes the asymptotic validity of the confidence set $CS^q(\hat{\phi}_q)$. A formal proof of the Theorem is provided in the Online Appendix. The proof of the second part closely follows the proof of Theorem~1 in \cite{AndrewsSoares2010a}.
However, a number of non-trivial modifications are required to account for the
potential rank reduction of $\tilde{S}(q)$ as a function of $q$.

\begin{theorem}
\label{t_asymptotics_q}
{\color{black}Suppose that Assumption~\ref{a_all} is satisfied.
(i) Then $d_{H}\big( F^{q}(\hat{\phi}_q),F^{q}(\phi_q)
\big) \stackrel{p}{\longrightarrow} 0.$ (ii) If $0 < \alpha < 1/2$, then the confidence set $CS^q(\hat{\phi}_q)$, defined in~(\ref{eq_idea_csq}),
is an asymptotically valid confidence set for $q$:}
\[
	\liminf_{T \longrightarrow \infty} \; \inf_{\rho \in {\cal R}} \; \inf_{q \in F^q(\phi_q(\rho))} \; P_\rho \{ q \in CS^q(\hat{\phi}_q) \} \ge 1-\alpha_.
\]
\end{theorem}

\subsection{Asymptotic Inference for $\theta$} 
\label{subsec_largesample_theta}

As discussed in Section~\ref{subsec_implementation_csthetaq}, we do not provide any new results on confidence intervals for impulse responses or variance decompositions {\em conditional on} the vector $q$. For these intervals, we rely on the existing literature. 
We use $CS^\theta_q(\hat{\phi}_\theta)$ to denote a confidence set for $\theta$ conditional on $q$. The following assumption is required for asymptotic inference about the parameter $\theta$.

\begin{assumption}
	\label{a_csthq} (i) The function $\theta = f(\Phi_\theta,q)$ is continuous in both its arguments.
	(ii) The set $CS^\theta_q(\hat{\phi}_\theta)$ satisfies:
	\[
	\lim \inf_{T}\inf_{\rho \in {\cal R}} \; \inf_{(\theta,q) \in F^{\theta,q}\left( \phi(\rho) \right)} P_{\rho}\left \{ \theta \in CS_{q}^{\theta}(\hat{\phi}_{\theta}) \right \} \geq 1-\alpha_{2},
	\]
where $F^{\theta,q}(\phi) =
\big\{\theta \in \Theta, q \in \mathbb{S}_n\; \big| \: q \in F^q(\phi_q), \, \theta = f(\Phi_\theta,q) \big\}$.
\end{assumption}

The first condition of Assumption \ref{a_csthq}(i) is quite weak and the two leading examples of $\theta$ in Section~\ref{subsec_idea_setid} satisfy this condition. 
The second condition of Assumption \ref{a_csthq}(ii) is a high-level condition that is needed for the asymptotic validify of the Bonferroni confidence set of $\theta$. The condition requires that the pointwise confidence set $CS_{q}^{\theta}(\hat{\phi}_{\theta})$ in $q$ is uniformly valid. In the leading examples of $\theta$, an impulse response of the form $\theta = \Phi_{\theta}'q$, the conditional confidence set $CS_{q}^{\theta}$ in (\ref{eq_idea_csthq}) satisfies the uniformity condition because the asymptotic normality of $\hat{\phi}_{\theta}$ in (\ref{eq_phihatcovergence}) holds uniformly in $\phi$. Combining the results of Theorem~\ref{t_asymptotics_q}(ii) with Assumption~\ref{a_csthq} leads to the following theorem:

\begin{theorem}
	\label{t_asymptotics_th}
	{\color{black} Suppose that Assumption~\ref{a_all} is satisfied.
	(i) {\color{black} If $\hat{\phi}_\theta \stackrel{p}{\longrightarrow} \phi_\theta$}, then $d_{H}\big( F^{\theta}(\hat{\phi}),F^{\theta}(\phi)
	\big) \stackrel{p}{\longrightarrow} 0$, where $\phi =[\phi_q',\phi_\theta']'$. (ii) Suppose that $0 < \alpha < 1/2$ and Assumption~\ref{a_csthq} is satisfied. Then the confidence set $CS^\theta(\hat{\phi})$, defined in~(\ref{eq_excsBFtheta}),
	is an asymptotically valid confidence set for $\theta$:}
	\[
\liminf_{T \longrightarrow \infty} \; \inf_{\rho \in {\cal R}} \;  \inf_{\theta \in F^\theta(\phi(\rho))} \;  P_\rho \big\{ \theta \in CS^\theta(\hat{\phi}) \} \ge 1 - \alpha.
	\]
\end{theorem}

\section{Extensions}
\label{sec_extensions}

We now discuss three extensions to the construction of $CS^q(\hat{\phi}_q)$: (i) models that use
both sign restrictions and zero restrictions to identify structural impulse responses, (ii) the identification
of multiple shocks, (iii) and the use of bootstrapped critical values instead of simulated asymptotic critical values.

\noindent {\bf Sign Restrictions Combined with Equality Restrictions.} Assumption~\ref{a_all}(i)
rules out that opposing sign restrictions are used to represent equality restrictions
on impulse responses. Nonetheless,
it is straightforward to sharpen the identified set by combining sign restrictions
with more traditional exclusion restrictions. In some applications,  the restriction that certain responses are zero on impact (zero restrictions) can be translated into a domain  restriction for $q$ that does not depend on any other reduced-form parameters. 
For instance, in the empirical analysis in Section~\ref{subsec_signzero} we will replace an unrestriced $4\times 1$ vector$q \in \mathbb{S}^n$  by the restricted vector $q=[0_{1 \times 2},q_2']'$, where $q_2$ is a $2 \times 1$ vector with $\|q_2\|=1$. In this case the previously developed methods can be applied without any modification.  

If the equality restrictions imposed on the impulse responses lead to restrictions on $q$ that depend on some of the reduced-form parameters, then they can be accommodated by generalizing the objective function $G(q;\phi,\tilde{W})$
in~(\ref{eq_idea_popobjfcn}) as follows. Define
\be
   \tilde{G}(q;\phi,\tilde{W}) = \min_{\mu \ge 0} \;
                      \left\| \left( \begin{array}{c}   S_{eq}(q)\phi \\
                                                        S(q) \phi - V(q) \mu
                                     \end{array} \right) \right\|^2_{\tilde{W}(q)},
   \label{eq_gpopdefeq}
\ee
where $S_{eq}(q) \phi$ corresponds to the responses that are restricted to be zero. Following the arguments in \cite{AndrewsSoares2010a}, it is straightforward albeit tedious to extend the proof of Theorem~\ref{t_asymptotics_q} to a mixture of equality and inequality conditions.\footnote{If we denote the matrix of zero-restricted orthogonalized responses
	by $\Phi_{q,eq}$, then the generalization of Assumption~\ref{a_all}(i) is: there do not
	exist vectors $\lambda > 0$ and $\lambda_{eq} \ge 0$ such that $\Phi_q \lambda + \Phi_{q,eq}\lambda_{eq} = 0$.
	The generalized analysis would use Motzkin's Transposition Theorem; see \cite{Border2007}.} The extension closely resembles the proof of Theorem 2(i) in the working paper version \cite{GSM2013} for the projection-based confidence set, which also involves a mix of equality and inequality conditions.
From a practioners perspective, the only other modification that is required, is to replace the limit objective function $\bar{\cal G}(\cdot)$ in (\ref{eq_calgbar}) that is used to simulate the critical value $c^{\alpha_1}(q)$ by the limit expression of $ \tilde{G}(q;\phi,\tilde{W})$ in (\ref{eq_gpopdefeq}).

\noindent {\bf Identifying Multiple Shocks.} Some authors use sign-restricted
SVARs to identify multiple shocks simultaneously. For instance, \cite{Peersman2005}
considers an $n=4$ dimensional VAR, composed of oil price inflation, output growth,
consumer price inflation, and nominal interest rates. He uses sign restrictions to identify
an oil price shock, aggregate demand and supply shocks, and a monetary policy shock.
To identify $n$ shocks, the unit vector $q$ has to be replaced by an orthogonal matrix,
and the restrictions will take the form
\[
     \tilde{S}(\Omega) \phi_q \ge 0
\]
for a suitably defined function  $\tilde{S}(\Omega)$. While all our
results easily generalize to multiple shocks (just replace $q$ by $\Omega$), the implementation becomes
computationally more difficult because the grid for the $n-1$ dimensional vector $q$ has to be replaced by a grid for
orthogonal matrix $\Omega$, which has $n(n-1)/2$ degrees of freedom.

\noindent {\bf Bootstrapped Critical Values Instead of Asymptotic Critical Values.} Our
simulated critical values rely on the Gaussian limit distribution of
$\sqrt{T} \hat{D}^{-1/2}(q) S(q) (\hat{\phi}_q - \phi_q)$, which is reflected in the vector $\hat{A}'(q)Z_m$
in the random function $\bar{\cal G}(\cdot)$ in~(\ref{eq_calgbar}). Alternatively, the
critical values could be constructed by replacing draws from $\hat{A}'(q)Z_m$
with draws from the bootstrap approximation of $\sqrt{T} \hat{D}^{-1/2}(q) S(q) (\hat{\phi}_q - \phi_q)$.
Bootstrap procedures for VAR impulse response functions are discussed, for instance, in \cite{Kilian1998} and \cite{KilianLuetkepohl2017}.

%
%

\section{Monte Carlo Illustrations}
\label{sec_mc}


In this section we conduct three Monte Carlo experiments to illustrate the properties of our proposed confidence sets. In these experiments $\theta$ is a scalar impulse response. 
During preliminary computations we noticed that the results for $\hat{B}(q) = I$ and $\hat{B}(q) = \hat{\Omega}^{-1}(q)$ were very similar.
Thus, we decided to subsequently report results for $\hat{B}(q) = I$ because in this case the critical values can be computed much faster. We will drop the $\hat{\phi}$ arguments from the confidence sets and report coverage probabilities and average lengths for $CS^q$ and $CS^\theta$.
Each Monte Carlo experiment involves the steps summarized in Table~\ref{t_mcsteps}, which are repeated $n_{sim}=5,000$ times.

\begin{table}
	\caption{Steps of Monte Carlo Experiments}
	\label{t_mcsteps}
	\begin{center}
		\begin{tabular}{ll} \hline \hline
			1. & Generate a sample of size $T$ from the data-generating process. \\
			2. & Compute $\hat{\phi}_q$, $\hat{\phi}_\theta$, and the bounds of $F^\theta(\hat{\phi}_q,\hat{\phi}_\theta)$. \\
			3. & Compute $\hat{\Lambda}_{qq}$ and $\hat{\Lambda}_{\theta \theta}$  using a parametric bootstrap approach. \\
			4. & Compute the $1-\alpha_1$ confidence set $CS^q$. \\
			5. & For each definition of $\theta$, compute the $1-\alpha_2$ confidence sets $CS^\theta_q$. \\
			6. & For each definition of $\theta$, compute the  $1-(\alpha_1+\alpha_2)$ confidence sets $CS^\theta$. \\
			 \hline
		\end{tabular}
	\end{center}
\end{table}

The three experiments differ with respect to the data generating process (DGP). Experiment~1 (Section~\ref{subsec_mc_exp1}) is based on the bivariate VAR(0) model in Section~\ref{subsec_idea_ex}. Experiment~2 (Section~\ref{subsec_mc_exp2}) features a bivariate VAR(1). The simulation designs for the Experiments~1 and~2
are obtained by fitting a VAR(0)
to data on U.S. inflation and GDP growth and
fitting first-order VARs to inflation and either output growth or linearly detrended log GDP. 
Finally, Experiment~3 (Section~\ref{subsec_mc_exp3}) mimics the four-variable VAR(2) fitted to U.S. data on output, inflation, interest rates, and money balances in the empirical analysis of Section~\ref{sec_empirical}.

\begin{table}[t!]
	\caption{Monte Carlo Design}
	\label{t_mcdesign}
	\begin{center}
		\begin{tabular}{ccccc} \hline \hline
			& Experiment 1   & \multicolumn{3}{c}{Experiment 2}\\
			& Design 1 & Design 2 & Design 3    & Design 4 \\
			& VAR(0)   & VAR(1)   & VAR(1)      & VAR(1)   \\ \hline
			$\Sigma_{11}^{tr}$       & 0.597  & 0.295  & 0283           & 0.210  \\
			$\Sigma_{21}^{tr}$       &-0.205  & -0.092 & -0.081          & -0.043 \\
			$\Sigma_{22}^{tr}$       & 0.812  & 0.795  & 0.817           & 0.542  \\ \hline
			$A_{1,11}$         &        & 0.873  & 0.806           & 0.450  \\
			$A_{1,12}$         &        & 0.003  & 0.032           & 0.014  \\
			$A_{1,21}$         &        & -0.229 & -0.278          & 0.060  \\
			$A_{1,22}$         &        & 0.230  & 0.985           & 0.953  \\ \hline
			$\lambda_1(A_1)$ &        & 0.871  & $0.89 - 0.03i$  & 0.955  \\
			$\lambda_2(A_1)$ &        & 0.231  & $0.89 + 0.03i$  & 0.498  \\ \hline
		\end{tabular}
	\end{center}
	{\footnotesize {\em Notes:} Designs are obtained by estimating a VAR(0) or VAR(1) of
		the form $y_t = A_0 + A_1 y_{t-1} + u_t$, $\EE[u_t u_t']= \Sigma_{tr} \Sigma_{tr}'$ using OLS.
        $\lambda_i(A_1)$
		is the $i$'th eigenvalue of $A_1$.
		$y_{1,t}$ is the log difference of the U.S. GDP deflator, scaled by 100 to convert
		into percentages. $y_{2,t}$ is either the log difference of U.S. GDP or deviations
		of log GDP from a linear trend, scaled by 100. Design 1:
		inflation and GDP growth, 1964:I to 2006:IV.
		Design 2: inflation and output deviations from trend, 1964:I to 2006:IV.
		Design 3: inflation and output growth, 1964:I to 2006:IV.
		Design 4: inflation and output deviations from trend, 1983:I to 2006:IV.}\setlength{\baselineskip}{4mm}
\end{table}

\subsection{Experiment 1}
\label{subsec_mc_exp1}

\noindent {\bf Design.} The parameterization of the DGP
$y_t \sim iidN(0,\Sigma_u)$
is provided in Table~\ref{t_mcdesign} in
the column labeled {\em Design 1}. We define $\theta$ as the response of $y_{1,t}$ to $\epsilon_{1,t}$.
Because $\Sigma_{21}^{tr}<0$ in our design,  the geometry of the Monte Carlo design corresponds to the left panel of Figure~\ref{f_VAR0IDset}. 
Thus, the upper bounds (in polar coordinates) of 
$F^q$ and $CS^q$ are $\pi/2$ and the 
lower bounds of $F^\theta$ and  $CS^\theta$ are zero, respectively. 
The identified set for $\theta$ is $F^\theta(\phi_{q,0},\phi_{\theta,0}) = [0,0.578]$.
Below, we report coverage probabilities for the lower bound of $F^q$ and the upper bound of $F^\theta$ because they are the least favorable parameter values in the respective identified sets. 
We consider sample sizes of $T=100$ and $T=500$. The grid ${\cal Q}$ for $q$
is obtained as follows: $q$ is transformed into polar coordinates $[\cos(\varphi),\sin(\varphi)]'$
and we choose $n_{\cal Q} = 315$ equally spaced grid points for $\varphi$ on the interval $(-\pi/2,\pi/2]$.
The number of bootstrap repetitions to obtain $\hat{\Lambda}_{qq}$ and $\hat{\Lambda}_{\theta \theta}$ is $n_{\Lambda}=1,000$ and the number
of simulations to obtain the critical value and $c^{\alpha_1}(q)$
is $n_Z = 500$.
Further details on the implementation are provided in the Online Appendix.

\begin{table}[t!]
\caption{Experiments 1 and 2: Single-Horizon Sign Restrictions}
\label{t_mcresults}
\begin{center}
\begin{tabular}{lcccccccc} \hline \hline
	                               & \multicolumn{2}{c}{Experiment 1} & \multicolumn{6}{c}{Experiment 2} \\
                                   & \multicolumn{2}{c}{Design 1} & \multicolumn{2}{c}{Design 2} & \multicolumn{2}{c}{Design 3} & \multicolumn{2}{c}{Design 4} \\
                                   & Coverage  &   Length    & Coverage & Length & Coverage & Length & Coverage & Length  \\ \hline
$F^q(\phi_q)$                          &      & $\frac{42}{100} \pi $ &      & $\frac{36}{100} \pi $ &      & $\frac{47}{100} \pi $ &      & $\frac{51}{100}\pi $ \\ 
$F^\theta(\phi)$                     &      & 0.579 &        & 0.233 &      & 0.226 &      & 0.094 \\
\hline
\multicolumn{9}{c}{Sample Size $T=100$} \\ \hline
$CS^q$                      & 0.938 & $\frac{47}{100}\pi $ & 0.936 & $\frac{81}{100}\pi $ & 0.932 & $\frac{57}{100}\pi $ & 0.940 & $\frac{67}{100}\pi $ \\
$CS^\theta$                 & 0.980 & 0.671 & 0.979 & 0.295 & 0.934 & 0.265 & 0.942 & 0.128\\  
$CS^{\phi_q}$                          & 0.879 &       & 0.865 &       & 0.865 &       & 0.871 \\  \hline
\multicolumn{9}{c}{Sample Size $T=500$} \\ \hline
$CS^q$                      & 0.930 & $\frac{44}{100}\pi$ & 0.936  & $\frac{44}{100}\pi$ & 0.932 & $\frac{51}{100}\pi$ & 0.936 & $\frac{56}{100}\pi$\\
$CS^\theta$                 & 0.990 & 0.622 & 0.991 & 0.265 & 0.963 & 0.244 & 0.958 & 0.110\\ 
$CS^{\phi_q}$                          & 0.909 &       & 0.894 &       & 0.901 &       & 0.904 &      \\ \hline
\end{tabular}
\end{center}
{\footnotesize {\em Notes:} {\em Length} refers to the average length
of the confidence intervals across Monte Carlo repetitions. For $F^q(\phi_q)$ and $CS^q$
we report the arc length, see  Figure~\ref{f_VAR0IDset}.
We let $\alpha_1 = \alpha_2 = 0.05$,
which implies that the nominal coverage probabilities are 95\% for $CS^q$ and 90\% for $CS^\theta$ and $CS^{\phi_q}$. The confidence interval for $\phi_q$ has a nominal coverage probability of 90\%.}\setlength{\baselineskip}{4mm}
\end{table}

\noindent {\bf Results.} Detailed results for the frequentist confidence intervals are summarized
in Table~\ref{t_mcresults}.  
Recall that the nominal
coverage probability for $\theta$ is 90\%. For $T=100$ the actual coverage probability
for the Bonferroni sets is 0.98.
As we increase the sample size to $T=500$, the length
of the confidence intervals shrinks, while the actual
coverage probabilities increases to 0.99.
It is instructive to also examine the coverage probabilities of $CS^q$ and the Wald confidence set for $\phi_q=\mbox{vech}(\Sigma_{tr})$, which we denote by
$CS^{\phi_q}$. The coverage probability for the reduced-form
parameter vector $\phi_q$ is 88\% for $T=100$ and approaches its nominal value of 90\% as
the sample size is increased to $T=500$. This increase in coverage probability for $\phi_q$
mirrors the increase in coverage probability for $\theta$.
The Bonferroni intervals are computed based on $\alpha_1=\alpha_2=0.05$, which implies that
the nominal coverage probability of $CS^q$ is 95\%.
The actual coverage probabilities for the nuisance parameter vector $q$ are slightly smaller, namely, around 93\%.
Overall, the Bonferroni-type marginalization
generates conservative confidence intervals for $\theta$.

\subsection{Experiment 2}
\label{subsec_mc_exp2}

\noindent {\bf Design.} We now add first-order autoregressive terms to the simulation design to introduce
persistence in the endogenous variables:
\[
     y_t = A_1 y_{t-1} + u_t, \quad u_t \sim iidN(0,\Sigma_u).
\]
The choices for $A_1$ and $\Sigma_u$ are
summarized in Table~\ref{t_mcdesign} under the headings {\em Design 2}, {\em Design 3},
and {\em Design 4}. The designs differ with respect to the persistence of the vector autoregressive
process. {\em Design 2} is the least persistent. The eigenvalues of $A_1$ are 0.871 and 0.231.
{\em Design 4} is the most persistent with eigenvalues 0.955 and 0.498.
We focus on responses at horizon $h=1$, which can be obtained from $\phi_q = \mbox{vec}\big( (A_1 \Sigma_{tr})' \big)$.
The structural parameter of interest, $\theta$, is defined as $\partial y_{1,t+1}/\partial \epsilon_{1,t}$. 
As in Experiment~1, we compute coverage probabilities for the lower bound of $F^q(\phi_q)$ and the upper bound of $F^\theta(\phi)$.
The grid ${\cal Q}$ for $q$ is obtained as follows: $q$ is transformed into polar coordinates $[\cos(\varphi),\sin(\varphi)]'$
and we choose $n_{\cal Q} = 629$ equally spaced grid points for $\varphi$ on the interval $(-\pi,\pi]$.
The remaining aspects of the design are the same as in Experiment~1. 

\noindent {\bf Sign Restrictions over a Single Horizon.} We impose the sign restrictions that $\partial y_{1,t+1}/\partial \epsilon_{1,t}$ and
$\partial y_{2,t+1}/\partial \epsilon_{1,t}$ are non-negative:
\[
    \phi_{q,1}q_1  +  \phi_{q,2}q_2 \ge 0 \quad \mbox{and} \quad  \phi_{q,3}q_1 +  \phi_{q,4}q_2 \ge 0.
\]
For now we do not impose sign restrictions on the responses
at impact or at horizons greater than $h=1$.
The geometry of the identified sets $F^q(\phi_q)$ and $F^\theta(\phi_q,\phi_\theta)$ and its projections is similar
to the geometry depicted in Figure~\ref{f_VAR0IDset}. The main difference is that
the second boundary of the identified set is given by the solution of $q_1\phi_{q,3} + q_2 \phi_{q,4} =0$
and $\|q\|=1$, instead of $q = [0,1]'$. Overall, the results for Experiment~2 reported in Table~\ref{t_mcresults} are qualitatively similar to those for
{\em Design~1}. The actual coverage probabilities of the confidence sets for $q$ are around 0.94 and therefore close to the nominal coverage probability of $1-\alpha_1=0.95$. The $\theta$ sets, on the other hand, are conservative. Their coverage probabilities range from 0.942 to 0.991, thereby exceeding the nominal level of 0.9. 

\begin{table}[t!]
	\caption{Experiment 2: Multiple-Horizon Sign Restrictions, Sample Size $T=100$}
	\label{t_mcresults2}
	\begin{center}
		\scalebox{0.87}{
			\begin{tabular}{lccccccccc} \hline \hline
				& \multicolumn{3}{c}{Design 2} & \multicolumn{3}{c}{Design 3} & \multicolumn{3}{c}{Design 4} \\
				& Coverage & Length  & Bind.Ineq & Coverage & Length & Bind.Ineq & Coverage & Length & Bind.Ineq \\ \hline
				\multicolumn{10}{c}{Restrictions: $h=0,1$} \\ \hline
				$F^q(\phi_q)$                   &         & $\frac{35}{100}\pi$ & &      & $\frac{43}{100}\pi$ & &      & $\frac{47}{100}\pi$ \\
				$CS^q$   & 0.953 & $\frac{48}{100} \pi$ & 1.29 & 0.976 & $\frac{50}{100} \pi$ & 1.97 & 0.953 & $\frac{53}{100} \pi$ & 2.06 \\
				$F^\theta(\phi)$                   &         & 0.265 & &       & 0.277 & &       & 0.209 \\        
				$CS^\theta$   &  0.982 & 0.307 & & 0.977 &  0.317 & & 0.949 & 0.236 & \\ \hline 
				\multicolumn{10}{c}{Restrictions: $h=0,\ldots,4$} \\ \hline
				$F^q(\phi_q)$                   &         & $\frac{6}{1000}\pi$ & &      & $\frac{37}{100}\pi$ & &      & $\frac{47}{100}\pi$ \\
				$CS^q$   & 0.985 & $\frac{40}{100} \pi$ & 7.78 & 0.989 & $\frac{49}{100} \pi$ & 4.48 & 0.979 & $\frac{56}{100} \pi$ & 7.83 \\
				$F^\theta(\phi)$                   &        & 0.006 & &       & 0.261 & &      & 0.208 \\
				$CS^\theta$                  &  1.000 & 0.277 & & 0.993 & 0.315 & & 0.949 & 0.235 \\ \hline
			\end{tabular}
		}
	\end{center}
	{\footnotesize {\em Notes:} {\em Length} either refers to the length of the population identified set or the average length
		of the confidence intervals across Monte Carlo repetitions. For $F^q(\phi_q)$ and $CS^q$
		we report the arc length, see  Figure~\ref{f_VAR0IDset}. {\em Bind.Ineq} is the average number of inequalities considered binding by the \cite{AndrewsSoares2010a} moment selection procedure. We let $\alpha_1 = \alpha_2 = 0.05$,
		which implies that the nominal coverage probabilities are 95\% for $CS^q$ and 90\% for $CS^\theta$. }\setlength{\baselineskip}{4mm}
\end{table}

\noindent {\bf Sign Restrictions over Multiple Horizons.} As before, we define $\theta$ as the contemporaneous impact of the shock
on $y_{1,t}$: $\theta =\partial y_{1,t}/\partial \epsilon_{1,t}$. 
However, we now restrict the signs of the impulse responses 
$\partial y_{i,t+h}/\partial \epsilon_{1,t} \ge 0$ for both variables $i=1,2$ over multiple periods: $h=0,1,\ldots,H$. This increases the number of inequality conditions.
Monte Carlo results are presented in Table~\ref{t_mcresults2}. 
The effect of adding sign restrictions differs across the three designs. In {\em Design~2} the lengths of the identified 
sets $F^q(\cdot)$ and $F^\theta(\cdot)$ shrink drastically: from $0.35 \pi$ and 0.265 for $H=1$ to $0.006 \pi$ and 0.007, respectively, for $H=4$. Under {\em Design~4} the sizes of the two identified sets remain constant as $H$ is increased from 1 to 4. {\em Design~2} is an intermediate case. Restricting impulse responses at multiple horizons essentially adds rays to Figure~\ref{f_VAR0IDset}. The location of the new rays relative to the $H=1$ rays determines whether the identified sets shrink or not.

For {\em Design 2} and {\em 3} the length of the confidence intervals for $q$ and $\theta$ are decreasing in the number of inequality restrictions, but they  do not shrink as quickly as the length of the identified sets. Simultaneously, the actual coverage probability for $q$ increases with the number of sign restrictions. In {\em Design~2} the coverage probability for $H=1$ is 0.953, which is close to the nominal coverage probability of $1-\alpha_1 = 0.95$. For $H=4$ the coverage probability increases to 0.985. While in population for any $H$ only one inequality is binding\footnote{An inspection of Figure~\ref{f_VAR0IDset} suggests that if more than one inequality condition is binding then it must be the case that the rays corresponding to the binding inequality conditions are identical.} at the boundary of the identified set for $q$, the average number of moment conditions deemed binding by the \cite{AndrewsSoares2010a} selection rule rises from 1.29 to 7.78 in {\em Design~2}. Recall that in order to guarantee a uniform asymptotic coverage probability, the selection rule has to classify too many rather than too few moment conditions as binding. This inflates the critical value as well as the coverage probability and makes the $q$ confidence set more conservative. We observe a similar pattern for {\em Design 3}.
The Bonferroni sets for $\theta$ are generally conservative across all designs and maximum horizons $H$. Under {\em Design~2} and~{\em 3} the actual coverage probabilities tend to increase as more restrictions are added. Nonetheless the average length decreases. This decrease is most pronounced for {\em Design~2}. Here the length shrinks from 0.307 ($H=1$) to 0.277 ($H=4$). 
Under {\em Design 4} the average length of the confidence interval and the coverage probability stay essentially constant as we vary the number of restrictions.

\subsection{Experiment 3}
\label{subsec_mc_exp3}

\begin{figure}[t!]
	\caption{Impulse Responses Bands and Coverage Probabilities}
	\label{f_exp3_baseline}
	\begin{center} 
		\begin{tabular}{cc}
			\multicolumn{2}{c}{Confidence Bands and Identified Sets} \\
			Variable 1 & Variable 2 \\
			\includegraphics[width=2.5in, trim={0.4in  2.5in 0.5in 3.25in}, clip]{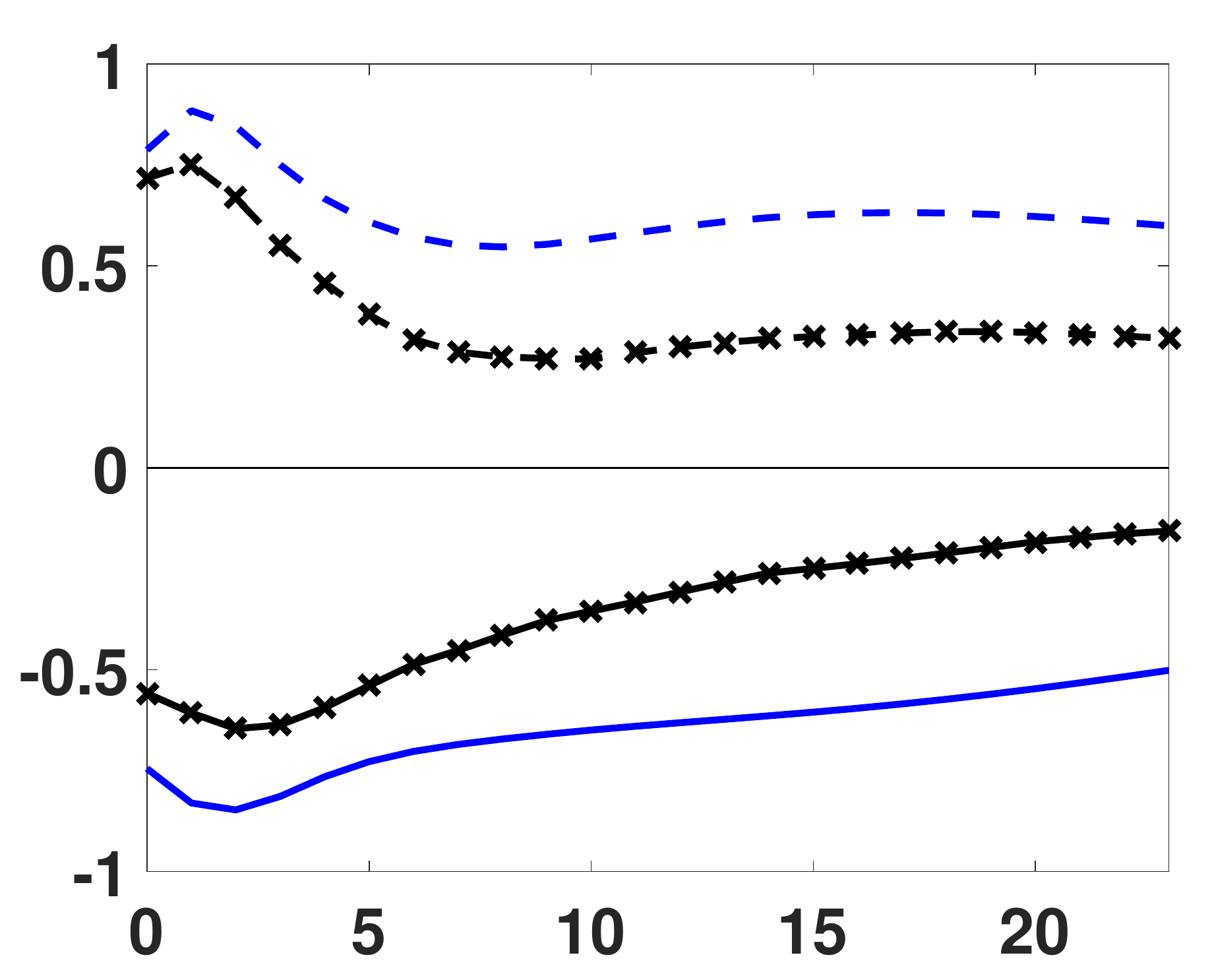} & 
			\includegraphics[width=2.5in, trim={0.4in  2.5in 0.5in 3.25in}, clip]{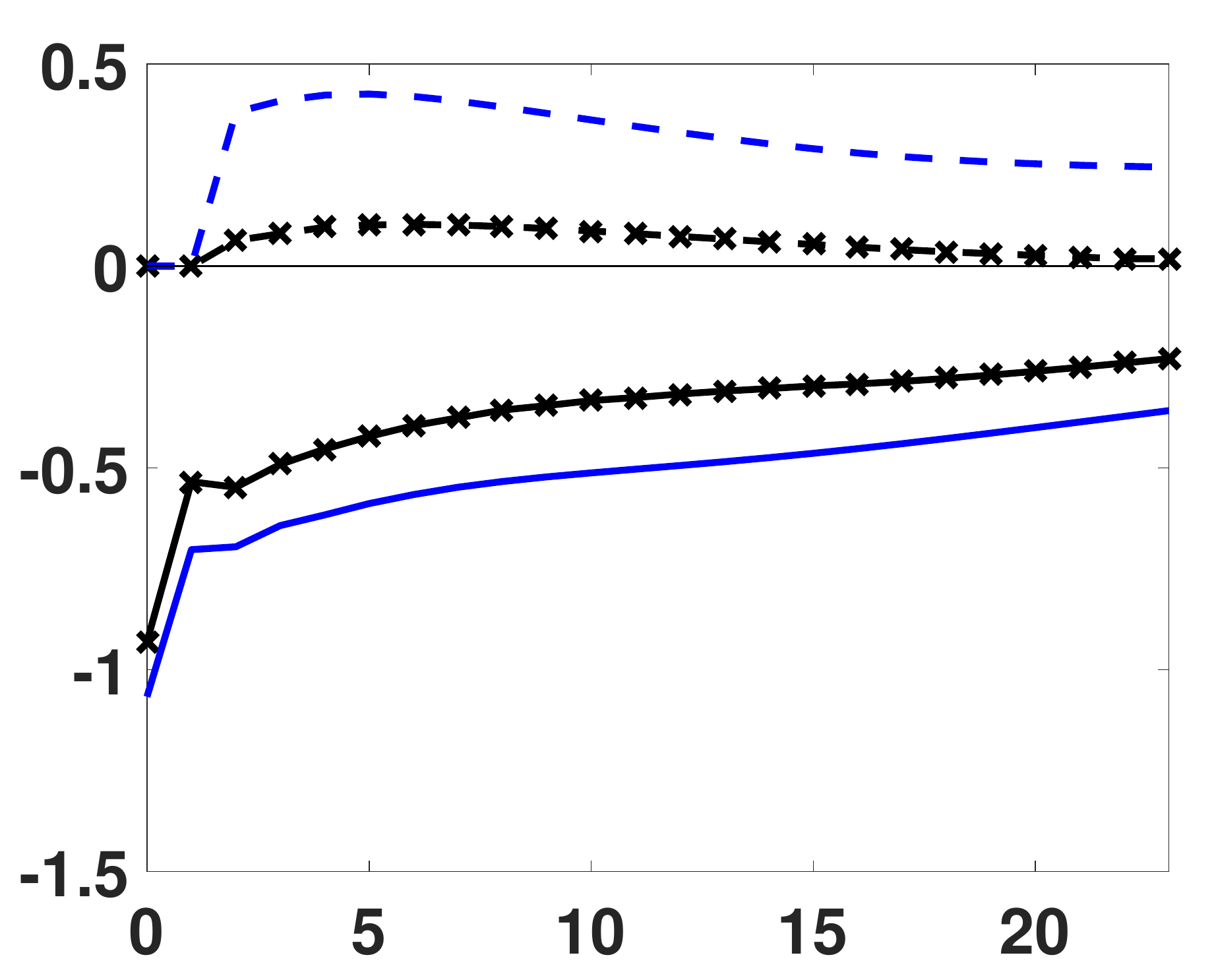}		
			\\
			\multicolumn{2}{c}{Coverage Probabilities} \\
			Variable 1 & Variable 2 \\
			\includegraphics[width=2.5in, trim={0.4in  2.5in 0.5in 3.25in}, clip]{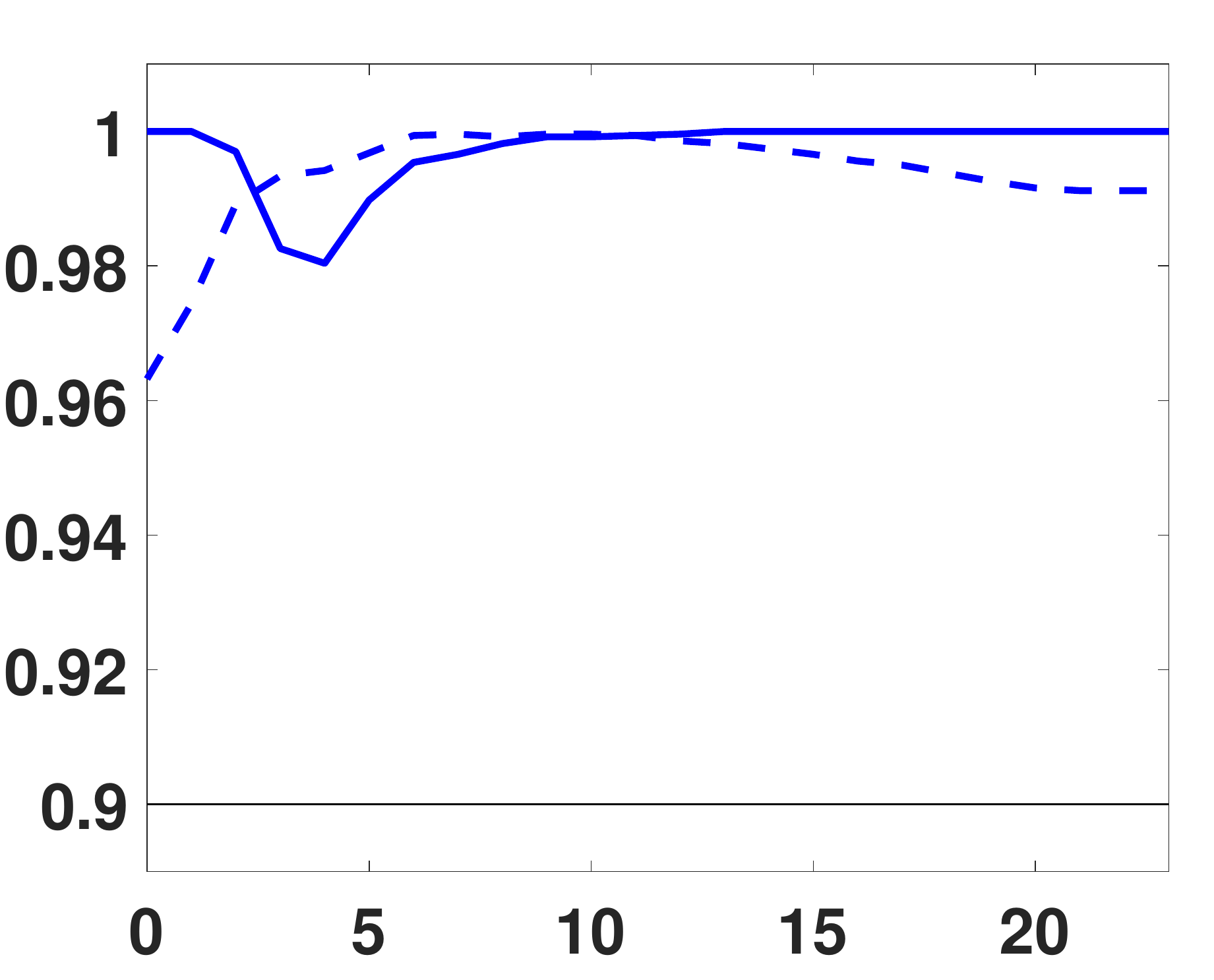} & 
			\includegraphics[width=2.5in, trim={0.4in  2.5in 0.5in 3.25in}, clip]{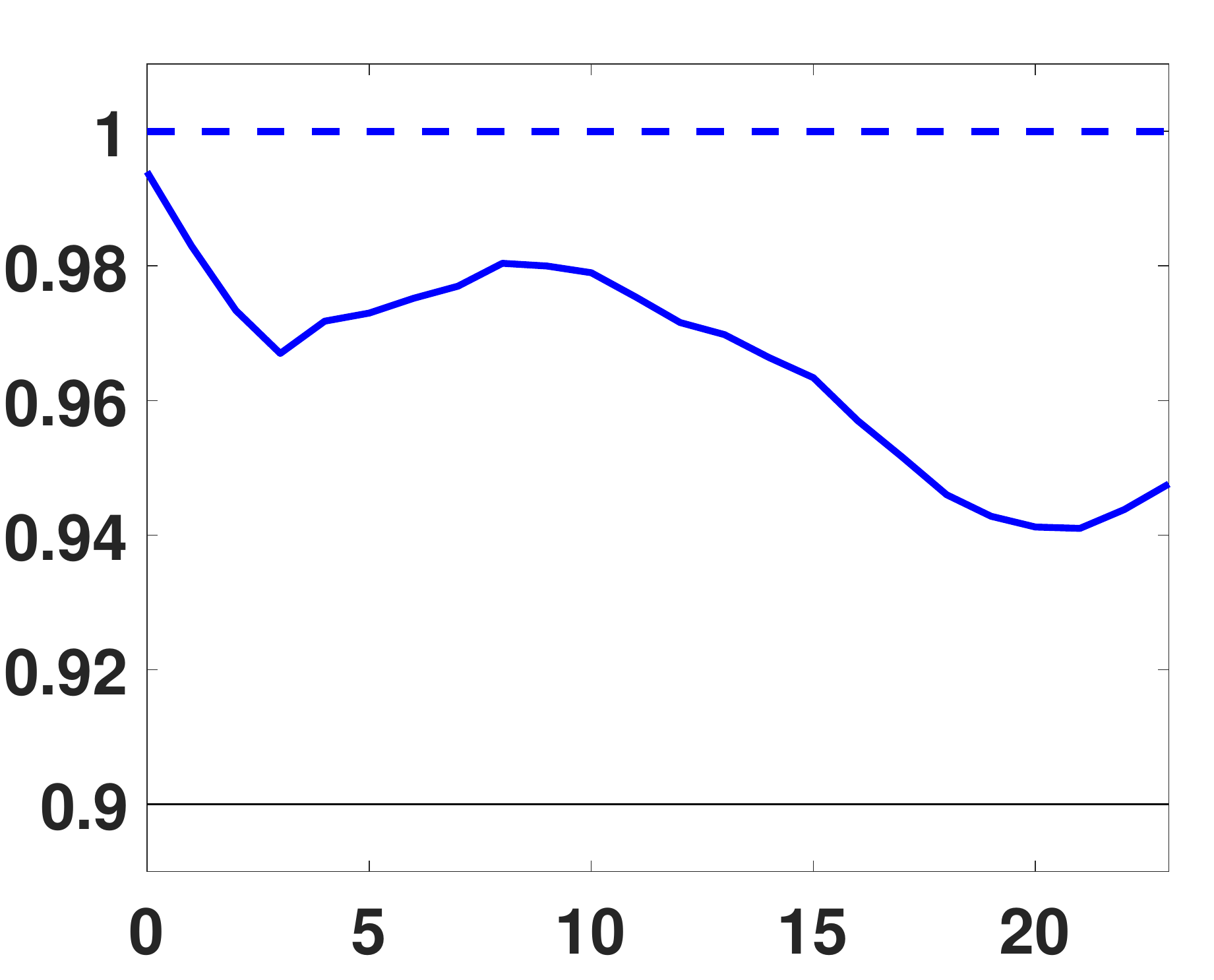}	
		\end{tabular}
	\end{center}
	{\footnotesize \emph{Notes:} Top panels: population identified set $F^\theta(\phi)$ (lines with crosses), averaged upper (dashed) and lower (solid) bounds of pointwise confidence sets, and a zero line (thin solid). Bottom panels: actual coverage probabilities for 90\% confidence sets at the lower (solid) and upper (dashed) bounds of the pointwise identified sets. The thin solid horizontal line indicates the nominal coverage probability. $\alpha_1=\alpha_2=0.05$.}\setlength{\baselineskip}{4mm}
\end{figure}

\noindent {\bf Design.} Finally, we consider a 
four-variable VAR(2) that mimics the model for per capita GDP (in deviations from a linear trend), inflation, the federal funds rate, and real money balances used in the empirical application in Section~\ref{sec_empirical}:
\begin{equation}
	y_{t}=c+A_{1}y_{t-1}+A_{2}y_{t-2}+u_{t}\text{ \ with }u_{t}\sim N\left(
	0,\Sigma \right).
\end{equation}
The reduced-form parameters are set equal to the empirical point estimates (reported in the Online Appendix). As in the application below, we consider the following sign restrictions:
\[
 \frac{\partial y_{2,t+h}}{\partial
 	\epsilon_{1,t}} \leq 0, \quad \frac{\partial y_{3,t+h}}{\partial
 	\epsilon_{1,t}} \geq 0, \quad \frac{\partial y_{4,t+h}}{\partial
 	\epsilon _{M,t}}\leq 0, \quad h=0,1.
\]
The sample size for the simulated data sets is $T=170$. We set the number of (randomly generated from a uniform distribution on the hypersphere) grid points for ${\cal Q}$ to $n_{\cal Q} = 20,000$.  The number of bootstrap repetitions to obtain $\hat{\Lambda}_{qq}$ and $\hat{\Lambda}_{\theta \theta}$ is $n_\Lambda=1,000$ and the number of simulations to obtain the critical value $c^{\alpha_1}(q)$ is $n_Z=1,000$. As in the previous experiments, we focus on $\hat{B}=I$. The implementation of the computations for $CS^\theta(I)$ follows the description in Section~\ref{sec_implementation}. 

\noindent {\bf Benchmark Results.} Baseline results for $\alpha_1=\alpha_2=0.05$ are plotted in Figure~\ref{f_exp3_baseline}. The top panels depict the upper bounds and the lower bounds for pointwise identified sets and confidence sets for the responses of Variables~1 and Variables~2 at horizons $h=0,1,\ldots,23$. The bounds of the confidence sets are averaged across Monte Carlo repetitions. An economic interpretation of the responses will be provided in Section~\ref{sec_empirical}. For now we focus on the widths of the confidence bands relative to the widths of the population identified sets and the coverage probabilities, which are depicted in the bottom panels. The identified sets have a considerable width that leaves the sign of the response of $y_{1,t}$ undetermined. The confidence bands are noticeably wider than the identified sets, which is a reflection of the sampling uncertainty associated with the estimators of the reduced-form parameters. The coverage probabilities in the bottom panels are computed for the upper bounds (dashed) and lower bounds (solid) of the pointwise identified sets. As in Experiments~1 and~2, the actual coverage probability is substantially larger than the nominal coverage probability of 90\% (indicated by the solid horizontal line), making the confidence bands conservative.

\begin{figure}[t!]
	\caption{The Effect of Varying $\alpha_1$, Fixed $\alpha = 0.10$}
	\label{f_exp3_A}
	\begin{center} 
		\begin{tabular}{cc}
			\multicolumn{2}{c}{Coverage Probability} \\
			Variable 1 & Variable 2 \\
			\includegraphics[width=2.5in, trim={0.4in  2.5in 0.5in 3.25in}, clip]{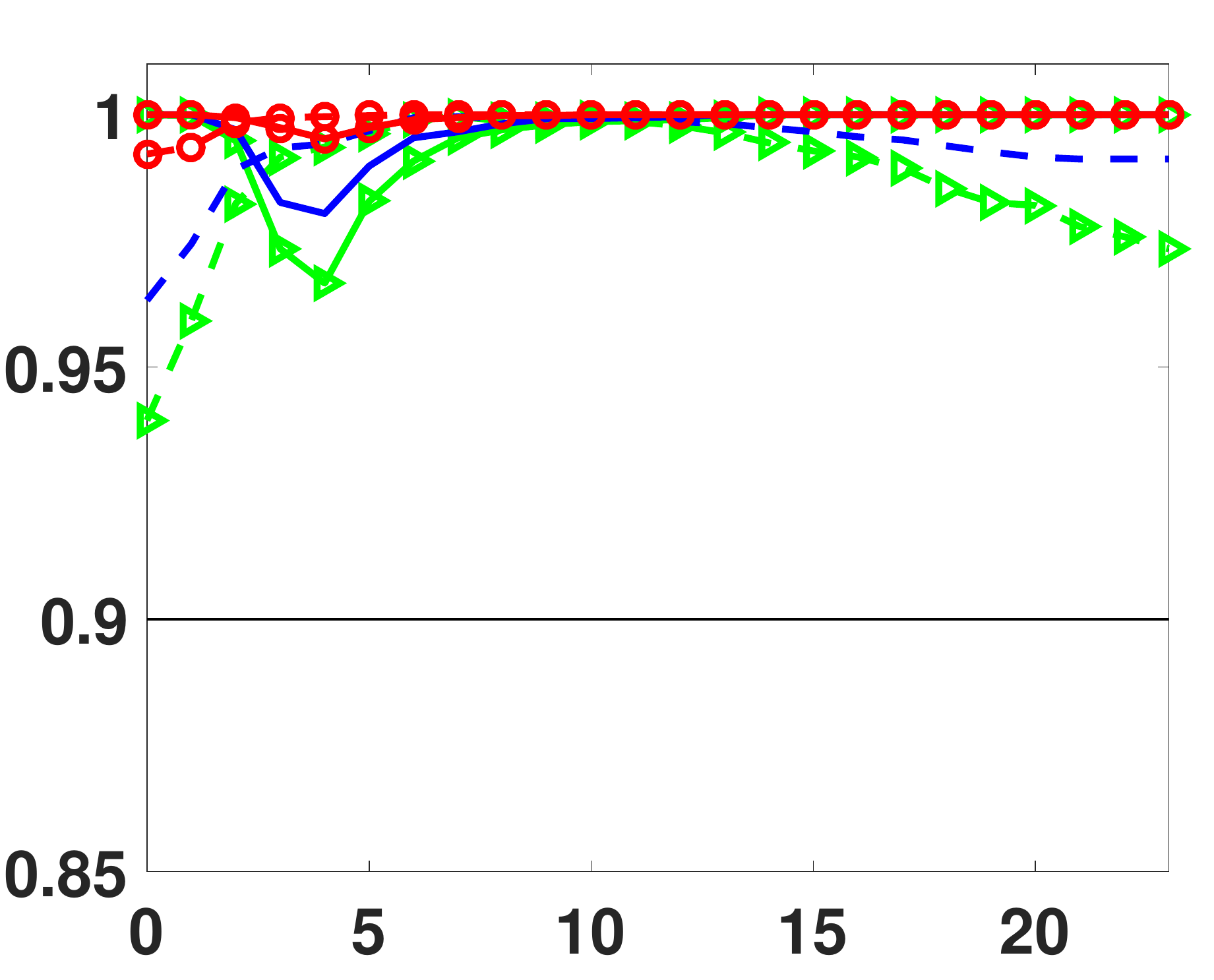} & 
			\includegraphics[width=2.5in, trim={0.4in  2.5in 0.5in 3.25in}, clip]{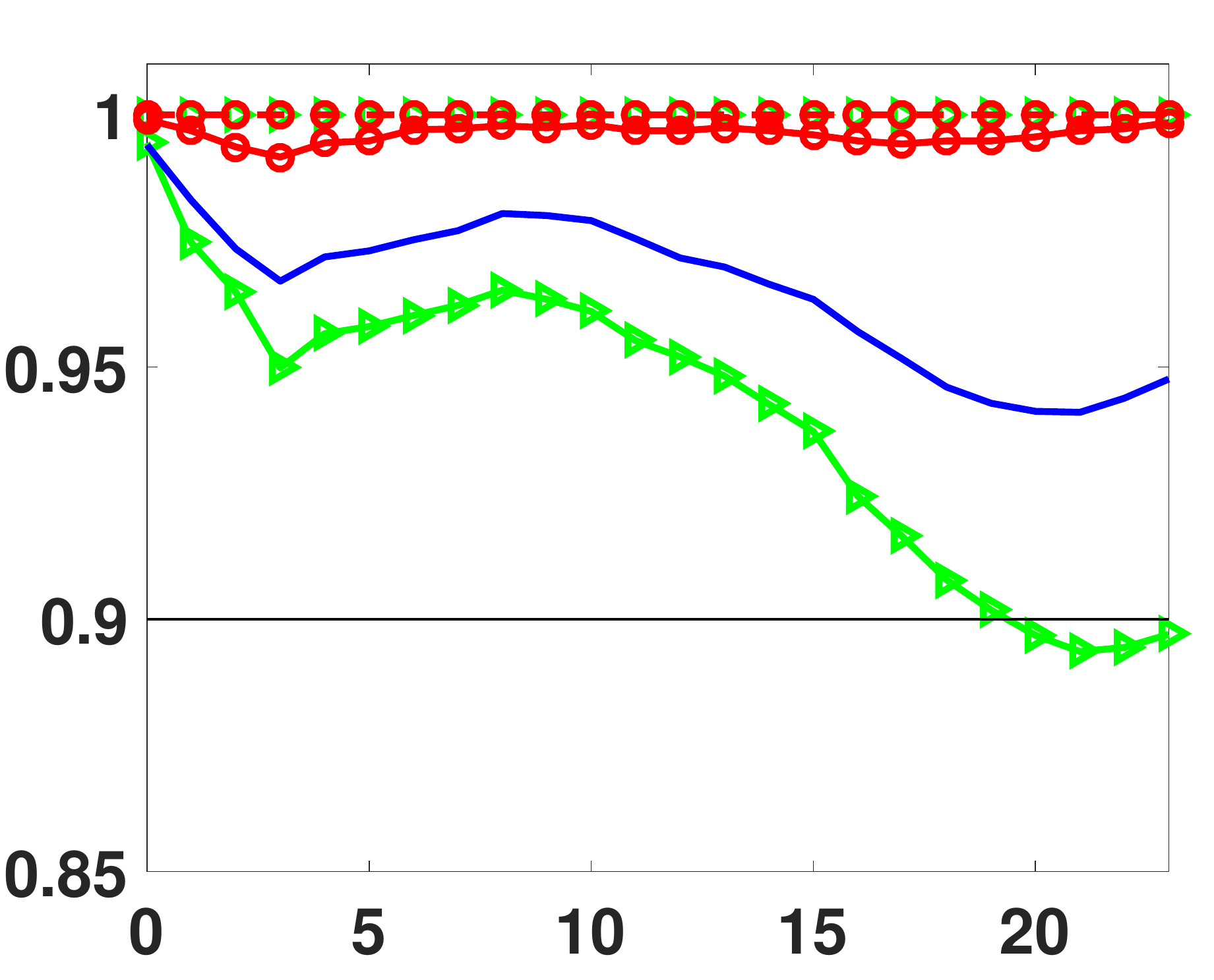}		
			\\
			\multicolumn{2}{c}{Average Interval Width} \\
			Variable 1 & Variable 2 \\
			\includegraphics[width=2.5in, trim={0.4in  2.5in 0.5in 3.25in}, clip]{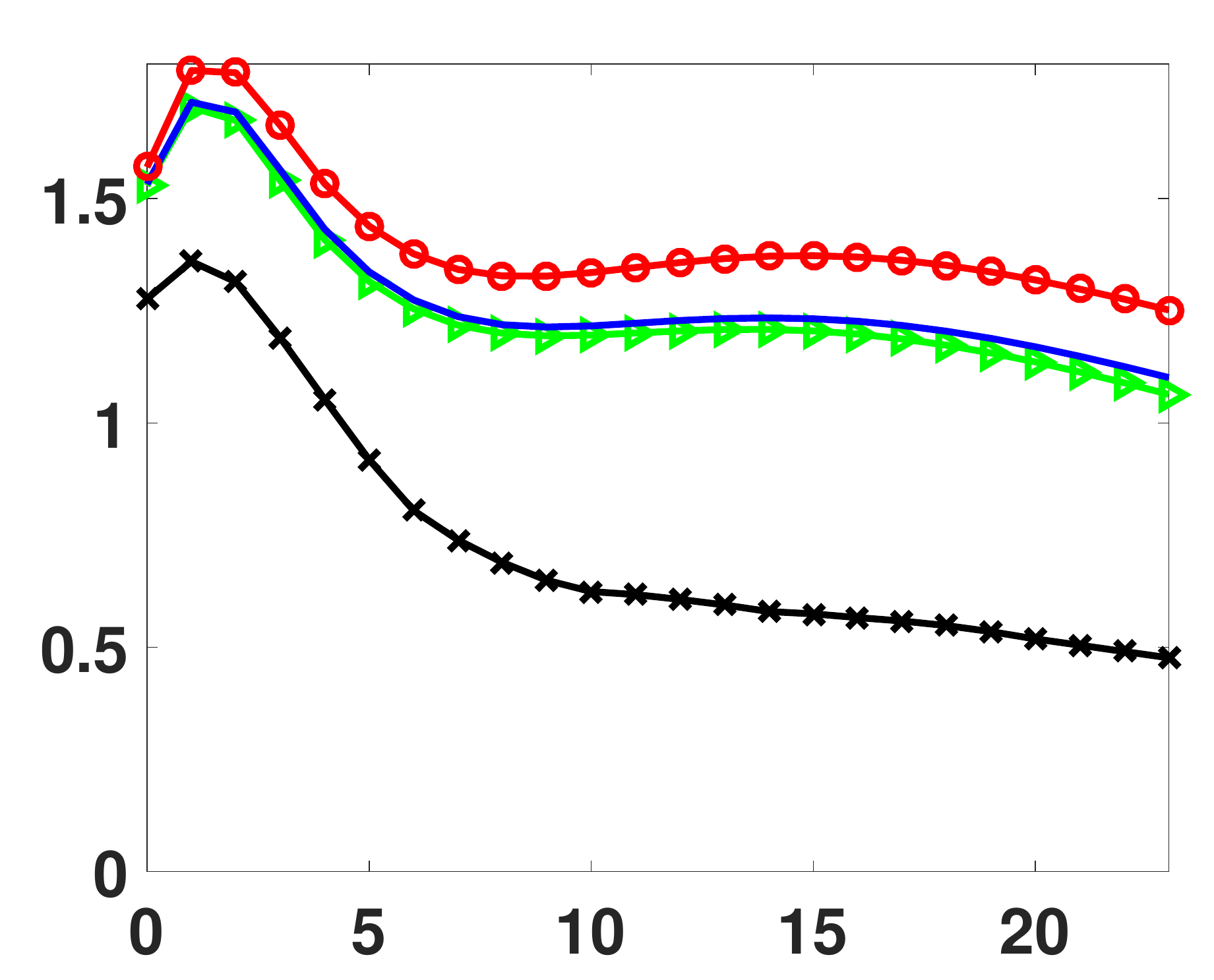} & 
			\includegraphics[width=2.5in, trim={0.4in  2.5in 0.5in 3.25in}, clip]{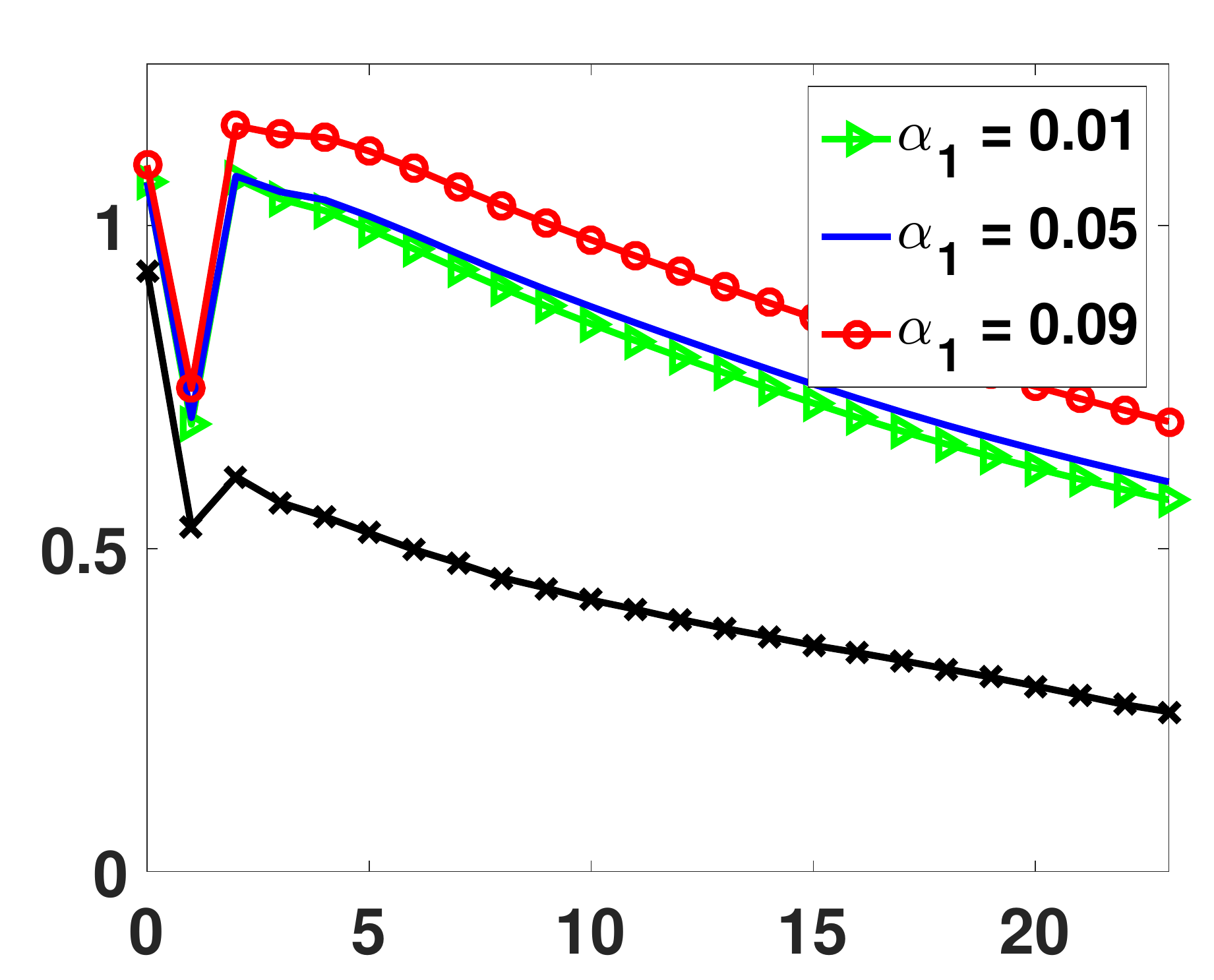}	
		\end{tabular}
	\end{center}
	{\footnotesize \emph{Notes:} Top panels: Thin horizontal line indicates the nominal coverage probability of 90\%, other lines represent actual coverage probabilities. Dashed lines refer to upper bounds and solid lines to lower bounds of pointwise identified sets; $\alpha_1=0.01$ is marked by triangles, $\alpha_1=0.05$ (baseline) has no line symbols,  and $\alpha_1=0.09$ is marked by circles. Bottom panels: average width of confidence bands (triangles, no line symbol, circles) and width of population identified set (crosses).}\setlength{\baselineskip}{4mm}
\end{figure}

\noindent {\bf Adjusting $\alpha_1$, Keeping $\alpha = 0.1$ Fixed.} The baseline choice of $\alpha_1 = 0.05$ has been 
arbitrary. Thus, it is worthwhile to explore what happens if we decrease or increase $\alpha_1$, which determines
the size of $CS^q$. Figure~\ref{f_VAR0IDset} provides some intuition for the potential outcomes of this experiment. In the left panel of the figure (labeled $\Sigma_{21}^{tr}<0$) the upper bound for the identified set $F^\theta(\cdot)$ is determined by the lower bound of $F^q(\cdot)$. Thus, increasing $\alpha_1$ and thereby decreasing the size of $F^q(\cdot)$ can potentially sharpen the confidence set for $\theta$, provided that the decrease in $F^q(\cdot)$ exceeds the increase in the conditional confidence set $CS^\theta_q$. Alternatively, if $\Sigma_{21}^{tr}>0$ (depicted in the right panel of Figure~\ref{f_VAR0IDset}), the upper bound of $F^\theta(\cdot)$ is determined by a value of $q$ that lies strictly in the interior of $F^q(\cdot)$. Thus, in order to shrink the confidence set for $\theta$ one should lower $\alpha_1$ and raise $\alpha_2$ so that $CS_q^\theta$ shrinks. 

Figure~\ref{f_exp3_A} depicts the coverage probabilities and the average interval width for three levels of $\alpha_1$: $\alpha_1=0.05$ which we used to generate the baseline results in Figure~\ref{f_exp3_baseline}, $\alpha_1=0.01$, and $\alpha_1=0.09$. {\color{black} The figure also depicts the width of the population identified set. The differences between the widths of the confidence intervals and the identified set can be interpreted as the excess lengths of the confidence intervals.}
It turns out that in this particular Monte Carlo design it is advantageous to reduce $\alpha_1$. Setting $\alpha_1=0.01$ reduces the coverage probability of the intervals and shrinks the width of the intervals. However, the effect is modest at best. Relative to the overall width of the identified sets and the baseline confidence bands, the reduction is very small. The actual coverage probability remains above 95\% except for the long horizon responses of $y_{2,t}$ which fall slightly below the nominal level of $\alpha=0.9$ for $h \ge 18$. As we saw in Table~\ref{t_mcresults}, the confidence set for the reduced-form VAR parameters can have an actual coverage probability that is less than its nominal coverage probability, which in turn tightens the confidence interval for $\theta$.

\begin{figure}[t!]
	\caption{The Effect of Varying $\alpha_2$, Fixed $\alpha_1 = 0.05$}
	\label{f_exp3_C}
	\begin{center} 
		\begin{tabular}{cc}
			\multicolumn{2}{c}{Coverage Probability} \\
			Variable 1 & Variable 2 \\
			\includegraphics[width=2.5in, trim={0.4in  2.5in 0.5in 3.25in}, clip]{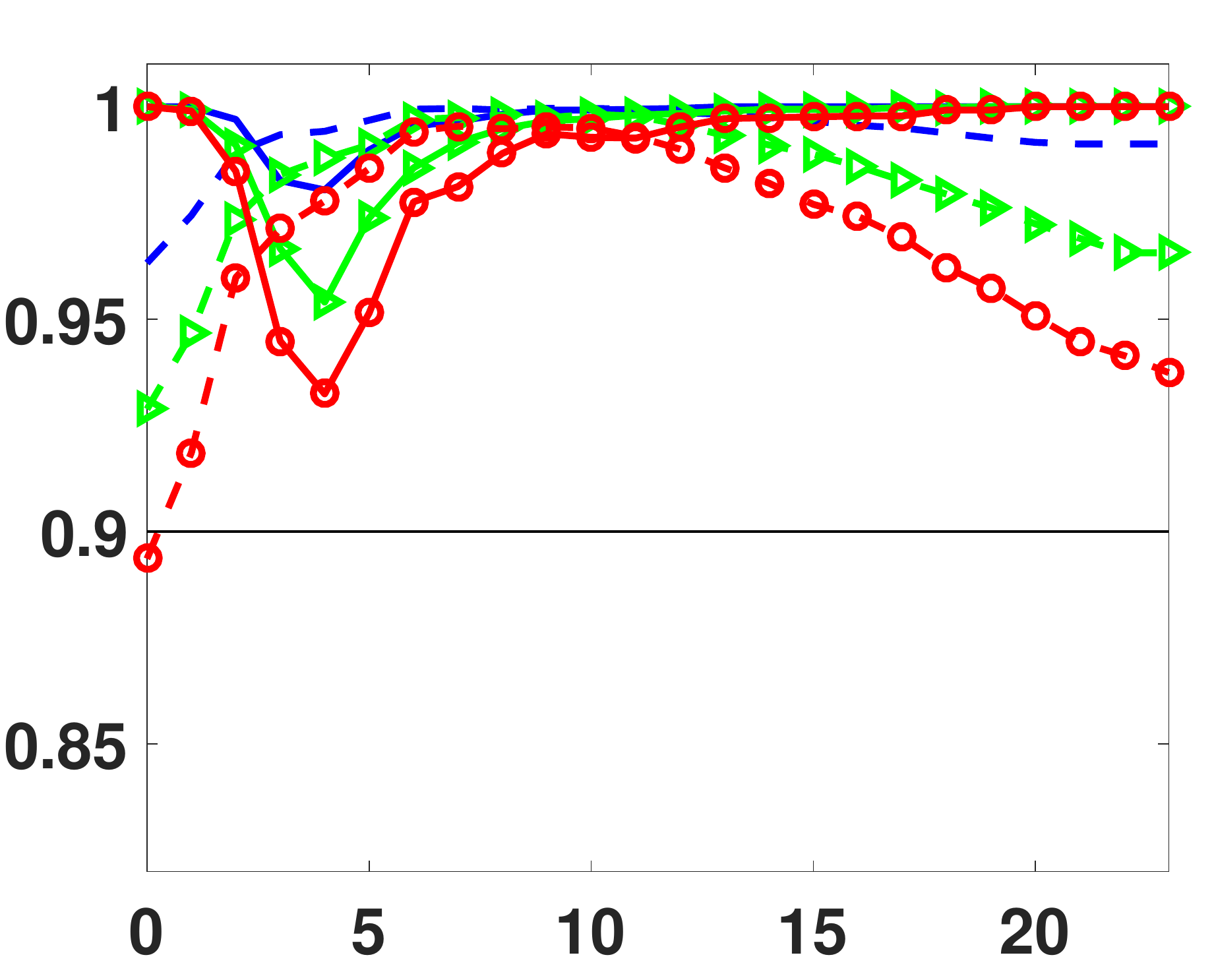} & 
			\includegraphics[width=2.5in, trim={0.4in  2.5in 0.5in 3.25in}, clip]{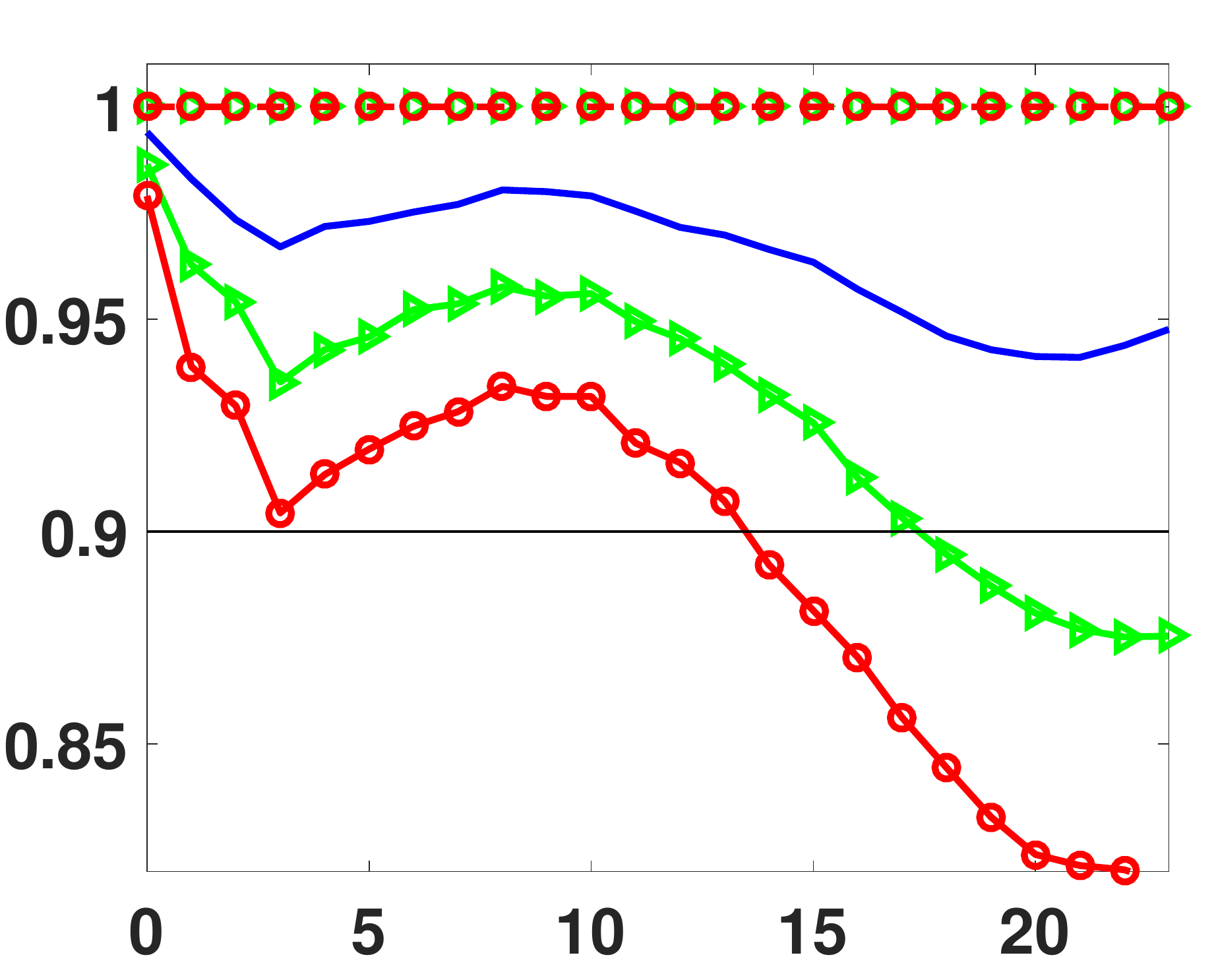}		
			\\
			\multicolumn{2}{c}{Average Interval Width} \\
			Variable 1 & Variable 2 \\
			\includegraphics[width=2.5in, trim={0.4in  2.5in 0.5in 3.25in}, clip]{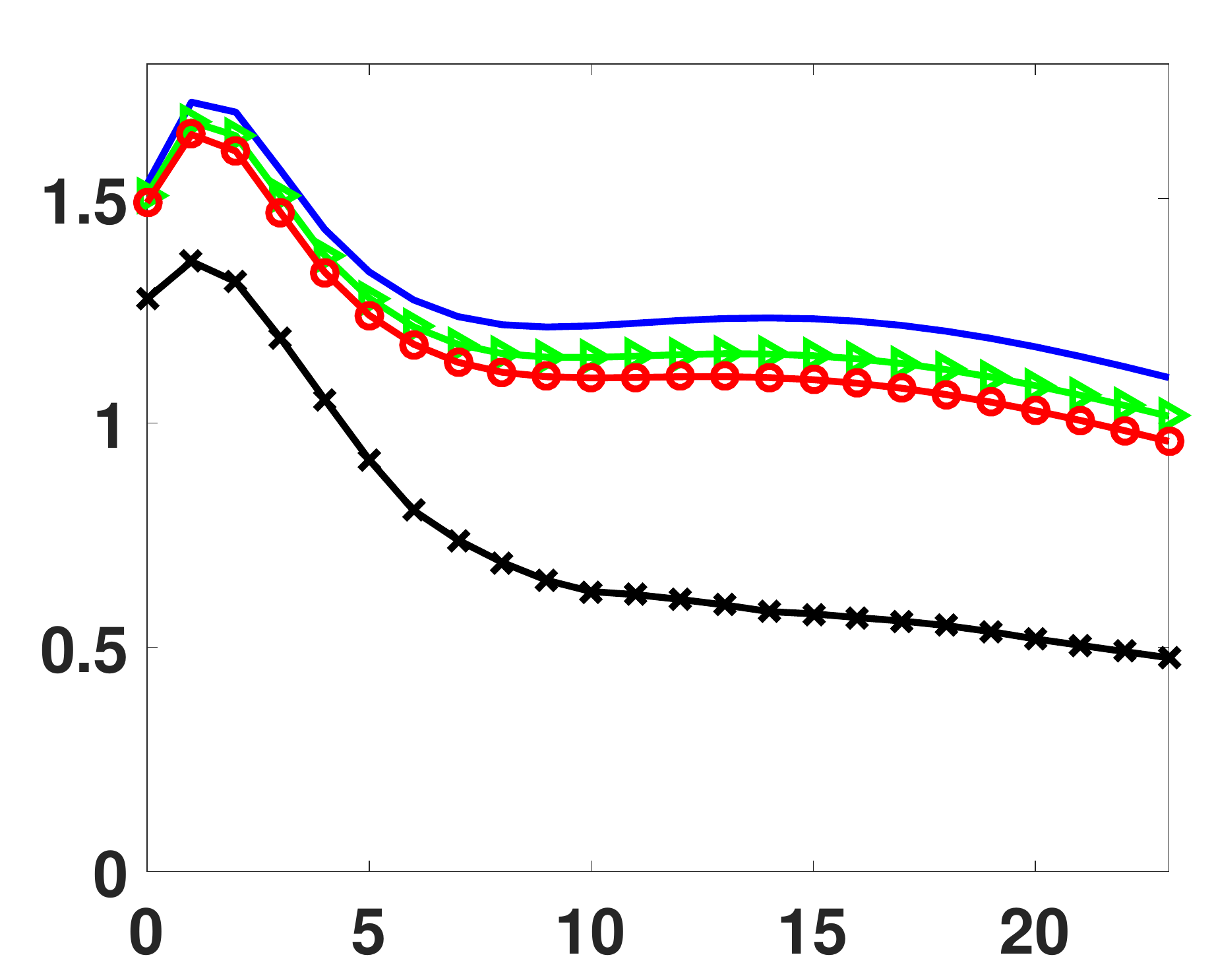} & 
			\includegraphics[width=2.5in, trim={0.4in  2.5in 0.5in 3.25in}, clip]{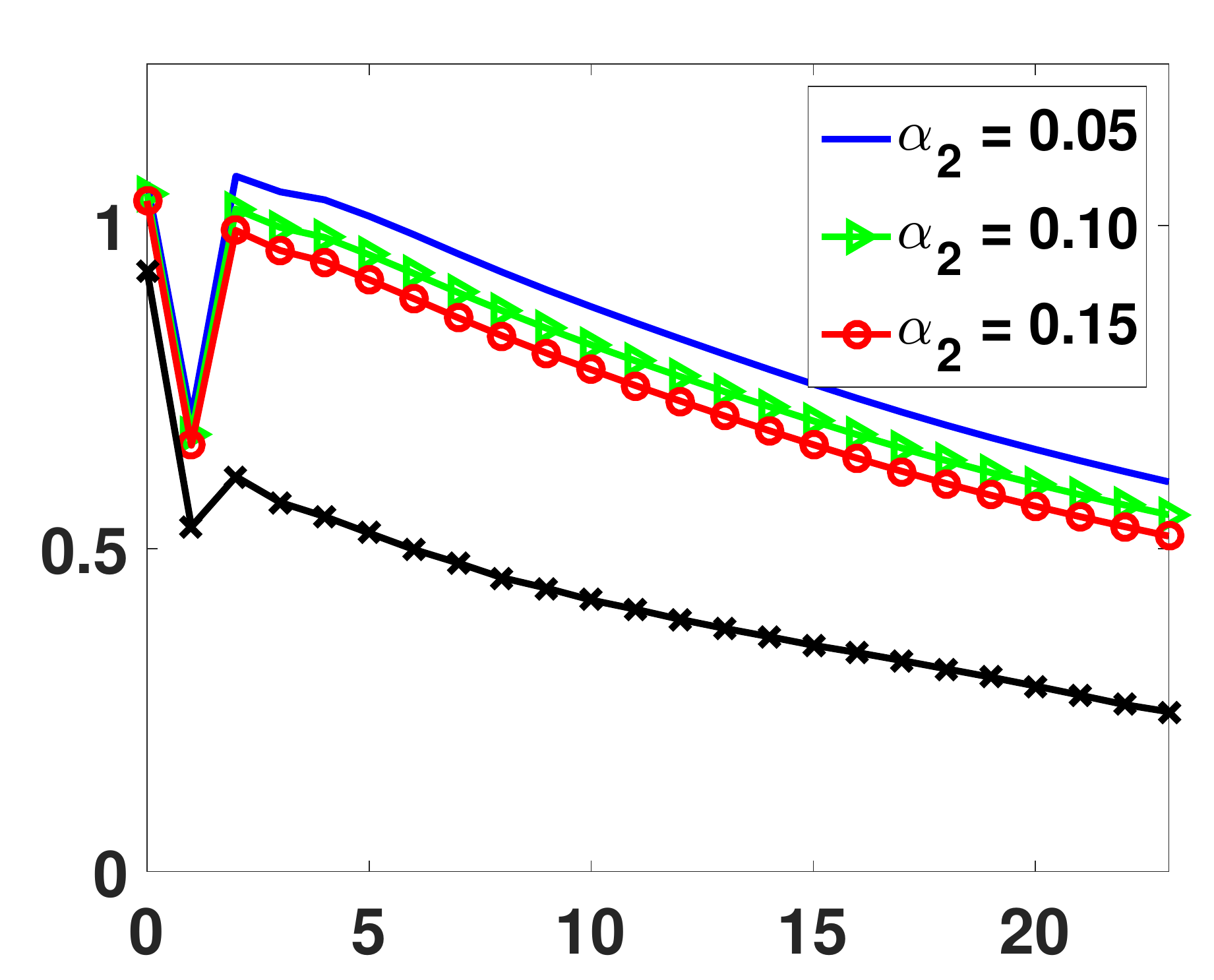}	
		\end{tabular}
	\end{center}
	{\footnotesize \emph{Notes:} Top panels: Thin horizontal line indicates the nominal coverage probability of 90\%, other lines represent actual coverage probabilities. Dashed lines refer to upper bounds and solid lines to lower bounds of pointwise identified sets; $\alpha_2=0.05$ (baseline) has no line symbols,  $\alpha_2=0.10$ is marked by triangles,  and $\alpha_2=0.15$ is marked by circles. Bottom panels: average width of confidence bands (no line symbol, triangles, circles) and width of population identified set (crosses).
	}\setlength{\baselineskip}{4mm}
\end{figure}

\noindent {\bf Adjusting $\alpha$, Keeping $\alpha_1$ Fixed.} Several authors devised methods to overcome the
conservativeness of Bonferroni confidence intervals by raising the nominal level $\alpha$ to target a desired
actual coverage probability $1-\alpha_*=0.90$, say. Examples of this approach are \cite{CampbellYogo2006} and, most recently, \cite{McCloskey2017}. The former paper reduces the size of the first-stage confidence interval by raising $\alpha_1$, keeping the size of the second-stage intervals, $CS^\theta_q$ in our notation constant. The latter paper 
proposes to reduce the size of the second-stage intervals, keeping $\alpha_1$ constant. In view of the results depicted in
Figure~\ref{f_exp3_A}, we informally follow McColskey's (2017) approach by increasing $\alpha_2$ from 0.05 (baseline) to 0.10 and 0.15, respectively. While, in principle, we could choose a different $\alpha_2$ for each variable and each horizon, the results depicted in Figure~\ref{f_exp3_C} are generated using the same $\alpha_2$ for each response. As expected, the actual coverage probability falls as we increase $\alpha_2$ (and thereby $\alpha=\alpha_1+\alpha_2$). For the $y_{1,t}$ responses the coverage probabilities remain above 90\%, while for the long-horizon responses of $y_{2,t}$ the coverage probability drops substantially below 90\% for $h \ge 15$ and $\alpha_2 =0.10$. The attainable reduction in the width of the confidence band is larger than in the case of fixed $\alpha$, but it remains small relative to the overall width of the bands.

%
%


\section{Empirical Illustration}
\label{sec_empirical}

We now apply the previously developed methods
to a four-variable VAR. The vector of observables consists of
per capita real GDP (in deviations from a linear trend), inflation, the federal funds rate, and
real money balances. We use quarterly U.S. data from 1965:I to 2006:IV, excluding the
2007-09 recession and the subsequent period of zero nominal interest rates. A detailed description of the data set is provided in
the Online Appendix. All VARs are estimated with $p=2$ lags, which is the preferred lag length according to BIC.
We will consider two set-identification
schemes for monetary policy shocks. The first scheme involves only sign
restrictions (Section~\ref{subsec_puresign}), whereas the second identification
is based on a combination of equality and sign restrictions (Section~\ref{subsec_signzero}).
As in Monte Carlo Experiment~3, we set $n_\Lambda = 1,000$, $n_{\cal Q}=20,000$, and $n_Z=1,000$.

In addition to computing Bonferroni confidence bands, we also
generate pointwise Bayesian credible intervals for the impulse responses,
which have been widely used in empirical research.
The Bayesian credible sets reported subsequently are
based on the VAR(p) given in~(\ref{eq_idea_varrf}) with Gaussian innovations $u_t \sim iid N(0,\Sigma_u)$.
Let $A = [ A_1, \ldots, A_p]'$ and define the unnormalized vector $\tilde{q}$
such that $q = \tilde{q} / \|\tilde{q}\|$.
If $\tilde{q} \sim N(0,I_n)$, then $q$ is uniformly distributed on the hypersphere.
Following \cite{Uhlig2005}, we use an improper prior of the form
\be
   p(A,\Sigma,\tilde{q}) \propto |\Sigma|^{-(n+1)/2} \exp\{
   -\tilde{q}'\tilde{q}/2 \} {\cal I} \left\{ \frac{\tilde{q}}{\|\tilde{q}\|} \in F^q \big( \phi(A,\Sigma) \big) \right\}.
   \label{eq_varprior}
\ee
We use the acceptance sampler described in \cite{Uhlig2005} to generate 50,000 draws from the posterior distribution of $(A,\Sigma,\tilde{q})$. These draws
are then converted into impulse responses and credible sets are computed
from the impulse response draws.

\begin{figure}[t!]
\caption{Pure Sign Restrictions over Horizons $h=0,1$}
\label{f_puresign_4v_h1}
\begin{center}
	\begin{tabular}{cc}
		Output & Inflation \\
        \includegraphics[width=2.5in,trim={0.4in  2.5in 0.5in 3.25in}, clip]{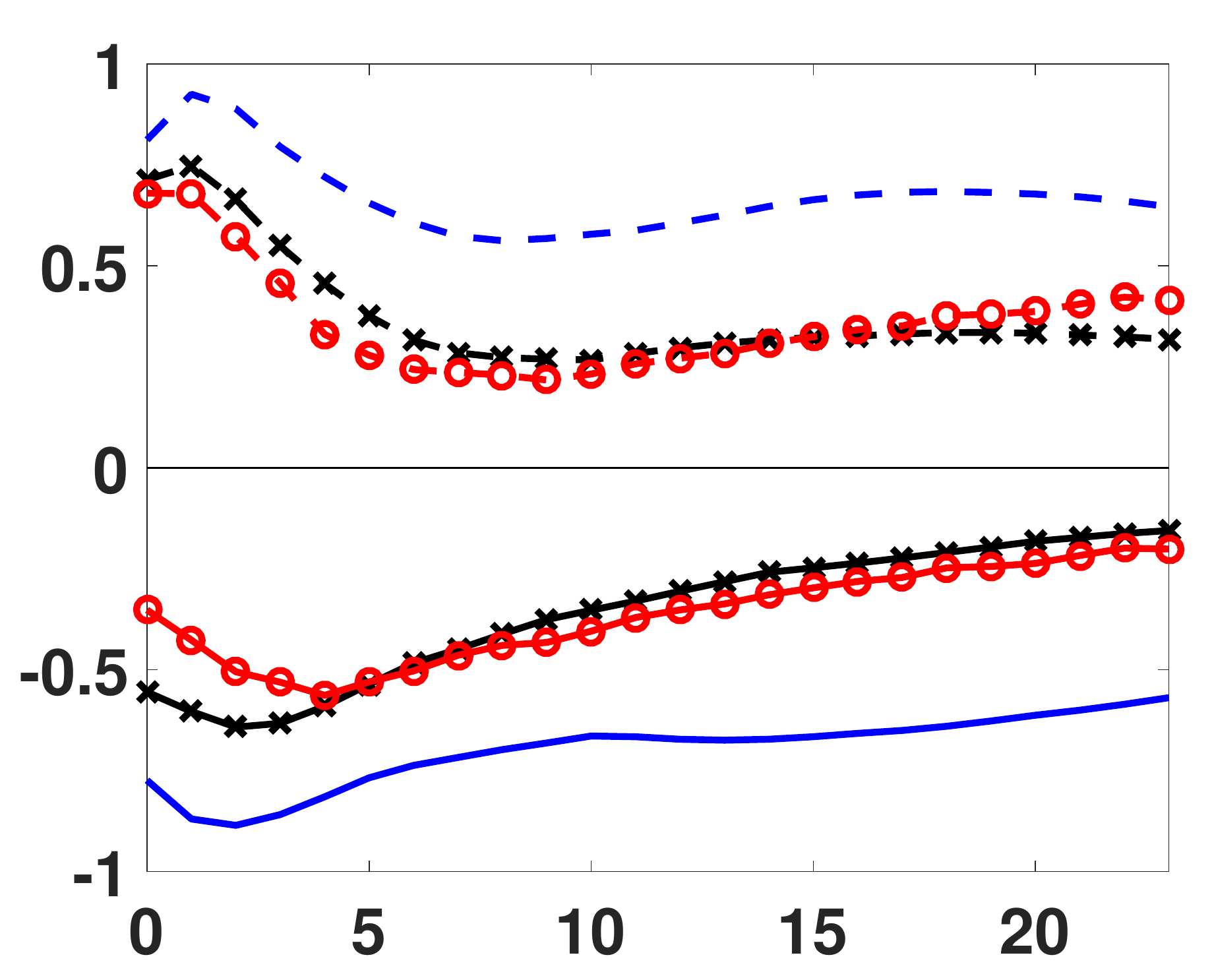} & 
        \includegraphics[width=2.5in,trim={0.4in  2.5in 0.5in 3.25in}, clip]{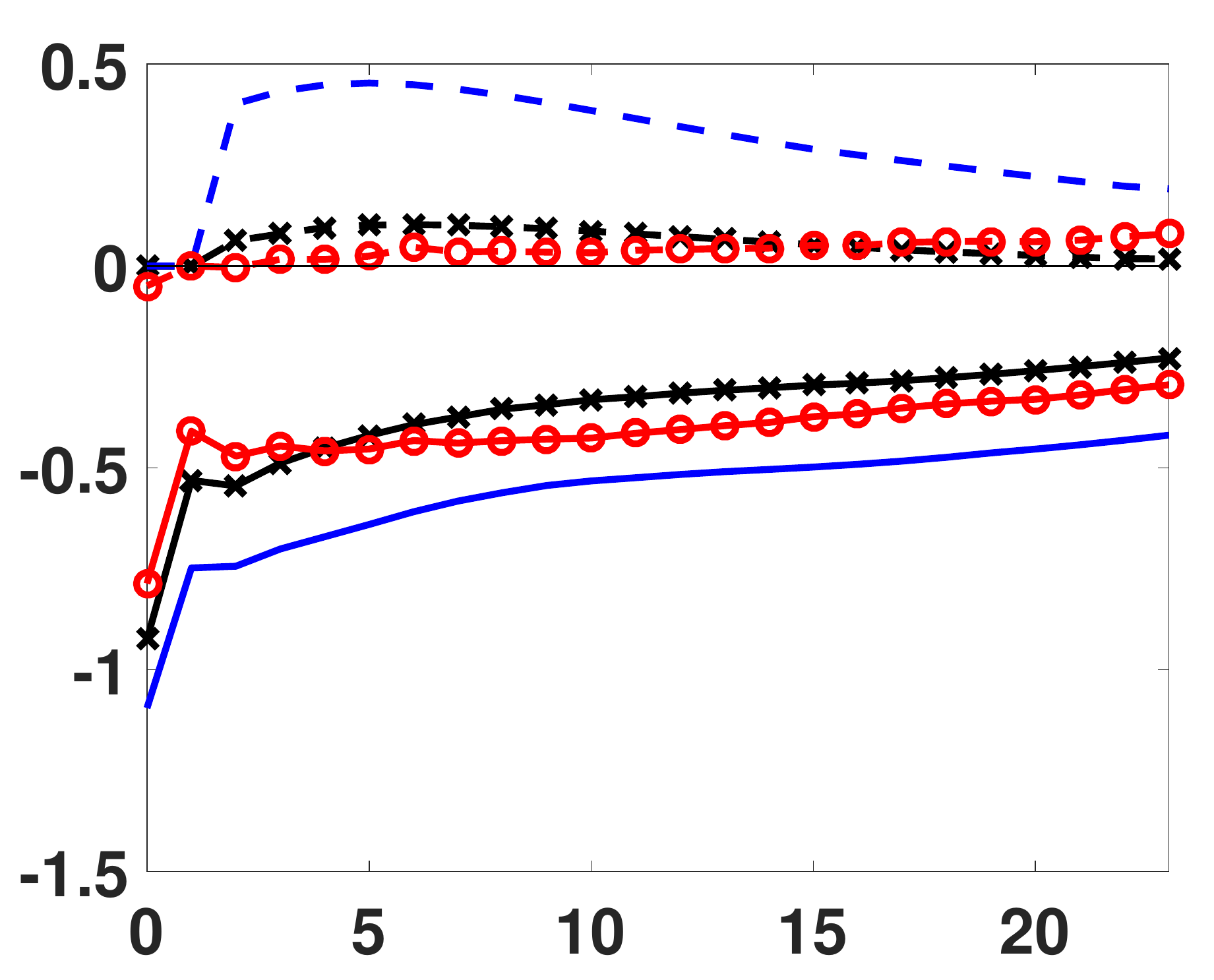} \\
        Interest Rates & Real Money \\
        \includegraphics[width=2.5in,trim={0.4in  2.5in 0.5in 3.25in}, clip]{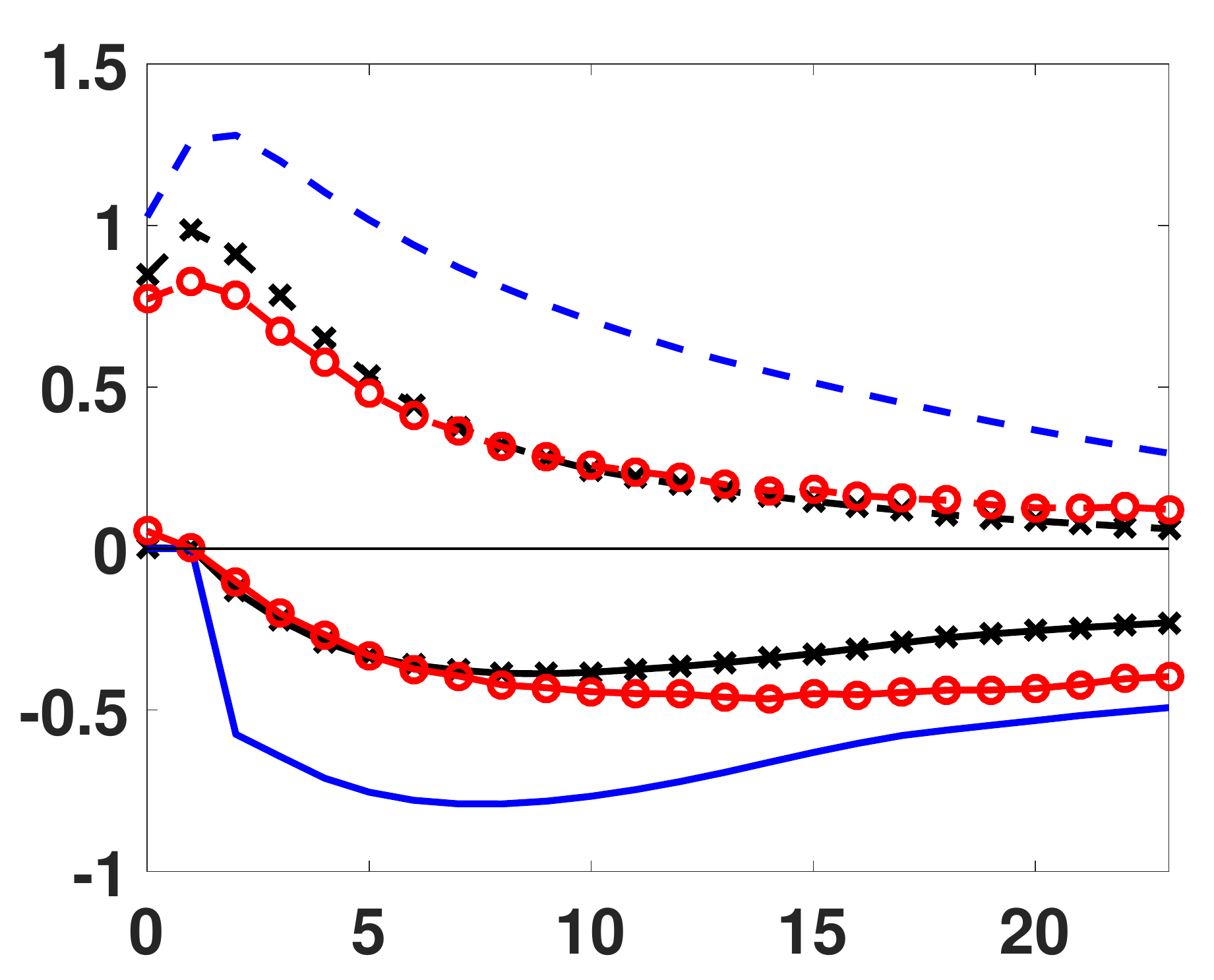} & 
        \includegraphics[width=2.5in,trim={0.4in  2.5in 0.5in 3.25in}, clip]{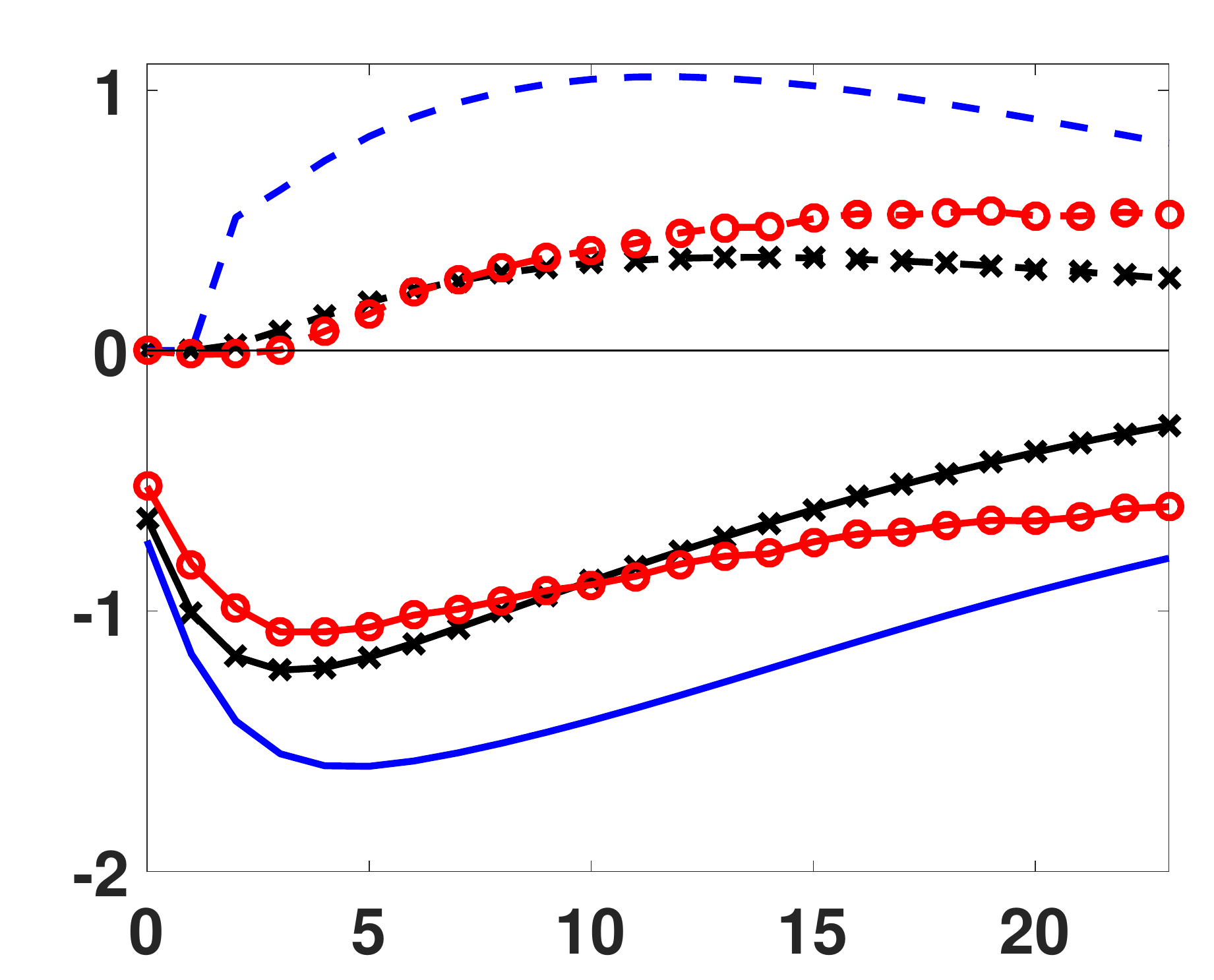}    
     \end{tabular}     
\end{center}
{\footnotesize \emph{Notes:} The figure depicts
90\% Bonferroni confidence bands $CS^\theta(I)$ (no line symbols) with $\alpha_1=\alpha_2=0.05$;
90\% Bayesian credible bands (circles); and the
estimated sets $F^\theta(\hat{\phi})$ (crosses). }\setlength{\baselineskip}{4mm}
\end{figure}

\subsection{Pure Sign Restrictions}
\label{subsec_puresign}

In order to make inference about the effects of a contractionary monetary
policy shock, we use the following sign restrictions to bound the identified
set: in periods $h=0,1$ (i) the interest rate response is weakly positive; (ii) the inflation response is weakly negative;
 and (iii) real money balances do not rise above
their steady-state level. These sign restrictions were also used in Monte Carlo Experiment~3 in Section~\ref{subsec_mc_exp3}.

Figure~\ref{f_puresign_4v_h1} depicts three bands: (pointwise) 90\% Bonferroni
confidence intervals (using a diagonal weight matrix) $CS^\theta(I)$, estimated sets $F^\theta(\hat{\phi})$,
and (pointwise) 90\% Bayesian credible sets.
The two most notable features of the bands are that the frequentist confidence bands (solid) are substantially
wider than the Bayesian credible bands (short dashes) and that the Bayesian credible bands approximately
coincide with the estimated set $F^\theta(\hat{\phi})$.
As explained in detail in \cite{MoonSchorfheide2012}, in a large sample, i.e., a sample in which
uncertainty about $\phi$ is small compared to the size of $F^\theta(\hat\phi)$, the Bayesian intervals lie
inside the estimated sets $F^\theta(\hat\phi)$ because in the limit essentially all of the probability mass is concentrated
on $F^\theta(\hat\phi)$ and a 90\% credible interval is always a subset of the support of the posterior
distribution. The frequentist interval, on the other hand, has to extend beyond the boundaries
of $F^\theta(\hat\phi)$ because it has to have, say, 90\% coverage probability for every element
of the identified set $F^\theta(\phi)$, including the boundary points. From a substantive
perspective, the use of sign restrictions leaves the direction of the output response undetermined.

\begin{figure}[t!]
\caption{Pure Sign Restrictions over Horizon $h=0,1,\ldots,8$}
\label{f_puresign_2v_h8}
\begin{center}
	\begin{tabular}{cc}
	\multicolumn{2}{c}{(i) Bonferroni ($\alpha_1=0.05$ and $\alpha_2=0.05$) vs. Bayes} \\
    Output & Inflation \\
	\includegraphics[width=2.5in,trim={0.4in  2.5in 0.5in 3.25in}, clip]{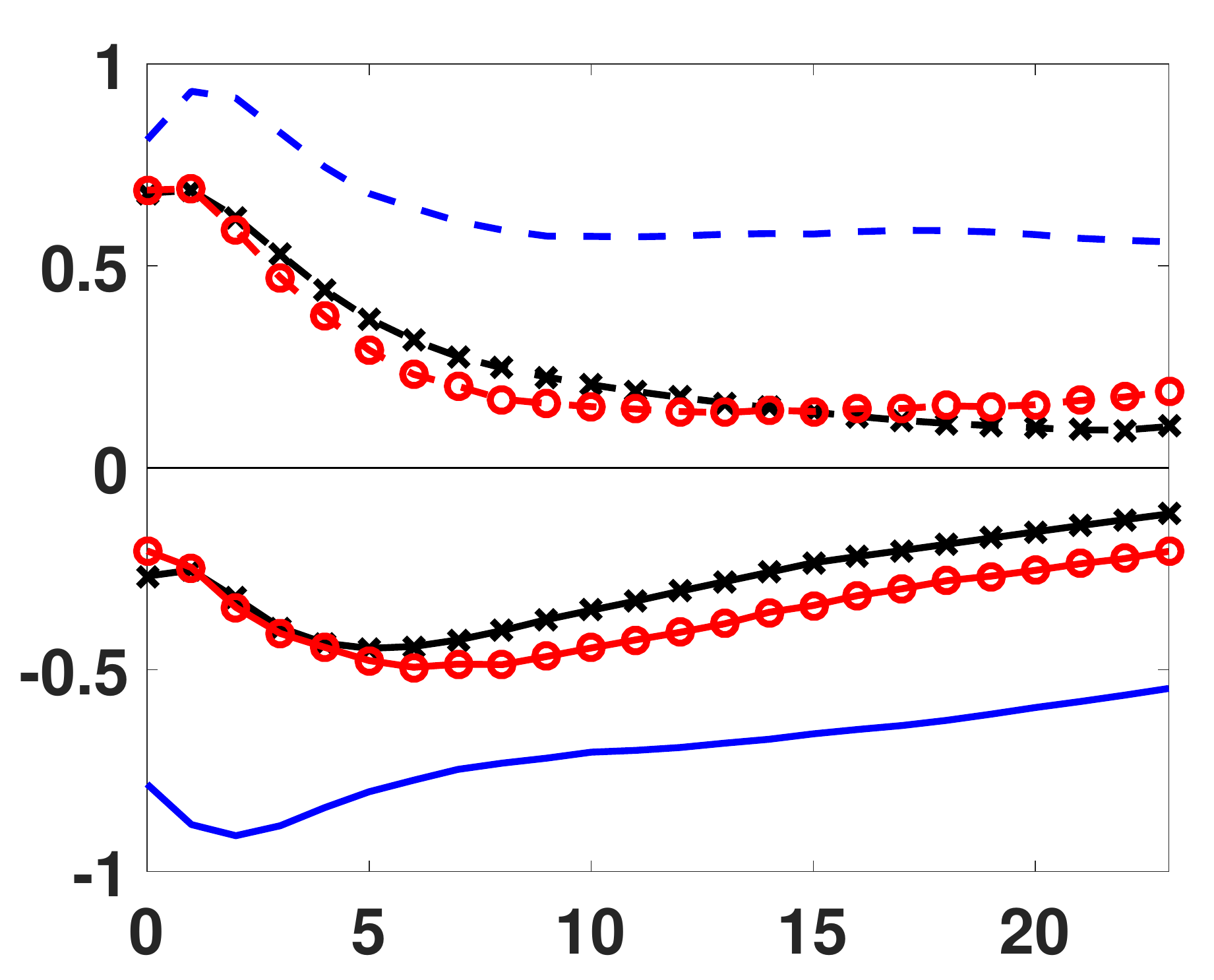} & 
	\includegraphics[width=2.5in,trim={0.4in  2.5in 0.5in 3.25in}, clip]{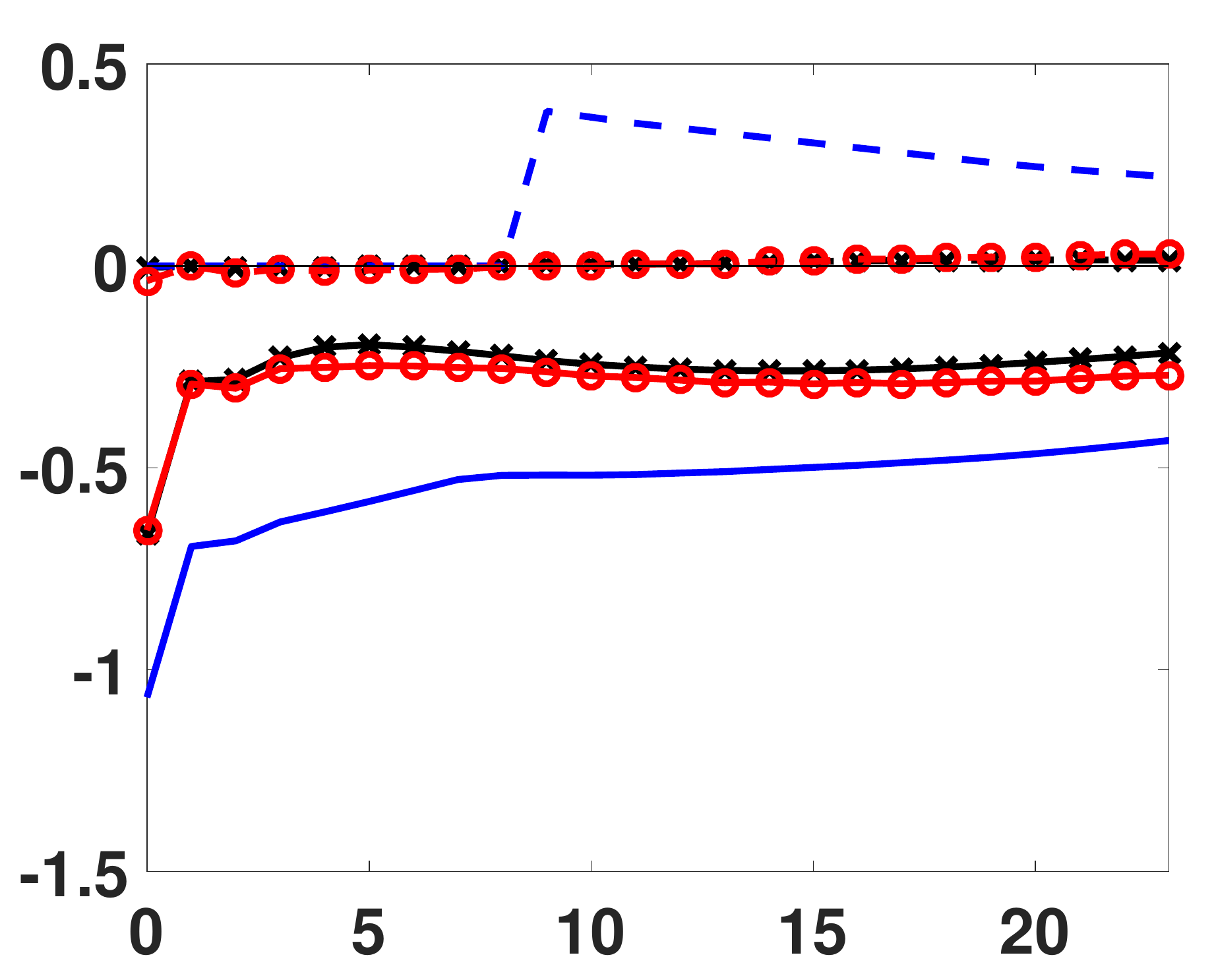} \\    
    \multicolumn{2}{c}{(ii) Effect of Varying $\alpha_2$ Given $\alpha_1=0.05$} \\
    Output & Inflation \\
    \includegraphics[width=2.5in,trim={0.4in  2.5in 0.5in 3.25in}, clip]{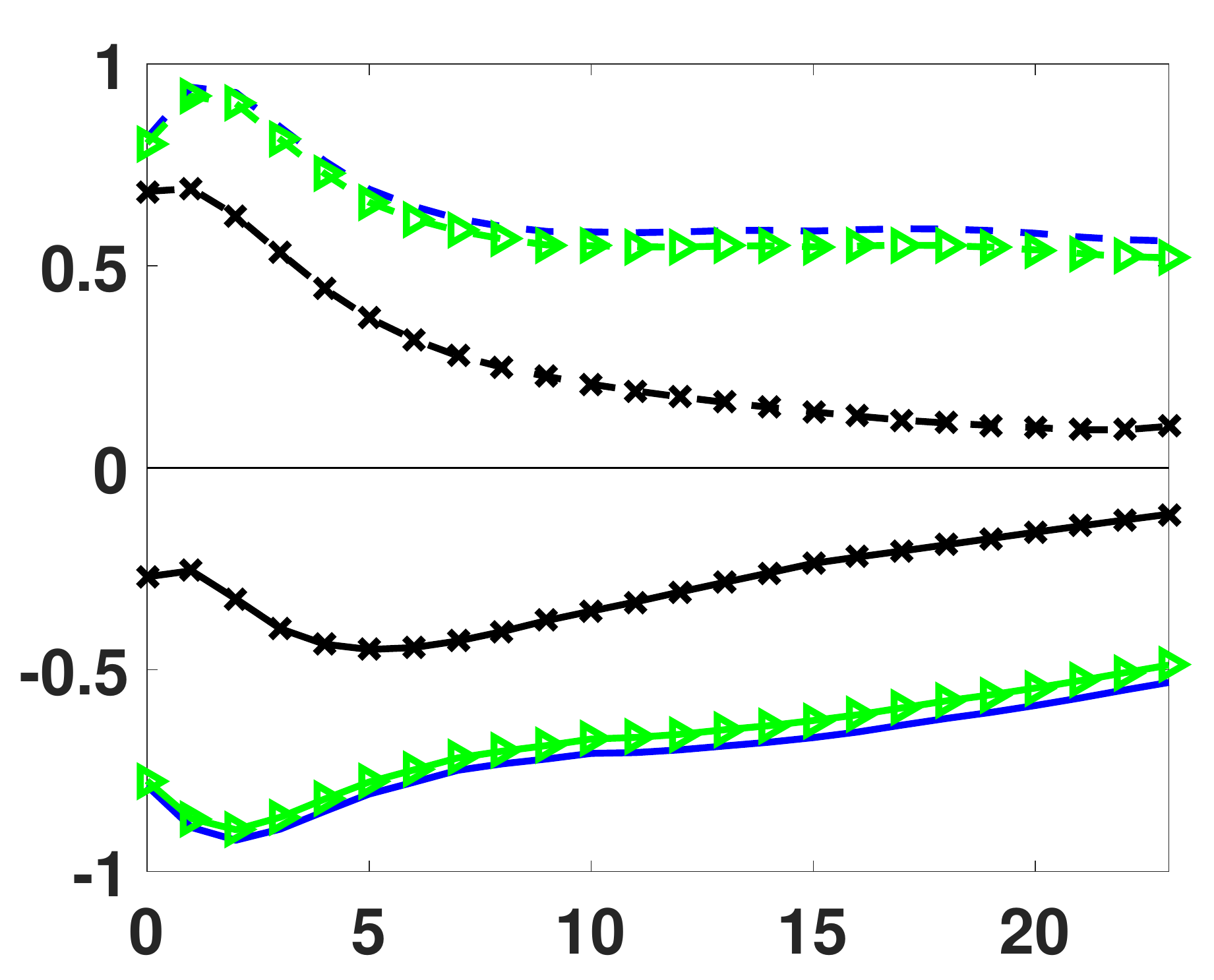} & 
    \includegraphics[width=2.5in,trim={0.4in  2.5in 0.5in 3.25in}, clip]{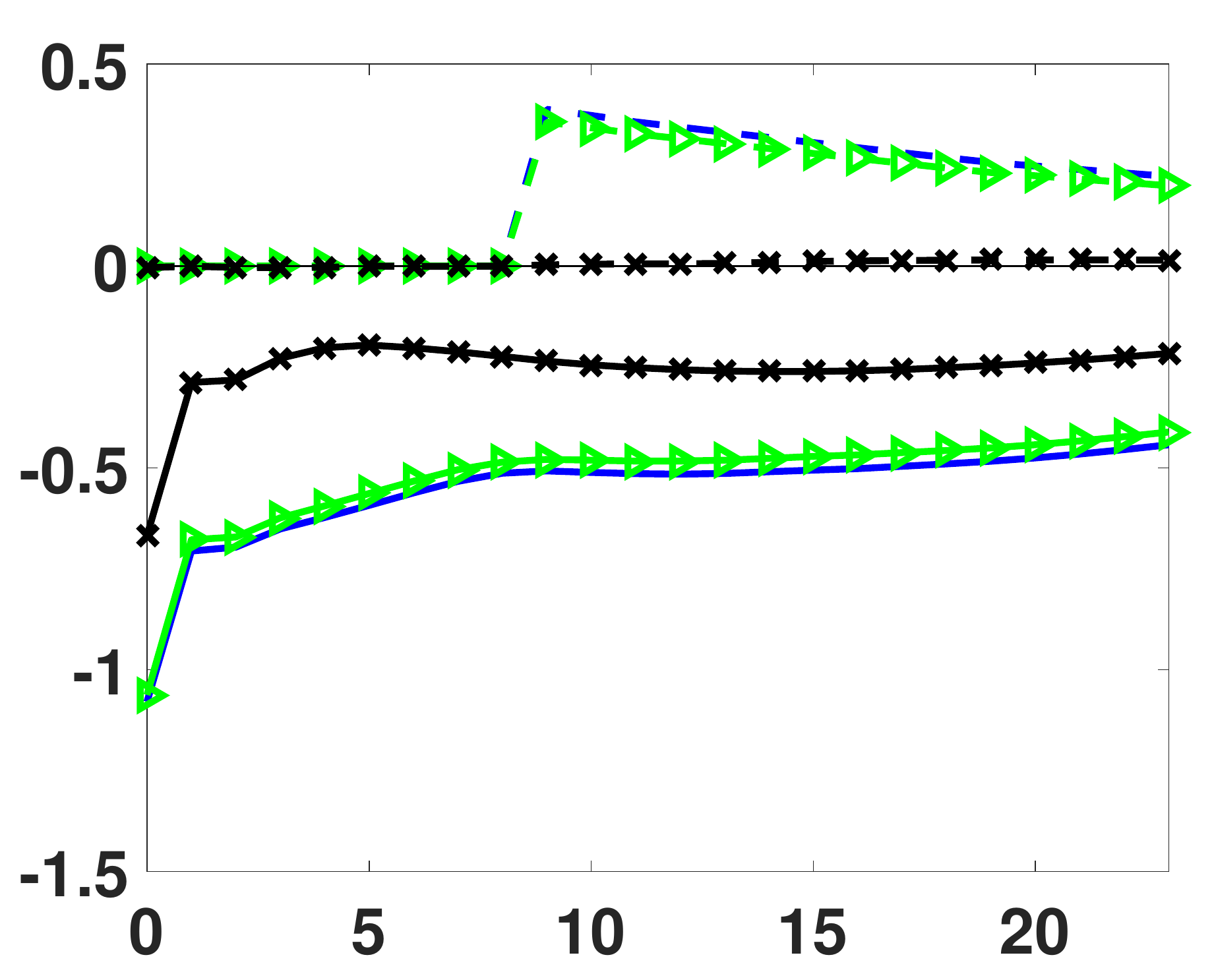} 
    \end{tabular}
\end{center}
{\footnotesize {\em Notes:} Top panels: 
	90\% Bonferroni confidence bands $CS^\theta(I)$ (no line symbols);
	90\% Bayesian credible bands (circles); and the
	estimated sets $F^\theta(\hat{\phi}_q,\hat{\phi}_\theta)$ (crosses).
	Botton panels: Bonferroni confidence bands $CS^\theta(I)$ (no line symbols) with $\alpha_2=0.05$;
	Bonferroni confidence bands $CS^\theta(I)$ (triangles) $\alpha_2=0.10$;
    and the estimated sets $F^\theta(\hat{\phi})$ (crosses).	
}\setlength{\baselineskip}{4mm}
\end{figure}

The top panels of Figure~\ref{f_puresign_2v_h8} show output and inflation responses obtained by requiring the the sign-restrictions hold for periods $h=0,1,\ldots,8$, keeping $\alpha_1=\alpha_2=0.05$. This modification increases the number of inequality restrictions from 6 to 27. As the number of sign restrictions increases, the width of the identified sets decreases.  As suggested by the Monte Carlo simulations, the width of the Bonferroni bands also decreases.
The bottom panels of Figure~\ref{f_puresign_2v_h8} show the effect of raising $\alpha_2$ from 0.05 to 0.10. According to the simulations in Section~\ref{subsec_mc_exp3}, this decrease in the nominal coverage probability brings the actual coverage probability closer to the desired coverage probability of 90\%. As a result the width of the confidence bands shrinks, but not by much. In fact, in percentage terms, the width reduction is very small. 

\subsection{Combining Sign Restrictions and Zero Restrictions}
\label{subsec_signzero}

A commonly used identification assumption for monetary
policy shocks is that private-sector variables such as output
and inflation cannot respond to changes in the federal funds rate within the period; see, for instance,
\cite{ChristianoEichenbaumEvans1999}. Because the initial impact of
the monetary policy shock is given by $\Sigma_{tr}q$ and
we ordered the elements of $y_t$ such that output and inflation appear before interest rates and real money
balances, the identification condition
implies that the first two elements of the vector $q$ have
to be equal to zero. Thus, we can reduce the dimension of the vector $q$ as follows: $q = [0, 0, \cos \varphi, \sin \varphi]'$,
where $\varphi \in [0,2\pi]$. This is more efficient than adding two equality conditions to the set of inequality conditions; see Section~\ref{sec_extensions}.
The zero restriction on
the instantaneous inflation response replaces the sign restriction
used in Section~\ref{subsec_puresign}. We maintain the other sign restrictions used
previously, that is, the interest rate responses for $h=0$ and $h=1$
are weakly positive and  
the inflation response in period $h=1$
as well as the real money balance responses in periods $h=0$ and $h=1$ are
weakly negative.

\begin{figure}[t!]
\caption{Combining Zero and Sign Restrictions over Horizons $h=0,1$}
\label{f_zerosign_2v_h1}
\par
\begin{center}
	\begin{tabular}{cc}
	Output & Inflation \\
	\includegraphics[width=2.5in,trim={0.4in  2.5in 0.5in 3.25in}, clip]{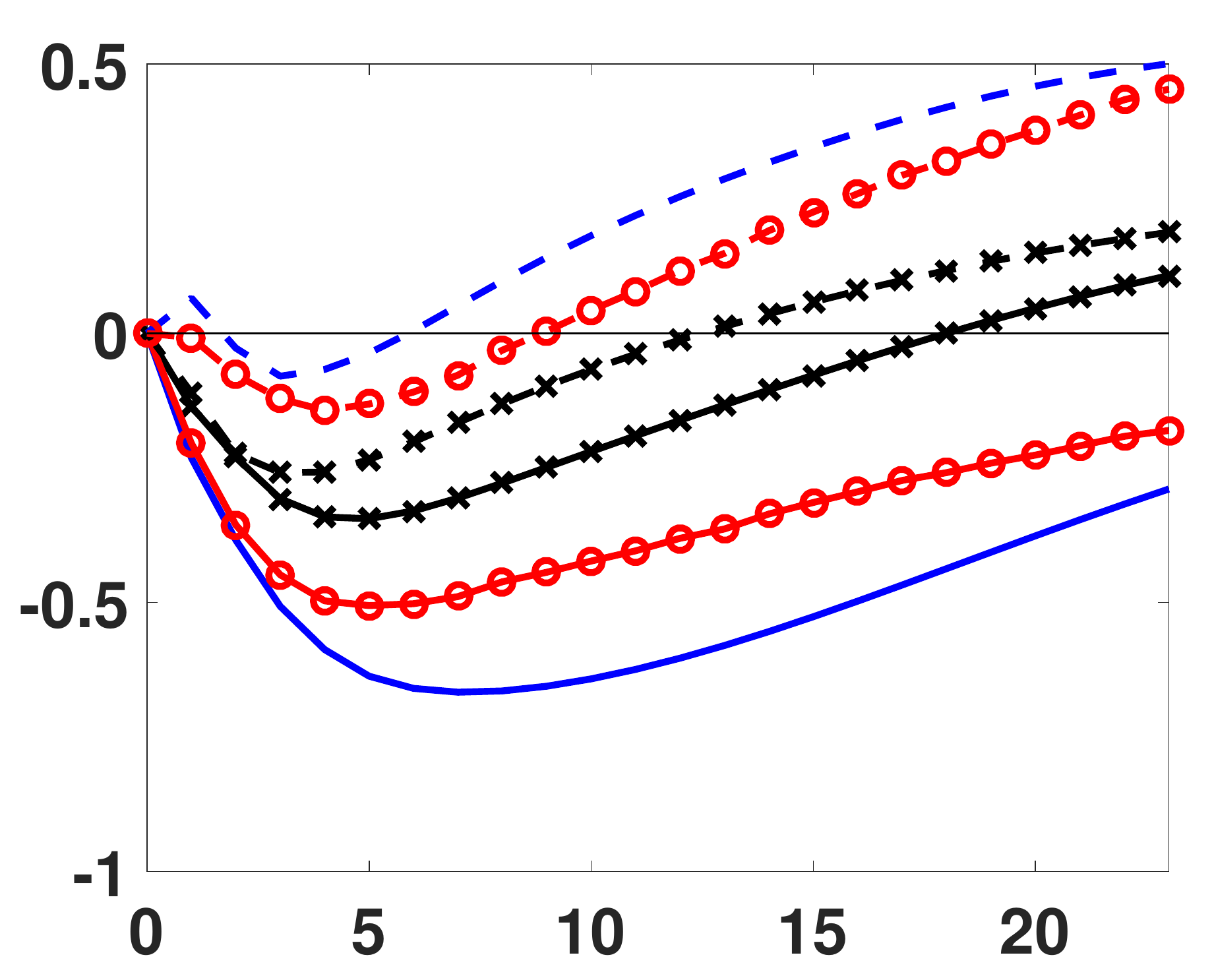} & 
	\includegraphics[width=2.5in,trim={0.4in  2.5in 0.5in 3.25in}, clip]{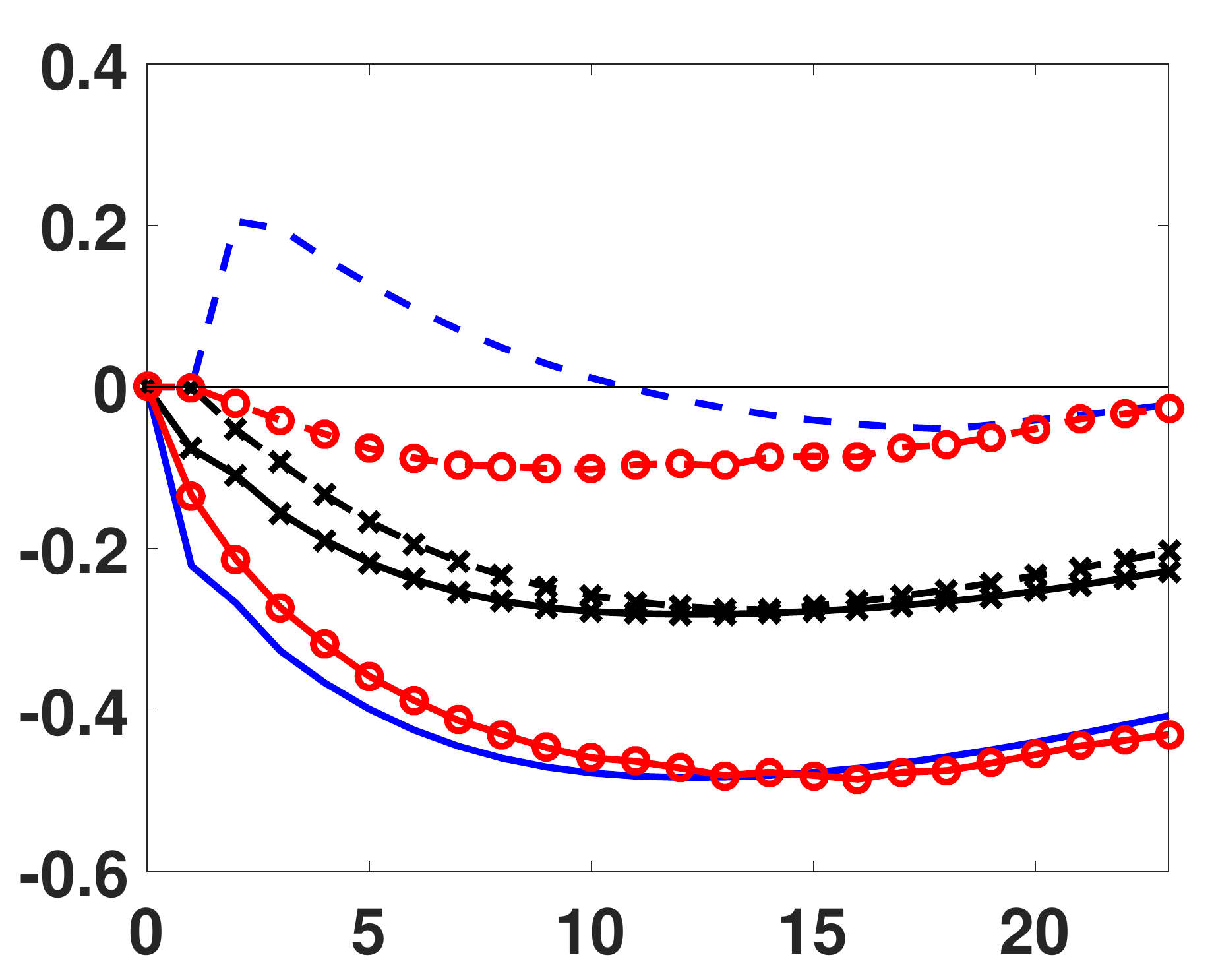} 	
	\end{tabular}
\end{center}
\par
{\footnotesize \emph{Notes:} The figure depicts
	90\% Bonferroni confidence bands $CS^\theta(I)$ (no line symbols) with $\alpha_1=\alpha_2=0.05$;
	90\% Bayesian credible bands (circles); and the
	estimated sets $F^\theta(\hat{\phi})$ (crosses).}\setlength{\baselineskip}{4mm}
\end{figure}

Impulse response bands are depicted in Figure~\ref{f_zerosign_2v_h1}.
A comparison
of $F^\theta(\hat{\phi})$ in Figures~\ref{f_puresign_4v_h1} and~\ref{f_zerosign_2v_h1} indicates
that the use of zero restrictions reduces the size of the identified set drastically.
For instance, if the zero restrictions are imposed, the inflation response is essentially
point identified for horizons exceeding 8 quarters. As a consequence, for output as well as medium- and long-run
inflation responses, the width of the frequentist and Bayesian coverage bands is now much more similar
than under the pure-sign-restriction scenario. However, some differences remain with
respect to the short-run inflation response. For the first two years, the frequentist
intervals cover both positive and negative inflation responses, whereas the Bayesian
credible intervals suggest that the inflation response is negative. With the zero
restrictions imposed, the direction of the output response is no longer ambiguous -- it is negative over the first two years.

%
%

\section{Conclusion}
\label{sec_conclusion}

With the exception of FRSW, the coverage bands for impulse responses of sign-restricted
SVARs that have been reported in the literature thus far were only meaningful from a Bayesian
perspective. The main contribution of our paper is to develop an easy-to-use 
frequentist method based on the Bonferroni approach to construct confidence intervals for impulse responses and other measures of the dynamic effects of structural shocks 
in VARs that are set-identified based on sign restrictions.
In the first stage, a confidence set for the vector of weights $q$ on the reduced-form impulse responses is obtained by inverting a point-wise hypothesis test for the moment inequalities implied by the sign restrictions. We employ the \cite{AndrewsSoares2010a} moment selection procedure to obtain critical values for this test that
are not diluted by non-binding inequality conditions. Our empirical application illustrates that in set-identified VARs, frequentist confidence bands can be substantially wider than Bayesian credible bands.
As a by-product, we establish the consistency of the plug-in estimator $F^\theta(\hat{\phi})$
of the identified set of impulse responses.
$F^\theta(\hat{\phi})$ is also useful from a Bayesian perspective.
Because in a Bayesian analysis, the prior distribution of the impulse response functions
conditional on the reduced-form parameters does not get updated, it is useful
to report the identified set and the prior conditional on some estimate of $\phi$, say, the posterior mean,
so that the audience can judge whether the conditional prior
distribution is highly concentrated in a particular area of the identified set.

\ifx\undefined\textsc
\newcommand{\textsc}[1]{{\sc #1}}
\newcommand{\emph}[1]{{\em #1\/}}
\let\tmpsmall\small
\renewcommand{\small}{\tmpsmall\sc}
\fi

\clearpage

\newpage

%
%

\clearpage

%
%

\begin{appendix}

\markright{Moon, Schorfheide, and Granziera: Online Technical Appendix}
\renewcommand{\thepage}{O-App.\arabic{page}}
\setcounter{page}{1}

\bc
{\Large {\bf Online Technical Appendix}}
\ec

\renewcommand{\theequation}{A.\arabic{equation}}
\setcounter{equation}{0}

\noindent This Online Appendix accompanies the paper ``Inference for VARs Identified with Sign Restrictions'' by
E. Granziera, H.R. Moon, and F. Schorfheide. 
In Section~\ref{appsec_proofs_derivations} we provide proofs for the theoretical results in Section~\ref{sec_largesample} of the main paper. Additional technical Lemmas are stated and proved in Section~\ref{appsec_technicallemma}. Section~\ref{appsec_derivationmc}
provides analytical derivations for the Monte Carlo experiment presented in Section~\ref{sec_mc} of the main text.
Section~\ref{appsec_empirical} contains additional information about the empirical analysis.

\setcounter{section}{0}

\section{Proofs of Main Results}
\label{appsec_proofs_derivations}

\renewcommand{\theequation}{A.\arabic{equation}}
\setcounter{equation}{0}

To simplify the notation in some of the proofs we eliminate $\rho$ from the formulas and
index the probability distribution by $\phi \in {\cal P}$ instead of $\rho \in {\cal R}$. Thus we write
\[
\inf_{\phi \in {\cal P}} \; \inf_{\theta \in F^\theta(\phi)} \; P_\phi \{ \theta \in CS^\theta(\hat{\phi}) \}
\]
instead of
\[
\inf_{\rho \in {\cal R}} \; \inf_{\theta \in F^\theta(\phi(\rho))} \; P_\rho \{ \theta \in CS^\theta(\hat{\phi}) \}.
\]
Reduced-form parameter sequences $\rho_T$ and $\phi(\rho_T)$ are simply abbreviated by $\phi_T$.

%
%

\subsection{Proof of Lemma~\ref{l_boundedsets}}

To simplify the notation we omit tildes and write $S_\theta(q)$, and $S(q)$ instead of
$\tilde{S}_\theta(q)$ and $\tilde{S}(q)$. 

\noindent {\bf Convexity:} Suppose $\theta_i \in F^\theta(\phi_q,\phi_\theta)$, $i=1,2$, and
$\theta_1 < \theta_2$.
Then there exist $q_i$ with $\|q_i\|=1$ and $\mu_i \ge 0$ such that
\be
S_\theta(q_i)\phi - \theta_i = 0, \quad S(q_i)\phi - \mu_i = 0.
\label{eqapp_idsetdef}
\ee
We distinguish two cases: $q_1 \not=-q_2$ and $q_1 =-q_2$.

{\em Case~(i):} Suppose that $q_1 \not= - q_2$. We now verify that for any $\lambda \in [0,1]$
$\theta = \lambda \theta_1 +(1-\lambda) (\theta_2) \in F^\theta(\cdot)$.
For $\tau \in [0,1]$ define
\[
q(\tau) = \frac{ \tau q_1 + (1-\tau)q_2}{\|\tau q_1 + (1-\tau)q_2 \|}, \quad H(\tau) = S_\theta(q(\tau)) \phi - \theta.
\]
The linearity of $S_\theta(q)$ with respect to $q$ and~(\ref{eqapp_idsetdef}) implies that
\begin{eqnarray*}
	H(\tau) &=& \frac{\tau  S_\theta(q_1)\phi}{\|\tau q_1 + (1-\tau)q_2 \|}
	+ \frac{(1-\tau)S_\theta(q_2)\phi}{\|\tau q_1 + (1-\tau)q_2 \|} - \lambda\theta_1 -(1-\lambda)\theta_2 \\
	&=& \frac{\tau  \theta_1}{\|\tau q_1 + (1-\tau)q_2 \|}
	+ \frac{(1-\tau) \theta_2}{\|\tau q_1 + (1-\tau)q_2 \|} - \lambda\theta_1 -(1-\lambda)\theta_2.
\end{eqnarray*}
Using $\|q_i\|=1$ we obtain
\begin{eqnarray*}
	H(0) &=& \theta_2 - \lambda \theta_1 -(1-\lambda)\theta_2 = \lambda(\theta_2-\theta_1) \ge 0 \\
	H(1) &=& \theta_1 - \lambda \theta_1 -(1-\lambda)\theta_2 = -(1-\lambda)(\theta_2-\theta_1) \le 0.
\end{eqnarray*}
Since $H(\tau)$ is continuous we deduce that there exists a $\tau^*$ such that $H(\tau^*) = 0$.
Now consider
\begin{eqnarray*}
	S(q(\tau^*))\phi &=& \frac{\tau^*  S(q_1)\phi}{\|\tau^* q_1 + (1-\tau^*)q_2 \|}
	+ \frac{(1-\tau^*)S(q_2)\phi}{\|\tau^* q_1 + (1-\tau^*)q_2 \|} \\
	&=& \frac{\tau^*  \mu_1}{\|\tau^* q_1 + (1-\tau^*)q_2 \|}
	+ \frac{(1-\tau^*)\mu_2}{\|\tau^* q_1 + (1-\tau^*)q_2 \|} \\
	&\ge& 0.
\end{eqnarray*}
The first equality follows from the linearity of $S(q)$, the second equality is implied by~(\ref{eqapp_idsetdef}),
and the inequality follows from $\mu_i \ge 0$. Thus, $\theta \in F^\theta(\phi_q,\phi_\theta)$.

\noindent {Case~(ii):} Suppose that $q_1 = - q_2$. The linearity of $S_\theta(q)$ implies that
$\theta_1 = - \theta_2$.
By assumption there exists
a $q_3 \not= q_1,-q_1$ with the property that $S(q_3)\phi \ge 0$.
Let $\theta_3 = S_\theta(q_3)\phi$. By construction, $\theta_3 \in F^\theta(\cdot)$.
If $\theta_3 = \theta_1$ ($\theta_3 = \theta_2$) we simply replace $q_1$ ($q_2$)
by $q_3$ and follow the steps outlined for Case~(i).
If $\theta_1 < \theta_3 < \theta_2$,  then the Case~(i) argument implies
that any $\theta$ in the intervals $[\theta_1, \theta_3]$ and $[\theta_3,\theta_2]$
and thereby any $\theta = \lambda \theta_1 + (1-\lambda)\theta_2$ is included
in the identified set. Finally, if $\theta_3 < \theta_1$ ($\theta_2 < \theta_3)$,
we deduce from Case~(i) that the interval $[\theta_3, \theta_2]$ $([\theta_1, \theta_3])$
is included in the identified set.

\noindent {\bf Boundedness:} We shall prove a slightly more general result. Throughout
the proof we omit tildes.
Suppose that $\tilde{\theta} \in F^\theta(\phi_q,\phi_\theta)$.
Since $F^\theta(\phi_q,\phi_\theta)$ is a multiple-value set, we assume without loss of generality that $\tilde{\theta} > 0$.
So, the sign restriction $\theta \ge 0$ is satisfied if it exists. 
Define 
\[
G^\theta(\theta; \phi_q,\phi_\theta)
= \min_{q = \|1 \|, \, \mu \ge 0}  \;
\| S_\theta(q) \phi_\theta - \theta \|^2 + \| S(q)\phi_q - \mu \|^2
\]
such that $G^{\theta}(\theta; \phi_q,\phi_\theta)=0$ if and only if $\theta \in F^\theta(\phi_q,\phi_\theta)$.
We now show by contradiction
that $F^\theta(\phi_q,\phi_\theta)$ has an upper bound. 

Suppose, to the contrary, that no such upper bound exists.
This guarantees the existence of a series $a_n > 0$ with $a_n \uparrow \infty$ such that $a_n \tilde{\theta}_n \in F^\theta(\phi_q,\phi_\theta)$ for each $n$. 
Consider the bound 
\[
G^\theta(a_n \tilde{\theta}; \phi_q,\phi_\theta) \ge \min_{q = \|1 \|} \; \| S_\theta(q) \phi_\theta - a_n \tilde{\theta} \|^2.
\]
Since $\| S_\theta(q)\phi_\theta \|$ is a continuous function of $q$ for fixed $\phi_\theta$ and the set
of $q$ is a compact unit sphere, there exists a finite constant $M$ such that $\| S_\theta(q) \phi_\theta \| < M$.
From this we deduce that
\[
\min_{q = \|1 \|} \; \| S_\theta(q) \phi_\theta - a_n \tilde{\theta} \|^2 \longrightarrow \infty,
\]
which contradicts the requirement $G^{\theta}(\theta; \phi_q,\phi_\theta)=0$.
The existence of a lower bound can be established by considering a sequence $-a_n$. Moreover, $\theta < 0$
can be handled by a straightforward modification of the argument. $\Box$

\subsection{Proof of Theorem~\ref{t_setidentified}}

{\color{black} 
	
Recall the definition $F^q(\Phi_q) = \big\{ q \in \mathbb{S}^n \, \big| \, \Phi_q' q \ge 0 \big\}$.
Thus, $\big\{ q \in \mathbb{S}^n \, | \, \Phi_q' q \gg 0, \big\} \subset F^q(\phi)$. The statement of the theorem 
follows once we have shown that there exists a non-empty, non-singleton, $n$-dimensional
subset $\mathbb{Q}$ of $\mathbb{S}^n$, such that $\Phi_q q \gg 0$ if $q \in \mathbb{Q}$.

\noindent {\bf Existence:} 
Suppose $\Phi_q'$ is an $r \times n$ matrix.
According to Gordan's Alternative Theorem -- see, for instance, Border (2007) --
exactly one of the two alternatives holds: (a) there exists an $x \in \mathbb{R}^n$
satisfying $\Phi_q' x^{*} \gg 0$; or (b) there exists an $r\times 1$ vector $z > 0$ satisfying $\Phi_q z = 0$. Assumption~\ref{a_all}(i)
rules out alternative (b). Thus, there exists an $x^{*}$ such that
\begin{equation}
\Phi_q' x^{*} \gg 0. \label{eq.ap.proof.thm1.existence}
\end{equation}
Notice that $x^{*}$ in (\ref{eq.ap.proof.thm1.existence}) is not zero. Then, $q^{*} := \frac{x^{*}}{\| x^{*} \| }$ satisfies the requirement $q^{*} \in \mathbb{Q}$ and $\Phi_q' q^{*} \gg 0$. 

\noindent {\bf Non-singleton:}  
We show that $F^q(\Phi_q)$ contains multiple elements by the method of contradiction.
For this, we define a function $f_{\Phi} : \mathbb{S}^n \rightarrow \mathbb{R}^r$ as $f_{\Phi}(q) := \Phi_q' q$ for $q \in \mathbb{S}^n$. Then, $f_{\Phi}(\cdot)$ is continuous on a compact set $\mathbb{S}^n$.    

Suppose that $q^{*}$ defined in the existence proof is the only element of $F^q(\Phi_q)$, that is, $F^q(\Phi_q) = \{ q^{*} \}$. This implies that $f_{\Phi}(q) \notin \mathbb{R}^r_{+}$ for all $q \in \mathbb{S}^{n}$ with $q \neq q^{*}$, where $ \mathbb{R}^r_{+}= \{ x \in \mathbb{R}^r : x \geq 0 \}$.
Let $\epsilon := \| \Phi_q' q^{*} \|_{\rm min}$, where the norm $\| x \|_{\rm min} := \min\{ |x_1|,...,|x_r| \}$ for $x \in \mathbb{R}^r$. Notice that $\Phi_q' q^{*} \gg 0$ implies $\epsilon > 0$.   
Consider an arbitrary $ q \in \mathbb{S}^{n}$ such that  $q \neq q^{*}$. Then, because $f_{\Phi}(q) \notin \mathbb{R}_+^r$ but $f_{\Phi}(q^{*}) \gg 0$, we have $ \| f_{\Phi}(q) - f_{\Phi}(q^{*}) \| \geq \epsilon$. Because $q$ was arbitrary, given our choice of $\epsilon > 0$ it is not possible to find
a $\delta > 0$ such that $ \| f_{\Phi}(q) - f_{\Phi}(q^{*}) \| \le \epsilon$ for $\| q - q^*\| \le \delta$. This contradicts the fact that $f_{\Phi}(q)$ is continuous at $q^{*}$.  
Therefore, we can deduce the that $F^{q}(\Phi_q)$ is not a singleton and contains multiple elements. $\Box$}

\subsection{Proof of Theorem~\ref{t_asymptotics_q}, Part (i)}

Recall the definition of the Hausdorff distance:
$d( A,B) = \max \; \left\{ d( A|B), \; d(B|A) \right\}$, where
$d(A|B) = \sup_{a\in A} \; d( a,B)$ and $d(a,B)=\inf_{b\in B}\; \left\Vert a-b\right\Vert$
We set $d( A,B) = \infty$ if either $A$ or $B$ is empty.
For any $\varepsilon >0,$ define an open ball around set $A\subset \mathbb{R}%
^{n}$ as $\mathbb{B}( A,\varepsilon ) =\left\{ b\in \mathbb{R}^{n}:d(
b|A) <\varepsilon \right\}.$

The proof of the theorem exploits the continuity of 
$F^q(\phi)$ with respect to $\phi$. The statement of the theorem is a consequence of
Lemma~\ref{l.continuity.hausdorff}, Lemma~\ref{l.continuity},
and the continuous mapping theorem. $\Box$

\begin{lemma}
	\label{l.continuity.hausdorff} Suppose that $F \left( \phi \right) $ is
	a non-empty compact-valued continuous correspondence. Then, $\phi \longrightarrow
	\phi ^{\ast }$ implies that $d\left( F \left( \phi \right) ,F
	\left( \phi ^{\ast }\right) \right) \longrightarrow 0.$
\end{lemma}

{\color{black} 
\noindent {\bf Proof of Lemma~\ref{l.continuity.hausdorff}}. Follows directly from 
Theorem 17.15 of \cite{AliprantisBorder2006}. $\Box$
}

\begin{lemma}
	\label{l.continuity}Suppose that Assumption~\ref{a_all}(i) is satisfied. Then,
	\begin{tlist}
		\item $F^{q}\left( \Phi_q\right) $ is compact for all $\Phi_q$;
		\item {\color{black}  $F^{q}\left( \Phi_q\right) $ is continuous at all $\Phi_{q}$.}
	\end{tlist}
\end{lemma}

\noindent {\bf Proof of Lemma~\ref{l.continuity}}:
For notational simplicity, we omit the subscript notation $q$ and write $\Phi_q$ as $\Phi$. 
Let $\mathbb{S}^{n}=\left\{ q\in \mathbb{R}^{n}:\left\Vert
q\right\Vert =1\right\} $ be the unit sphere in $\mathbb{R}^{n}.$
Recall from Theorem~\ref{t_setidentified} that $F^q(\Phi)$ is nonempty.

\noindent {\bf Part (i):} We show that $F^{q}\left( \Phi\right) $ is bounded and
closed.

\noindent {\em Boundedness}: It is straightforward since $F^{q}\left( \Phi\right)
\subset \mathbb{S}^{n}.$

\noindent {\em Closedness}: Consider any sequence $q_{j}\in F^{q}\left( \Phi\right) $
such that $q_{j}\longrightarrow q$, where $\|q_j\|=1$ and $\|q \|=1$. Then, $0\leq \Phi' q_{j}\longrightarrow
\Phi' q,$ so that it should be $\Phi' q \geq 0.$ This implies that $%
q_{0}\in F^{q}\left( \Phi\right) ,$ as required for closedness.

\noindent {\bf Part (ii):} We show $F^{q}\left( \Phi\right) $ is upper hemi-continuous
(UHC) and lower hemi-continuous (LHC) at $\Phi.$

\noindent {\em UHC}: Since $F^{q}\left( \Phi\right) $ is non-empty and compact-valued,
the UHC of $F^{q}\left( \Phi\right) $ at $\Phi$ follows if we show that for
every sequence $\Phi_{j}\longrightarrow \Phi$ and $q_{j}\in F^{q}\left(
\Phi_{j}\right),$ there exists a subsequence $q_{j_{i}}$ of $q_{j}$ such that $%
q_{j_{i}}\longrightarrow q \in F^{q}\left( \Phi \right) $. (See Border (2010)
Proposition 20). Since $\left\{ q_{j}\right\} \subset \mathbb{S}^{n}$ and $%
\mathbb{S}^{n}$ is compact, we can choose a convergent subsequence $q_{j_{i}}
$ such that $q_{j_{i}}\longrightarrow q.$ Then, $0\leq
\Phi_{j_{i}}'q_{j_{i}}\longrightarrow \Phi'q,$ and it follows that $%
\Phi'q\geq 0.$ This implies that $q \in F^{q}\left( \Phi \right) ,$
as required.\smallskip

\noindent {\em LHC}: $F^{q}\left( \Phi\right) $ is LHC at $\Phi$ if and only if for
any sequence $\left\{ \Phi_{j}\right\} $ with $\Phi_{j}\longrightarrow \Phi$ and $%
q \in F^{q}\left( \Phi \right) ,$ there exists a sequence $q_{j}\in
F^{q}\left( \Phi_{j}\right) $ with $q_{j}\longrightarrow q.$ 
We re-order and partition
the matrix $\Phi_{0}$ to $\Phi = \left[ \Phi_{1},\Phi_{2}\right],$ where $\Phi_{1}'q=0$ and $\Phi_{2}'q \gg 0.$

For a matrix $A,$ we denote the $l^{th}$ column of $A$ as $\left( A \right)_l.$
By Gordan's Alternative Theorem, see Border (2007),
Assumption~\ref{a_all}(i) implies that there exists a $\xi^* \in \mathbb{R}^m$
such that
\[
\Phi_{1}' \xi^* \gg 0.
\]
Let
\[
\xi = \frac{1}{\min_l \; (\Phi_1)_l' \xi^*} \xi^*
\]
such that for all $l$
\[
(\Phi_{1})_l' \xi > 1.
\]
Set $\epsilon _{j,l}=\left\Vert \left(
\Phi_{j}-\Phi \right)_{l}\right\Vert $ and $\epsilon _{j}=\max_{l}\left\{
\epsilon _{j,l}\right\} ,$ and define
\[
q_{j}=\frac{q +\epsilon _{j}\xi }{\left\Vert q + \epsilon _{j}\xi
	\right\Vert }.
\]%
Notice that $q_{j}$ is well defined when $\epsilon _{j}$ is small enough
because $q \in \mathbb{S}^{n}$ and as
a result, $q \neq \epsilon _{j}\xi $ when $\epsilon _{j}$ is small and $%
\xi $ is fixed.

\noindent Case (i): Suppose $\left( \Phi_{l} \right)'q =0.$ Then, when $j$ is large
so that $\left( \Phi \right)_{l}'\xi -1\geq \epsilon _{j}\|\xi\|$, we have
\begin{eqnarray*}
	\left( \Phi_{j}\right)_{l}'q_{j} &=&\left( \Phi_{r,j}-\Phi_{r}\right)
	_{l}'q_{j}+\left( \Phi \right)_{l}'q_{j} \\
	&=&\frac{1}{\left\Vert q + \epsilon _{j}\xi \right\Vert }\left\{ \left(
	\Phi_{j}-\Phi \right)_l'q + \epsilon _{j}\left( \Phi_{j}-\Phi \right)_
	{l}'\xi +\left( \Phi \right)_{l}'q + \epsilon _{j}\left( \Phi \right)_{l}'\xi \right\}  \\
	&\geq &\frac{1}{\left\Vert q + \epsilon _{j}\xi \right\Vert }\left\{
	-\left\Vert \left( \Phi_{j}-\Phi \right)_{l} \right\Vert \left\Vert
	q \right\Vert -\epsilon _{j}\left\Vert \left( \Phi_{j}-\Phi \right)_{l}\right\Vert \left\Vert \xi \right\Vert +\epsilon _{n}\left(\Phi \right)_{l}'\xi \right\}  \\
	&\geq &\frac{1}{\left\Vert q + \epsilon _{j}\xi \right\Vert }\left(
	-\epsilon _{j}-\epsilon _{j}^{2}\|\xi\|+\epsilon _{j}\left( \Phi \right)_{l}'\xi
	\right)  \\
	&= &\frac{1}{\left\Vert q + \epsilon _{j}\xi \right\Vert }\epsilon
	_{j}\left( \left( \Phi \right)_{l}'\xi -1-\epsilon _{j}\|\xi\|\right)  \\
	&\geq &0.
\end{eqnarray*}

\noindent Case (ii): Suppose $\left( \Phi \right)_{l}'q > 0.$ Then, since $\left\Vert
\left( \Phi \right)_{l}\right\Vert \leq M$ (compact parameter set), we
have
\begin{eqnarray*}
	\left( \Phi_{j}\right)_{l}'q_{j}
	&=&\left( \Phi_{j}-\Phi \right)_{l}'q_{j}+\left( \Phi \right)_{l}'q_{j} \\
	&=&\frac{1}{\left\Vert q + \epsilon _{j}\xi \right\Vert }\left\{ \left(
	\Phi_{j}-\Phi \right)_{l}'q + \epsilon _{j}\left( \Phi_{j}-\Phi \right)_
	{l}'\xi +\left( \Phi \right)_{l}'q + \epsilon _{j}\left( \Phi \right)_{l}'\xi \right\}  \\
	&\geq &\frac{1}{\left\Vert q + \epsilon _{j}\xi \right\Vert }\left\{
	-\left\Vert \left( \Phi_{j}-\Phi \right)_{l}\right\Vert \left\Vert
	q \right\Vert -\epsilon _{j}\left\Vert \left( \Phi_{j}-\Phi \right)_
	{l}\right\Vert \left\Vert \xi \right\Vert +\left( \Phi \right)_{l}'q - \epsilon _{j}\left\Vert \left( \Phi \right)_{l}\right\Vert
	\left\Vert \xi \right\Vert \right\}  \\
	&\geq &\frac{1}{\left\Vert q + \epsilon _{j}\xi \right\Vert }\left( \left(
	\Phi \right)_{l}'q -\epsilon _{j}-\epsilon _{j}^{2}M-\epsilon
	_{j}M^{2}\right)  \\
	&\geq &0,
\end{eqnarray*}%
when $j$ is large. The last inequality holds since $\left( \Phi \right)_{l}'q > 0.$

From these, we can deduce that
\[
\Phi_{j}'q_{j}\geq 0.
\]%
Also, since $\epsilon _{j}\longrightarrow 0,$ we have
\[
q_{j}\longrightarrow q.
\]%
Then, we have all the required results for the LHC. $\Box$

\subsection{Proof of Theorem~\ref{t_asymptotics_q}, Part (ii)}

We closely follow the proofs of Theorem 1 and Lemma 2 of Andrews and Soares (2010). The main modification is to accommodate the reduced rank possibility
of $\Sigma \left( q\right) $ and $D\left( q\right)$. The proof makes use of various lemmas that
are stated and proved in Section~\ref{appsec_technicallemma} below. 
To simplify the notation we eliminate $\rho$ from the formulas and
index the probability distribution by $\phi \in {\cal P}$ instead of $\rho \in {\cal R}$. We also skip the subscription notation $q$ and write, for example, $\phi_q, \hat{\phi}_q, \Lambda_{qq}, D_q$ as $\phi, \hat{\phi}, \Lambda,D$, respectively. Thus, we write
\[
\inf_{\phi \in {\cal P}} \; \inf_{q \in F^q(\phi)} \; P_\phi \{ q \in CS^q(\hat{\phi}) \} \quad \mbox{instead of} \quad
\inf_{\rho \in {\cal R}} \; \inf_{q \in F^q(\phi(\rho))} \; P_\rho \{ q \in CS^q(\hat{\phi}_q) \}.
\]
Reduced-form parameter sequences $\rho_T$ and $\phi(\rho_T)$ are simply abbreviated by $\phi_T$.

We need to show
\begin{equation}
\lim \inf_{T}\inf_{\phi \in {\cal P}}\inf_{q \in
	F^{q}\left( \phi \right) }P_{\phi}\left \{ q \in CS^{q}\right \} \geq 1-\alpha .  \label{eq.desired.result.thm}
\end{equation}%
Let
\[
AsyCP=\lim \inf_{T}\inf_{\phi \in {\cal P} }\inf_{q \in
	F^{q}\left( \phi \right) }P_{\phi }\left \{ q \in CS^{q}\right \} .
\]%
Then, there exists sequences $\left \{ \phi _{T},q_{T}\right \} $
such that $q_{T} \in F^{q}\left(\phi_{T}\right) $ and
\[
AsyCP=\lim \inf_{T}P_{\phi_{T}}\left\{ q_{T} \in
CS^{q}\right \} .
\]
Furthermore, there exists a subsequence of $\left \{ T\right \} ,$ $\left \{
T^{\prime }\right \} \subset \left \{ T\right \} ,$ such that
\[
AsyCP=\lim_{T^{\prime }}P_{\phi _{T^{\prime }}}\left \{ q_{T^{\prime }} \in CS^{q}\right \} .
\]
In what follows we show that there exists a subsubsequence, say $\left \{
T^{\prime \prime }\right \} \subset \left \{ T^{\prime }\right \}$, such that
\begin{equation}
\lim_{T^{\prime \prime }}P_{\phi _{T^{\prime \prime }}}\left \{ q_{T^{^{\prime \prime }}} \in CS^{q}\right \} \geq 1-\alpha .  \label{eq.thm.desired.cp}
\end{equation}%
Then, the desired result $\left(\ref{eq.desired.result.thm}\right)$ follows and the proof of the theorem is complete.

Define 
\[
\mu \left( q,\phi \right) =S\left( q\right) \phi
\] 
and decompose
\[
\Sigma(q) =S(q) \Lambda S(q) ^{\prime} = S(q) LL' S(q) ^{\prime}
= D^{1/2}(q) \Omega(q) D^{1/2}.
\]
Moreover, let
\[
A( q) =L^{\prime }S^{\prime }(
q) D^{-1/2}( q).
\]
To simplify the notation we suppress the dependence of matrices on $\phi$.
The matrix $\Omega \left( q\right) =A^{\prime }\left( q\right) A\left( q\right) $
is a correlation matrix and $D^{1/2}\left( q\right) $ is a diagonal matrix
of standard deviations. Also, recall that $W\left( q\right) =D^{1/2}\left( q\right)
B\left( q\right) D^{1/2}\left( q\right) ,$ where either $B\left( q\right)
=\Omega ^{-1}\left( q\right) $ or $B\left( q\right) =I.$
The proof is completed in three steps.

\noindent {\bf Step 1: Choosing the subsequence $T^{\prime \prime}$.}
We choose a subsequence $T^{\prime \prime }$ from $T^{\prime }$ along
which the subsequent conditions are satisfied. This is done sequentially by choosing a subsequence that satisfies criterion (i), and then, step-by-step choosing subsequences of the subsequences to satisfy the next criterion until all five conditions are satisfied:

\begin{tlist}
	\item $\phi _{T^{\prime \prime }}\longrightarrow \phi $.
	\item $r\left(
	q_{T^{\prime \prime }}\right) =r,$ $V\left( q_{T^{\prime \prime }}\right) =V$ for all $T^{\prime
		\prime }.$
	\item For $j=1,...,r,$ the slackness (recall that $\mu_j = \big[S(q)\phi_q \big]_j {\cal I}\big\{ \big[S(q)\phi_q \big]_j \ge 0  \big\}$) in
	inequality $j$ converges to
	\[
	\begin{array}{rcc}
	\sqrt{T^{\prime \prime }}\mu _{j}\left( q_{T^{\prime \prime }},\phi
	_{T^{\prime \prime }}\right) & \longrightarrow & h_{j} \\
	\kappa _{T^{\prime \prime }}^{-1}D_{jj}^{-1/2}\left( q_{T^{\prime \prime
		}}\right) \sqrt{T^{\prime \prime }}\mu _{j}\left( q_{T^{\prime \prime
	}},\phi _{T^{\prime \prime }}\right) & \longrightarrow & \pi _{j}%
	\end{array}%
	\]%
	such that one of the following is true: (a) $h_{j}<\infty $ and $\pi _{j}=0;$
	(b) $h_{j}=\infty $ and $\pi _{j}<\infty ;$ (c) $h_{j}=\infty $ and $\pi
	_{j}=\infty .$
	\item The sequence $A\left( q_{T^{\prime \prime }}\right) $
	has a full rank limit, denoted by $A.$
\end{tlist}

{\color{black} We can satisfy condition~(i) because the reduced-form parameter set $\mathcal{R}$
is assumed to be compact (Assumption~\ref{a_all}(i)) and the function $\phi(\rho)$ is continuously differentiable (Assumption~\ref{a_all}(ii)). As remarked in the main text, this implies that the parameter set for $\phi$ is also compact. If condition~(i) holds, then we obtain:}

\begin{tlist}
	\setcounter{pkt}{4}
	\item The convergence\ $\phi _{T^{\prime \prime }}\longrightarrow \phi $
	implies that $\Lambda \left( \phi _{T^{\prime \prime }}\right) \longrightarrow
	\Lambda \left( \phi \right) $ since $\Lambda \left( \phi \right) $ is
	continuous by Assumption~\ref{a_all}(v). Also, $\hat{\Lambda}%
	(\hat{\phi}_{T^{\prime \prime }}) \stackrel{p}{\longrightarrow} \Lambda $ by
	Assumption~\ref{a_all}(v).
\end{tlist}

Condition~(ii) is satisfied since $r( q_{T^{\prime }})$
and $V( q_{T^{\prime }})$ are sequences that take only a finite
number of discrete values. Condition~(iii) is satisfied because the range of
the sequences of interest is $[ 0,\infty ] $ and by a similar
argument used in the proof of Theorem~1 of Andrews and Soares (2010b).
Roughly speaking, in Case~(iii)-(a) the slackness is small and the
selection criterion regards the inequality asymptotically as binding. In
Case~(iii)-(c) the slackness is large and the selection criterion regards
the inequality as non-binding and (iii)-(b) is an intermediate case.
Condition~(iv) is satisfied according to Lemma~B~\ref{l_lemmaB5}.
If Condition~(iv) is satisfied, then the following conditions
also hold ((vii) is a consequence of Lemma~B~\ref{l_lemmaB5}):

\begin{tlist}
	\setcounter{pkt}{5}
	\item $\Omega \left( q_{T^{\prime \prime }}\right) \longrightarrow
	A^{\prime }A>0$ and $B\left( q_{T^{\prime \prime }}\right) \longrightarrow B>0,$
	where $B=\left( A^{\prime }A\right) ^{-1}$ if $B\left( q\right) =\Omega
	^{-1}\left( q\right) $ and $B=I$ if $B\left( q\right) =I.$
	\item $\hat{\Omega}\left( q_{T^{\prime \prime }}\right) \stackrel{p}{\longrightarrow}
	A^{\prime }A>0$ and $\hat{B}\left( q_{T^{\prime \prime }}\right) \stackrel{p}{\longrightarrow}
	B>0,$ where $B=\left( A^{\prime }A\right) ^{-1}$ if $\hat{B}\left( q\right) =%
	\hat{\Omega}^{-1}\left( q\right) $ and $B=I$ if $\hat{B}\left( q\right) =I.$
\end{tlist}

We now reorder the rows of $S\left( q_{T^{\prime \prime }}\right) $ such
that $\pi _{j}=0$ for rows $j=1,\ldots,r_{1}$ and $\pi _{j}>0$ for rows $%
j=r_{1}+1,\ldots,r.$ Along the sequence $T''$, the last $r_2=r-r_1$ restrictions correspond to non-binding moment inequalities. In the subsequent steps we show that the inequality selection procedure used in the critical-value computation in~(\ref{eq_cqalpha}) asymptotically underestimates $r_2$ (and thereby overestimates $r_1$), which makes the critical value asymptotically conservative to achieve the uniform coverage requirement.


\noindent {\bf Step 2: Constructing an upper bound for the critical value $c^{\alpha }\left( q\right)$ in~(\ref{eq_cqalpha}).}
For notational simplicity we use sequence notation $\left \{ T\right \} $
for the subsubsequence $\left \{ T^{\prime \prime }\right \} $ in Step~1.
Recall the definitions
\[
\xi_{j,T}(q_{T})=D_{jj}^{-1/2}(q_{T})%
\sqrt{T}\mu _{j}(q_{T},\hat{\phi}) \quad \mbox{and} \quad
\hat{\xi}_{j,T}(q_{T})=\hat{D}_{jj}^{-1/2}(q_{T})%
\sqrt{T}\mu _{j}(q_{T},\hat{\phi}).
\]
Let $\varphi_{T}(q_T)$ and $\hat{\varphi}_T$
be vectors with elements
\[
\varphi_{j,T}(q_{T})=\left \{
\begin{array}{cc}
\infty & \text{if }\xi_{j,T}(q_{T})\geq \kappa _{T} \\
0 & \text{otherwise}%
\end{array}%
\right.
\quad \mbox{and} \quad
\hat{\varphi}_{j,T}(q_{T})=\left \{
\begin{array}{cc}
\infty & \text{if }\hat{\xi}_{j,T}(q_{T})\geq \kappa _{T} \\
0 & \text{otherwise}%
\end{array}%
\right.,
\]%
respectively. Moreover, define $\varphi_{T}^{\ast }(q_{T})$
and $\hat{\varphi}_{T}^{\ast }(q_{T})$ with elements
\[
\varphi_{j,T}^{\ast }(q_{T})=\left \{
\begin{array}{cc}
\varphi_{j,T}(q_{T}) & \text{if }\pi _{j}=0 \\
\infty & \text{otherwise}%
\end{array}%
\right.
\quad \mbox{and} \quad
\hat{\varphi}_{j,T}^{\ast }(q_{T})=\left \{
\begin{array}{cc}
\hat{\varphi}_{j,T}(q_{T}) & \text{if }\pi _{j}=0 \\
\infty & \text{otherwise}%
\end{array}%
\right. ,
\]%
where, according to Case~(iii) in Step~1,
\[
\pi_j = \lim \;
\kappa_{T}^{-1}D_{jj}^{-1/2}(q_{T}) \sqrt{T} \mu_{j} (q_{T},\phi_{T}) .
\]
Finally, define the vector $\pi_*$ with elements
\[
\pi _{j}^{\ast }=\left \{
\begin{array}{cc}
0 & \text{if }\pi _{j}=0 \\
\infty & \text{otherwise}%
\end{array}%
\right. .
\]%

To characterize the critical values, define
the objective function
\[
\mathcal{\bar{G}} \big( q_{T};A(\cdot) ,B(\cdot),\varphi(\cdot) \big) = \min_{v\geq -\varphi(q_T) } \; \left \Vert A\left(
q_{T}\right) ^{\prime }Z_{m}-v\right \Vert _{B\left( q_{T}\right) }^{2}.
\]%
Note that the notation in the proof is slightly different from the notation in the main
text. In~(\ref{eq_calgbar}) of the main text we defined
$\mathcal{\bar{G}}\big(q;\hat{B}(q),M_{\hat{\xi}}(q)\big)$, which corresponds
to $\mathcal{\bar{G}}\big( q_{T}; \hat{A}(\cdot),\hat{B}(\cdot), \hat{\varphi}(\cdot) \big)$
in this proof. We let
\be
c_{T}^{\alpha }\big( A\left( \cdot \right) ,B\left( \cdot \right) ,\varphi(\cdot)
\big)  = \left( 1-\alpha \right) \, \text{quantile of }\mathcal{%
	\bar{G}}\left( q_{T};A\left( \cdot \right) ,B\left( \cdot
\right) ,\varphi(\cdot) \right) .
\label{eq_cTalphafcn1}
\ee
To cover the special case $r=r_{2}>0$, i.e., all the
inequality conditions are non-binding, we adopt the convention that
\be
c_{T}^{\alpha }\left( A\left( \cdot \right) ,B\left( \cdot \right) ,\varphi(\cdot)
\right) =0
\label{eq_cTalphafcn2}
\ee
if $\varphi(q_T) =\hat{\varphi}_{T}^{\ast }(q_{T})$ or $\varphi(q_T) = \pi^*$.
The critical value $c^\alpha(q)$ in~(\ref{eq_cqalpha}) in the main text can be expressed
as
\[
c^{\alpha }\left( q_{T}\right) =c_{T}^{\alpha }\left( \hat{A}\left(
q_{T}\right) ,\hat{B}\left( q_{T}\right) ,\hat{\varphi}_{T}\left(
q_{T}\right) \right) .
\]%
Notice by definition that
\[
\hat{\varphi}_{T}^{\ast }(q_{T})\geq \hat{\varphi}_{T}(q_{T}).
\]%
This implies that
\begin{equation}
c_{T}^{\alpha }\left( \hat{A}\left( q_{T}\right) ,\hat{B}\left( q_{T}\right)
,\hat{\varphi}_{T}^{\ast }\left( q_{T}\right) \right) \leq c_{T}^{\alpha
}\left( \hat{A}\left( q_{T}\right) ,\hat{B}\left( q_{T}\right) ,\hat{\varphi}%
_{T}\left( q_{T}\right) \right) .  \label{eq.ap.thm.cvinequality}
\end{equation}

\noindent {\bf Step 3: Establish the asymptotic coverage probability.}
Along the sequence defined in Step~1 we will show the desired
result
\[
AsyCP=\lim_{T}P_{\phi _{T}}\left \{ \left( q_{T}\right) \in
CS^{q}\right \} \geq 1-\alpha.
\]%
We consider two different cases: (i) some inequalities are ``binding'', i.e., $r_1>0$;
(ii) all inequalities are ``non-binding'', i.e., $r_1=0$. 

\noindent {\bf Step 3(i).} Suppose that  $r_1>0$. By Lemma~B~\ref{l_lemmaB2}
and $\left( \ref{eq.ap.thm.cvinequality}\right),$ we have
\begin{eqnarray*}
	AsyCP &=&\lim_{T}P_{\phi _{T}}\left \{ q_{T} \in
	CS^{q}\right \}  \\
	&=&\lim_{T}P_{\phi _{T}}\left \{ G\left( q_{T};\hat{\phi},\hat{W}%
	\left( \cdot \right) \right) \leq c_{T}^{\alpha }\left( \hat{A}\left(
	q_{T}\right) ,\hat{B}\left( q_{T}\right) ,\hat{\varphi}_{T}\left(
	q_{T}\right) \right) \right \}  \\
	&=&\lim_{T}P_{\phi _{T}}\left \{ G\left( q_{T};\hat{\phi},W\left(
	\cdot \right) \right) +o_{p}\left( 1\right) \leq c_{T}^{\alpha }\left( \hat{A%
	}\left( q_{T}\right) ,\hat{B}\left( q_{T}\right) ,\hat{\varphi}_{T}\left(
	q_{T}\right) \right) \right \} \text{ } \\
	&\geq &\lim_{T}P_{\phi _{T}}\left \{ G\left( q_{T};\hat{\phi}%
	,W\left( \cdot \right) \right) +o_{p}\left( 1\right) \leq c_{T}^{\alpha
	}\left( \hat{A}\left( q_{T}\right) ,\hat{B}\left( q_{T}\right) ,\hat{\varphi}%
	_{T}^{\ast }\left( q_{T}\right) \right) \right \} .\text{ }
\end{eqnarray*}%
By using an argument similar to that used in showing (A.10) of Andrews and
Guggenberger (2009), it can be shown that
\begin{eqnarray*}
	\lefteqn{ G\left( q_{T};\hat{\phi},W\left( \cdot \right) \right)
		+o_{p}\left( 1\right) } \\
	&=&\min_{v\geq -D_{R}^{-1/2}(q_{T})\sqrt{T}\mu
		(q_{T},\phi _{T})}\left \Vert A(q_{T})^{\prime }L^{-1}\sqrt{T}\left( \hat{\phi%
	}-\phi _{T}\right) -v\right \Vert _{B\left( q_{T}\right)
}^{2}+o_{p}\left( 1\right)  \\
&\Longrightarrow &\min_{v\geq -h}\left \Vert A^{\prime }Z_{m}-v\right \Vert
_{B} \\
&\leq &\min_{v\geq -\pi ^{\ast }}\left \Vert A^{\prime
}Z_{m}-v\right \Vert _{B}.
\end{eqnarray*}%
The last inequality holds because $h\geq \pi ^{\ast }.$ (This is true
because $\pi _{j}=0$ implies that $h_{j}<\infty $ and $\pi _{j}^{\ast }=0,$
while $\pi _{j}>0$ implies that $h_{j}=\pi _{j}^{\ast }=\infty .)$ According
to Lemma~B~\ref{l_lemmaB9},
\[
c_{T}^{\alpha }\left( \hat{A}\left( q_{T}\right) ,\hat{B}\left( q_{T}\right)
,\hat{\varphi}_{T}^{\ast }\left( q_{T}\right) \right) \stackrel{p}{\longrightarrow}
c_{T}^{\alpha }\left( A,B,\pi ^{\ast }\right) .
\]%
Since $r>r_{2},$ $c_{T}^{\alpha }\left( A,B,\pi ^{\ast }\right)
>0.$ Also, the distribution function of $\min_{v\geq -\pi ^{\ast
	}}\left \Vert A^{\prime }Z_{m}-v\right \Vert _{B}$ is continuous near the
	$\left( 1-\alpha \right) ^{th}$ quantile. (See page 6 of Andrews and Soares (2010).) Then, we have the required result:
	\begin{eqnarray*}
		AsyCP &=&\lim_{T}P_{\phi _{T}}\left \{ q_{T} \in
		CS^{q}\right \}  \\
		&\geq &\lim_{T}P_{\phi _{T}}\left \{ G\left( q_{T};\hat{\phi}%
		,W\left( \cdot \right) \right) \leq c_{T}^{\alpha }\left( \hat{A}\left(
		q_{T}\right) ,\hat{B}\left( q_{T}\right) ,\hat{\varphi}_{T}^{\ast }\left(
		q_{T}\right) \right) +o_{p}\left( 1\right) \right \}  \\
		&\geq &P\left \{ \min_{v\geq -\pi ^{\ast }}\left \Vert A^{\prime
		}Z_{m}-v\right \Vert _{B}\leq c_{T}^{\alpha }\left( A,B,\pi ^{\ast
	}\right) \right \}  \\
	&=&1-\alpha .
\end{eqnarray*}

\noindent {\bf Step 3(ii).} Suppose that $r_1=0$. In this case, $h_{j}=\infty $ and $\pi _{j}>0$ for all $j=1,...,r.$
Then, we have $\hat{\varphi}_{T}^{\ast }\left( q_{T}\right) =\pi =\infty $
for all $T.$ Recall the definitions that $c_{T}^{\alpha }\left( \hat{A}%
\left( q_{T}\right) ,\hat{B}\left( q_{T}\right) ,\hat{\varphi}_{T}^{\ast
}(q_{T})\right) =c_{T}^{\alpha }\left( A,B,\pi ^{\ast }\right) =0.$ Then,
by Lemma~B~\ref{l_lemmaB2} and $\left( \ref{eq.ap.thm.cvinequality}\right) ,$ we
have
\begin{eqnarray*}
	AsyCP &=&\lim_{T}P_{\phi _{T}}\left \{ q_{T} \in
	CS^{q}\right \}  \\
	&=&\lim_{T}P_{\phi _{T}}\left \{ G\left( q_{T};\hat{\phi},\hat{W}%
	\left( \cdot \right) \right) \leq c_{T}^{\alpha }\left( \hat{A}\left(
	q_{T}\right) ,\hat{B}\left( q_{T}\right) ,\hat{\varphi}_{T}\left(
	q_{T}\right) \right) \right \}  \\
	&=&\lim_{T}P_{\phi _{T}}\left \{ G\left( q_{T};\hat{\phi},W\left(
	\cdot \right) \right) +o_{p}\left( 1\right) \leq c_{T}^{\alpha }\left( \hat{A%
	}\left( q_{T}\right) ,\hat{B}\left( q_{T}\right) ,\hat{\varphi}_{T}\left(
	q_{T}\right) \right) \right \} \text{ } \\
	&\geq &\lim_{T}P_{\phi _{T}}\left \{ G\left( q_{T};\hat{\phi}%
	,W\left( \cdot \right) \right) +o_{p}\left( 1\right) \leq c_{T}^{\alpha
	}(A,B,\pi ^{\ast })=0\right \} .\text{ }
\end{eqnarray*}%
By using the same argument used in (S1.23) on page 7 of Andrews and Soares
(2010), we can deduce that
\[
\lim_{T}P_{\phi _{T}}\left \{ G\left(q_{T};\hat{\phi},W\left(
\cdot \right) \right) +o_{p}\left( 1\right) \leq 0\right \} \geq 1-\alpha. \quad \Box
\]%

\subsection{Proof of Theorem~\ref{t_asymptotics_th}, Part (i)}

The proof of the theorem exploits the continuity of 
$F^q(\phi_q)$ with respect to $\phi_q$. The statement of the theorem is a consequence of
Lemma~\ref{l.continuity.hausdorff}, Lemma~\ref{l.continuity2},
and the continuous mapping theorem. $\Box$

\begin{lemma}
	\label{l.continuity2}Suppose that Assumptions~\ref{a_all}(i) and~\ref{a_csthq}(ii) are satisfied. Then,
	\begin{tlist}
		\item $F^{\theta}( \Phi_q,\Phi_\theta ) $ is compact {\color{black} for all $(\Phi_{q},\Phi_{\theta})$};
		\item $F^{\theta}( \Phi_q,\Phi_\theta ) $ is continuous {\color{black} at all  $(\Phi_{q},\Phi_{\theta})$}.
	\end{tlist}
\end{lemma}

\noindent {\bf Proof of Lemma~\ref{l.continuity2}, Part (i).}
Since $F^\theta(\Phi_q,\Phi_\theta) \subset \mathbb{R}^k$, for the required result, we show that $F^\theta(\Phi_q,\Phi_\theta)$ is closed and bounded.

\noindent {\bf Boundedness:}  The set $\{ \theta=f(\Phi_\theta,q) \; : \; \| q \| = 1 \}$ is compact because $f(\cdot)$
is continuous in both of its arguments by Assumption~\ref{a_csthq}(i) and the domain of $q$, $\mathbb{S}^n$, is compact. Since $F^\theta(\Phi_q,\Phi_\theta) \subset \{ \theta=f(\Phi_\theta,q) \; : \; \| q \| = 1 \}$, we deduce that $F^\theta(\Phi_q,\Phi_\theta)$ is bounded.

\noindent {\bf Closedness:} Consider any sequence $\theta_j \in F^\theta(\Phi_q,\Phi_\theta)$, $j=1,2,\ldots$, such that $\theta_j \longrightarrow \theta $. We show that $\theta \in F^\theta(\Phi_q,\Phi_\theta)$, that is, we need to find a $q \in F^q(\Phi_q)$ such that $\theta = f(\Phi_\theta,q)$. Then, the desired result follows.

For $\theta_j \in F^\theta(\Phi_q,\Phi_\theta)$, by definition we can choose a $q_j \in F^{q}(\Phi_q)$ such that $\theta_j = f(\Phi_\theta,q_j)$. Since $\{q_j\} \in \mathbb{S}^n$ and $\mathbb{S}^n$ is compact, we can choose a convergent subsequence $q_{j_i}$ such that $q_{j_i} \longrightarrow q$. Then it follows from the continuity of $f(\cdot)$ that $f(\Phi_\theta,q_{j_i}) \longrightarrow f(\Phi_\theta,q)$. Since the subsequence $\theta_{j_i}$ also converges to $\theta$, we have $f(\Phi_\theta,q) = \theta$. By definition of $F^\theta(\Phi_q,\Phi_\theta)$, then, we have $\theta \in F^\theta(\Phi_q,\Phi_\theta)$, as required for closedness.

\noindent {\bf Part (ii).} According to Assumption~\ref{a_csthq}(i) the function $f(\Phi_\theta,q)$ is continuous. Then,
the product correspondence
\[
\tilde{F}^q(\Phi_q,\Phi_\theta) = F^q(\Phi_q) \times \Phi_\theta
\]
is continuous by Proposition 34 of \cite{Border2010}. Notice that the correspondence $F^\theta(\Phi_q,\Phi_\theta)$ is a composite of $f(\cdot)$ and $\tilde{F}^q(\cdot)$:
\[
F^\theta(\Phi_q,\Phi_\theta) = \bigcup_{q \times \Phi_\theta \in \tilde{F}^q(\Phi_q,\Phi_\theta)} \; f(\Phi_\theta,q).
\]
Since both $f(\cdot)$ and $\tilde{F}^q(\cdot)$ are continuous, by Theorem 12.23 of Aliprantis and Border (2006), $F^\theta(\Phi_q,\Phi_\theta)$ is continuous. $\Box$

\subsection{Proof of Theorem~\ref{t_asymptotics_th}, Part (ii)}

Let $F^{\theta,q}(\phi_q,\phi_\theta) =
 \big\{\theta \in \Theta, q \in {\mathbb S^n} \; : \: q \in F^q(\phi_q), \, \theta = f(\Phi_\theta,q) \big\}$. Notice that
\begin{eqnarray*}
	\lefteqn{ \lim \inf_{T}  \inf_{\rho \in {\cal R}} \inf_{ \theta  \in F^{\theta
				}\left( \phi_q(\rho), \phi_\theta(\rho) \right) } P_{\rho}\left \{ \theta \in CS^{\theta}(\hat{\phi}_q,\hat{\phi}_\theta)\right \} }  \\
	&\geq &\lim \inf_{T}  \inf_{\rho \in {\cal R}}  \inf_{\left( \theta ,q\right) \in F^{\theta
			,q} \left( \phi_q(\rho),\phi_\theta(\rho) \right) } P_{\rho } \left \{ q \in CS^{q}(\hat{\phi}_{q}),\text{ }%
	\theta \in
	CS_{q}^{\theta}(\hat{\phi}_{\theta}) \right \}  \\
	&\geq &\lim \inf_{T} \inf_{\rho \in {\cal R} } \inf_{q\in F^{q} \left( \phi_q(\rho) \right)}
	 P_{\rho} \left \{ q\in CS^{q}(\hat{\phi}_q)\right \} \\
	&&+ \lim \inf_{T} \inf_{\rho \in {\cal R}} \; \inf_{(\theta,q)\in F^{\theta,q} \left( \phi_q(\rho), \phi_\theta(\rho) \right)} P_{\rho} \left \{ \theta \in CS_{q}^{\theta}(\hat{\phi}_{\theta})\right \} -1.
\end{eqnarray*}%
Recall that $\theta = f(\Phi_\theta,q)$.
According to Theorem~\ref{t_asymptotics_q}(ii)
\begin{equation}
\lim \inf_{T} \inf_{\rho \in {\cal R}} \inf_{q\in F^{q} \left( \phi_q(\rho) \right) }  P_{\rho 
 } \left \{ q\in CS^{q}(\hat{\phi}_q) \right \} \geq 1-\alpha_{1}
\label{bonferroni.desired.1}
\end{equation}%
and
	\begin{equation} 
	\lim \inf_{T} \inf_{\rho \in {\cal R}} \; \inf_{(\theta,q) \in F^{\theta,q} \left( \phi_q(\rho),\phi_\theta(\rho) \right)  }  P_{\rho} \left \{ \theta \in CS_{q}^{\theta}(\hat{\phi}_{\theta}) \right \} \geq 1-\alpha_{2}.
	\label{bonferroni.desired.2}
	\end{equation}
holds according to Assumption~\ref{a_csthq}.  $\Box$

%
%
%

\section{Additional Technical Lemmas}
\label{appsec_technicallemma}

Throughout this section we use the following notation.
When $A$ is a matrix, $\lambda _{\max}( A)$ and $\lambda_{\min}(A)$ are the largest
and the smallest eigenvalues of $A$, respectively.
We denote $A_{k}$
as the $k^{th}$ column vector of $A$; $A^{j}$ as the $j^{th}$ row vector of $%
A$; and $A_{jk}$ as the $\left( j,k\right) ^{th}$ element of $A.$
Throughout the proofs, we sometimes omit the $\phi_T$ argument from the asymptotic
covariance matrix $\Lambda = LL'$ and use the notation $\Lambda_{T},$ $%
\hat{\Lambda}_{T},$ $L_{T},$ and $\hat{L}_{T}$ for simplicity.
We also often omit the $q_T$
argument for some of the matrices that depend on $q_T$ and simply write, say, $S_T$, $D_{T}$,
$\hat{D}_{T}$, $A_{T}$, $\hat{A}_{T}$, $\Omega _{T}$, and $\hat{\Omega}_{T}$ for short.

%
%

%
%

%
%

\begin{applemma}
\label{l_lemmaB2}
Suppose that Assumption~\ref{a_all} is
satisfied. Consider the sequence $\left \{ \left( \phi _{T},q_{T}\right) \right \} $ with $q_{T}\in
F^{q}\left( \phi _{T}\right) $ that satisfies conditions~(i) and~(iv) in the
proof of Theorem~\ref{t_asymptotics_q}(ii). Then,%
\[
G\left(q_{T};\hat{\phi},\hat{W}\left( \cdot \right) \right)
-G\left(q_{T};\hat{\phi},W\left( \cdot \right) \right)
=o_{p}\left( 1\right) .
\]
\end{applemma}

\noindent {\bf Proof of Lemma B~\ref{l_lemmaB2}.}
According to condition~(ii) in the proof of Theorem~\ref{t_asymptotics_q}(ii)
$V\left(q_{T}\right) =V$ for all $T.$
If $V=0,$ i.e., $S(q_{T}) =0$ for all $T,$ it is trivial to
deduce the required result because by definition
\[
 G\left( q_{T};\hat{\phi},%
\hat{W}(\cdot) \right) =G\left( q_{T};\hat{\phi}%
,W(\cdot) \right) =0.
\]
Now suppose that $V\neq 0$. Notice that $S_{T}\phi _{T}\geq 0$ and $D_{T}^{-1/2}$ and $\hat{D}%
_{T}^{-1/2}$ are well defined since $S_{T}$ is a full (row) rank matrix and $%
\Lambda _{T},\hat{\Lambda}_{T}>0.$ We now consider the two cases (i) $B_{T}=\hat{B}%
_{T}=I$ and (ii) $B_{T}=\Omega_{T}^{-1}$ and $\hat{B}_{T}=\hat{\Omega}_{T}^{-1}$ separately.

\noindent {\em Case (i): $B_{T}=\hat{B}_{T}=I$.} Write%
\begin{eqnarray*}
G\left( q_{T};\hat{\phi},\hat{W}\left( \cdot \right) \right)
&=&\min_{\mu \geq 0} \; T\left \Vert \hat{D}_{T}^{-1/2}S_{T}\hat{\phi}-\hat{D}%
_{T}^{-1/2}
V_{T}\mu%
 \right \Vert ^{2} \\
&=&
\min_{v\geq -\sqrt{T}\hat{D}_{T}^{-1/2}\mu \left( q_{T},\phi _{T}\right)
}\left \Vert \hat{D}_{T}^{-1/2}S_{T}\sqrt{T}\left( \hat{\phi}-\phi
_{T}\right) - v
\right \Vert ^{2} 
\end{eqnarray*}%
and
\begin{eqnarray*}
G\left( q;\hat{\phi},W\left( \cdot \right) \right) 
&=&
\min_{v\geq -\sqrt{T}D_{T}^{-1/2}\mu \left( q,\phi \right) }\left \Vert
D_{T}^{-1/2}S_{T}\sqrt{T}\left( \hat{\phi}-\phi _{T}\right) -v \right \Vert ^{2} ,
\end{eqnarray*}%
where $\mu \left(q_{T},\phi_{T}\right) =S_{T}\phi _{T}$.
Define
\begin{eqnarray*}
v_{T}( \hat{\Lambda}_{T})  &=&\mbox{argmin}_{v\geq -\sqrt{T}\hat{D}%
_{T}^{-1/2}\mu \left( q_{T},\phi _{T}\right) }\left \Vert \hat{D}%
_{T}^{-1/2}S_{T}\sqrt{T}\left( \hat{\phi}-\phi _{T}\right) -
v \right \Vert ^{2} \\
v_{T}\left( \Lambda _{T}\right)  &=&\mbox{argmin}_{v\geq -\sqrt{T}%
D_{T}^{-1/2}\mu \left( q,\phi \right) }\left \Vert D_{T}^{-1/2}S_{T}\sqrt{T}%
\left( \hat{\phi}-\phi _{T}\right) - v\right \Vert ^{2}.
\end{eqnarray*}%
Recall that $A'=D^{-1/2} S L$ and therefore $D^{-1/2}S = A' L^{-1}$. Then,%
\begin{eqnarray*}
\lefteqn{ G\left( q_{T};\hat{\phi},\hat{W}\left( \cdot \right) \right)
-G\left( q_{T};\hat{\phi},W\left( \cdot \right) \right) } \\
&\leq &\left \Vert \hat{D}_{T}^{-1/2}S_{T}\sqrt{T}\left( \hat{\phi}-\phi
_{T}\right) - v_{T}\left( \Lambda _{T}\right)
 \right \Vert ^{2}
-\left \Vert D_{T}^{-1/2}S_{T}\sqrt{T}\left( \hat{\phi}-\phi _{T}\right)
-v_{T}\left( \Lambda _{T} \right)
 \right \Vert ^{2} \\
&\leq &\left \Vert \hat{D}_{T}^{-1/2}S_{T}\sqrt{T}\left( \hat{\phi}-\phi
_{T}\right) -D_{T}^{-1/2}S_{T}\sqrt{T}\left( \hat{\phi}-\phi _{T}\right)
\right \Vert ^{2} \\
&\leq &\left \Vert \left( \hat{A}_{T}-A_{T}\right) ^{\prime }\hat{L}_{T}^{-1}%
\sqrt{T}\left( \hat{\phi}-\phi _{T}\right) \right \Vert +\left \Vert
A_{T}^{\prime }\left( \hat{L}_{T}^{-1}-L_{T}^{-1}\right) \sqrt{T}\left( \hat{%
\phi}-\phi _{T}\right) \right \Vert  \\
&=&o_{p}\left( 1\right).
\end{eqnarray*}%
The last equality holds because $\hat{A}_{T}-A_{T}=o_{p}\left( 1\right) ,$ $%
A_{T}=O\left( 1\right) ,$ $\hat{L}_{T}^{-1}-L_{T}^{-1}=o_{p}\left( 1\right) ,
$ $L_{T}^{-1}=O\left( 1\right) ,$ $\sqrt{T}\left( \hat{\phi}-\phi
_{T}\right) =O_{p}\left( 1\right) ,$ and $\hat{B}_{T}\stackrel{p}{\longrightarrow} B>0$
according to Lemma~B~\ref{l_lemmaB5} and Assumption~\ref{a_all}(v-vi).

\noindent {\em Case (ii): $B_{T}=\Omega_T^{-1}$ and $\hat{B}_{T}=\hat{\Omega}_T^{-1}$.} In this case, we can write%
\begin{eqnarray*}
 G\left( q_{T};\hat{\phi},\hat{W}\left( \cdot \right) \right)
&=&\min_{\mu \geq 0}T \; \left \Vert S_{T}\hat{\phi}- V_T \mu \right \Vert _{\hat{\Sigma}_{T}^{-1}}^{2} \\
&=&\min_{v\geq -\sqrt{T}\mu \left( q_{T},\phi _{T}\right) }\left \Vert S_{T}%
\sqrt{T}\left( \hat{\phi}-\phi _{T}\right) - 
v
\right \Vert _{\hat{\Sigma}_{T}^{-1}}^{2} 
\end{eqnarray*}%
and
\begin{eqnarray*}
G\left( \theta ,q;\hat{\phi},W\left( \cdot \right) \right) 
&=&
\min_{v\geq -\sqrt{T}\mu \left( q_{T},\phi _{T}\right) }\left \Vert S_{T}%
\sqrt{T}\left( \hat{\phi}-\phi _{T}\right) -
v \right \Vert _{\Sigma _{T}^{-1}}^{2},
\end{eqnarray*}%
where $\mu \left( q_{T},\phi _{T}\right) =S_{T}\phi _{T}$.
Define
\begin{eqnarray*}
v_{T}(\hat{\Lambda}_{T}) &=& \mbox{argmin}_{v\geq -\sqrt{T}\mu \left(
q_{T},\phi _{T}\right) }\left \Vert S_{T}\sqrt{T}\left( \hat{\phi}-\phi
_{T}\right) -v \right \Vert _{\hat{\Sigma}_{T}^{-1}}^{2} \\
v_{T}(\Lambda_{T}) &=& \mbox{argmin}_{v\geq -\sqrt{T}\mu \left(
q_{T},\phi _{T}\right) }\left \Vert S_{T}\sqrt{T}\left( \hat{\phi}-\phi
_{T}\right) - v \right \Vert _{\Sigma _{T}^{-1}}^{2}.
\end{eqnarray*}%
Then,%
\begin{eqnarray*}
\lefteqn{G\left( \theta _{T},q_{T};\hat{\phi},\hat{W}\left( \cdot \right) \right)
-G\left( \theta _{T},q_{T};\hat{\phi},W\left( \cdot \right) \right) }\\
&\leq &\left \Vert S_{T}\sqrt{T}\left( \hat{\phi}-\phi _{T}\right) -
v_{T}\left( \Lambda _{T}\right)\right \Vert _{\hat{\Sigma}_{T}^{-1}}^{2}-\left \Vert S_{T}\sqrt{T}%
\left( \hat{\phi}-\phi _{T}\right) - 
v_{T}\left( \Lambda _{T}\right) \right \Vert _{\Sigma _{T}^{-1}}^{2} \\
&=&\left[ S_{T}\sqrt{T}\left( \hat{\phi}-\phi _{T}\right) - 
v_{T}\left( \Lambda _{T}\right)\right] ^{\prime }\Sigma _{T}^{-1/2}\left[ \Sigma _{T}^{1/2}\hat{%
\Sigma}_{T}^{-1}\Sigma _{T}^{1/2}-I_{r}\right] \\
&&\times \Sigma _{T}^{-1/2}\left [ S_{T}\sqrt{T}\left( \hat{\phi}-\phi
_{T}\right) - v_{T}\left( \Lambda _{T}\right) \right ] \\
&\leq &\left \Vert S_{T}\sqrt{T}\left( \hat{\phi}-\phi _{T}\right) - 
v_{T}\left( \Lambda _{T}\right)\right \Vert _{\Sigma _{T}^{-1}}^{2}\left \Vert \Sigma _{T}^{1/2}%
\hat{\Sigma}_{T}^{-1}\Sigma _{T}^{1/2}-I_{r}\right \Vert \\
&=&I\times II,\text{ say.}
\end{eqnarray*}%
For term $I,$ notice that since $\mu \left( q_{T},\phi _{T}\right) \geq 0,$
we have%
\[
I =\min_{v\geq -\sqrt{T}\mu \left( q_{T},\phi _{T}\right) }\left \Vert
S_{T}\sqrt{T}\left( \hat{\phi}-\phi \right) - 
v \right \Vert _{\Sigma _{T}^{-1}}^{2}
\leq \left \Vert A_{T}{}^{\prime }L_{T}^{^{-1}}\sqrt{T}\left( \hat{\phi}%
-\phi _{T}\right) \right \Vert _{\Omega _{T}^{-1}}^{2}=O_{p}\left( 1\right).
\]
The last equality holds since $A_{T}{}^{\prime }L_{T}^{^{-1}}\sqrt{T}%
\left( \hat{\phi}-\phi _{T}\right) =O_{p}\left( 1\right) $ and $\Omega
_{T}^{-1}=\left( A_{T}^{\prime }A_{T}\right) ^{-1} \longrightarrow \left(
A^{\prime }A\right) >0$ by condition (vi) in the proof of Theorem~\ref{t_asymptotics_q}(ii).
Since $\Sigma _{T}=S_{T}L_{T}L_{T}^{\prime
}S_{T}^{\prime }$, term II can be bounded as follows:
\begin{eqnarray*}
II &=&\left \Vert \Sigma _{T}^{1/2}\left( \hat{\Sigma}_{T}^{-1}-\Sigma
_{T}^{-1}\right) \Sigma _{T}^{1/2}\right \Vert \\
&=&\left \Vert \Sigma _{T}^{-1/2}\left( \Sigma _{T}-\hat{\Sigma}_{T}\right)
\hat{\Sigma}_{T}^{-1/2}\hat{\Sigma}_{T}^{-1/2}\Sigma _{T}^{1/2}\right \Vert
\\
&=&\left \Vert \Sigma _{T}^{-1/2}S_{T}\left( \Lambda _{T}-\hat{\Lambda}%
_{T}\right) S_{T}^{\prime }\hat{\Sigma}_{T}^{-1/2}\hat{\Sigma}%
_{T}^{-1/2}\Sigma _{T}^{1/2}\right \Vert \\
&=&\left \Vert \left( \Sigma _{T}^{-1/2}S_{T}L_{T}\right) \left(
L_{T}^{\prime }\hat{L}_{T}^{\prime -1}-L_{T}^{-1}\hat{L}_{T}\right) \left(
\hat{L}_{T}S_{T}^{\prime }\hat{\Sigma}_{T}^{-1/2}\right) \hat{\Sigma}%
_{T}^{-1/2}\Sigma _{T}^{1/2}\right \Vert \\
&\leq &\left \Vert \Sigma _{T}^{-1/2}S_{T}L_{T}\right \Vert \left \Vert
L_{T}^{\prime }\hat{L}_{T}^{\prime -1}-L_{T}^{-1}\hat{L}_{T}\right \Vert
\left \Vert \hat{L}_{T}S_{T}^{\prime }\hat{\Sigma}_{T}^{-1/2}\right \Vert
\left \Vert \hat{\Sigma}_{T}^{-1/2}\Sigma _{T}^{1/2}\right \Vert \\
&=&O\left( 1\right) o_{p}\left( 1\right) O_{p}\left( 1\right) O_{p}\left(
1\right).
\end{eqnarray*}%
The last line holds because
\begin{eqnarray*}
\left \Vert \Sigma _{T}^{-1/2}S_{T}L_{T}\right \Vert ^{2} &=&tr\left(
L_{T}^{\prime }S_{T}^{\prime }\left( S_{T}L_{T}L_{T}^{\prime }S_{T}^{\prime
}\right) ^{-1}S_{T}L_{T}\right) =l \\
\left \Vert \hat{L}_{T}S_{T}^{\prime }\hat{\Sigma}_{T}^{-1/2}\right \Vert
&=&tr\left( \hat{L}_{T}^{\prime }S_{T}^{\prime }\left( S_{T}\hat{L}_{T}\hat{L%
}_{T}^{\prime }S_{T}^{\prime }\right) ^{-1}S_{T}\hat{L}_{T}\right) =l \\
\left \Vert L_{T}^{\prime }\hat{L}_{T}^{\prime -1}-L_{T}^{-1}\hat{L}%
_{T}\right \Vert &=&o_{p}\left( 1\right) \text{ under Assumption~\ref{a_all}(vi). }
\end{eqnarray*}%
Moreover,
\begin{eqnarray*}
\left \Vert \hat{\Sigma}_{T}^{-1/2}\Sigma _{T}^{1/2}\right \Vert ^{2}
&=&\left \Vert \hat{\Sigma}_{T}^{-1/2}\left( S_{T}L_{T}\right) \left [
L_{T}^{\prime }S_{T}^{\prime }\left( S_{T}L_{T}L_{T}S_{T}^{\prime }\right)
^{-1}\right] \Sigma _{T}^{1/2}\right \Vert ^{2} \\
&=&\left \Vert \hat{\Sigma}_{T}^{-1/2}\left( S_{T}\hat{L}_{T}\right) \left(
\hat{L}_{T}^{-1}L_{T}\right) \left( L_{T}^{\prime }S_{T}^{\prime }\Sigma
_{T}^{-1/2}\right) \right \Vert ^{2} \\
&\leq &\left \Vert \hat{\Sigma}_{T}^{-1/2}S_{T}\hat{L}_{T}\right \Vert
^{2}\left \Vert \hat{L}_{T}^{-1}L_{T}\right \Vert ^{2}\left \Vert
L_{T}^{\prime }S_{T}^{\prime }\Sigma _{T}^{-1/2}\right \Vert ^{2} \\
&=&O_{p}\left( 1\right) O_{p}\left( 1\right) O\left( 1\right) =O_{p}\left(
1\right). 
\end{eqnarray*}
This completes the proof for Case (ii). $\Box$

%
%

%
%

%
%

\begin{applemma}
\label{l_lemmaB5}
Suppose that a converging sequence $\left\{ \phi _{T},q_{T}\right\} $ satisfies the rank condition $r\left( q_{T}\right) =r>0$
and $V\left( q_{T}\right) $ is a non-zero constant selection matrix for all $%
T.$ Then, there exists a subsequence $\left\{ T^{\prime }\right\} \subset
\left\{ T\right\} $ such that along the subsequence, we have (i)%
\begin{equation*}
D^{-1/2}\left( q_{T^{\prime }}\right) S\left( q_{T^{\prime }}\right) L\left(
\phi _{T^{\prime }}\right) \longrightarrow A,
\end{equation*}%
where $A$ is a full rank matrix, and (ii)
\begin{equation*}
\hat{D}^{-1/2}\left( q_{T^{\prime }}\right) S\left( q_{T^{\prime }}\right)
\hat{L}( \hat{\phi}_{T^{\prime }}) =D^{-1/2}\left( q_{T^{\prime
}}\right) S\left( q_{T^{\prime }}\right) L\left( \phi _{T^{\prime }}\right)
+o_{p}\left( 1\right) .
\end{equation*}
\end{applemma}

\noindent {\bf Proof of Lemma B~\ref{l_lemmaB5}.}
Part (i):  Recall that
$S_T = V_T \tilde{S}_T$. The rank reduction of $\tilde{S}_{T} $ is caused only by zero rows (see Section~\ref{subsec_idea_rankreductions}).
Moreover, according to condition~(ii) in the proof of Theorem~\ref{t_asymptotics_q}(ii)
the non-zero row selection matrix is $V_{T}$ constant over $T$.
Thus, we can construct an index set $\mathcal{J}$
of non-zero rows of $\tilde{S}_{T}.$ By construction, the size of $\mathcal{J}$ is $l$ and
\[
   S_T = \big[ \tilde{S}^j_T \big]_{j \in {\cal J}}.
\]
In turn we obtain
\[
  D^{-1/2}_{T} S_{T} L_{T}
    = D_{T}^{-1/2} \big[ \tilde{S}_{T}^j L_{T} \big]_{j \in {\cal J}}.
\]
Recall from the definition of $L$ and $D$ that (omitting the $T$ subscripts)
\[
    S L L' S' = D^{1/2} \Omega D^{1/2} \quad \mbox{and} \quad D^{-1/2} S LL'S' D^{-1/2} = \Omega,
\]
where $\Omega$ is a correlation matrix with ones on its diagonal. Thus, $D_{ii}^{-1/2}$ normalizes
the length of the $i$'th row of the matrix $(SL)$ to one. Therefore,
\[
   D^{-1/2}_{T^{\prime }} S_{T^{\prime }} L_{T^{\prime}}
     = \left[ \frac{\tilde{S}_{T}^j L_{T}}{ \|\tilde{S}_{T}^j L_{T}\|}  \right]_{j \in {\cal J}}
     = \left[ \frac{\tilde{S}_{T}^{j}}{\left\Vert \tilde{S}_{T}^{j}L_{T}\right\Vert }\right] _{j\in \mathcal{J}}L_{T}.
\]

By construction, $\tilde{S}_{T}^{j}\neq 0$ for all $T$ and $j\in \mathcal{J}.$ Since $L_{T}>0$,
it follows that $\tilde{S}_{T}^{j}L_{T}\neq 0$ for all $T$ and $j\in \mathcal{J}$. In turn,
$\| \tilde{S}_{T}^{j}L_{T}\| >0$ for all $T$ and $%
j\in \mathcal{J}$ and $\tilde{S}_{T}^{j}L_{T}/ \|\tilde{S}_{T}^{j}L_{T}\|$ is well defined for all $T$ and $j\in \mathcal{J%
}.$ Notice that $\left\{ \tilde{S}_{T}^{j}L_{T}/\|\tilde{S}_{T}^{j}L_{T}\|\right\}_{T}$ is a sequence on a unit sphere,
which is compact. We can then choose a subsequence $\left\{ T^{\prime
}\right\} $ such that $\tilde{S}_{T^{\prime }}^{j}L_{T^{\prime }}/\|
\tilde{S}_{T^{\prime }}^{j}L_{T^{\prime }}\| $ converges for all $j\in
\mathcal{J}$. Thus, we can write
\[
D^{-1/2}_{T^{\prime }} S_{T^{\prime }} L_{T^{\prime}} =
  \left[ \frac{\tilde{S}_{T^{\prime }}^{j}}{\left\Vert \tilde{S}_{T^{\prime
  }}^{j}L_{T^{\prime }}\right\Vert }\right] _{j\in \mathcal{J}}L_{T^{\prime }}
 \longrightarrow A.
\]

To obtain the desired result, it remains to be shown that $A$ is full rank. Since
$L_{T^{\prime }}^{-1}\longrightarrow L^{-1}>0,$ it suffices to show that the limit
\begin{equation}
   AL^{-1} = \lim_{T' \longrightarrow \infty} \;
   \left[\frac{\tilde{S}_{T^{\prime }}^{j}}{\left\Vert \tilde{S}_{T^{\prime }}^{j}L_{T^{\prime
}}\right\Vert }\right] _{j\in \mathcal{J}}
\label{eq.ap.lemmaB5.limit}
\end{equation}%
has full rank. Recall that $\tilde{S}\left( q\right)  =\left( I\otimes q^{\prime }\right)
\bar{S} \phi_q$. By construction, the non-zero rows of $%
\tilde{S}_{T^{\prime }}^{j}$ are orthogonal to each other because $\left\{
\tilde{S}_{T^{\prime }}^{j}\right\} _{j\in \mathcal{J}}$ is composed of rows
$\left( +/-\right) \left( I^{j}\otimes q\right) \bar{S} \phi_q$ that are
orthogonal to each other. This implies that each row of the limit $AL^{-1}$
is non-zero and orthogonal, which delivers the required result.

\noindent Part (ii): Consider the subsequence $\left\{ T^{\prime }\right\} $ in the
proof of Part (i). Since $\hat{L}_{T'} > 0$ and $\tilde{S}^j_{T'} \not= 0$ for all $T'$,
\[
   \big\| \tilde{S}_{T'}^j \hat{L}_{T'} \big\| > 0
\]
for all $T'$. We will now show that
\begin{equation*}
\frac{\tilde{S}_{T^{\prime }}^{j}\hat{L}_{T^{\prime }}}{\left\Vert \tilde{S}_{T^{\prime
}}^{j}\hat{L}_{T^{\prime }}\right\Vert }=\frac{\tilde{S}_{T^{\prime
}}^{j}L_{T^{\prime }}}{\left\Vert \tilde{S}_{T^{\prime }}^{j}L_{T^{\prime
}}\right\Vert }+o_{p}\left( 1\right)
\end{equation*}%
for all $j\in \mathcal{J}.$ Since it could be the case that $\| \tilde{S}_{T'}^j L_{T'} \| \longrightarrow 0$,
we provide a detailed argument. Write
\[
\frac{\tilde{S}_{T^{\prime }}^{j}\hat{L}_{T^{\prime }}}{\left\Vert \tilde{S}_{T^{\prime
}}^{j}\hat{L}_{T^{\prime }}\right\Vert }-\frac{\tilde{S}_{T^{\prime
}}^{j}L_{T^{\prime }}}{\left\Vert \tilde{S}_{T^{\prime }}^{j}L_{T^{\prime
}}\right\Vert }
=\frac{\tilde{S}_{T^{\prime }}^{j}L_{T^{\prime }}}{\left\Vert \tilde{S}_{T^{\prime
}}^{j}L_{T^{\prime }}\right\Vert }\left( \frac{\left\Vert \tilde{S}_{T^{\prime
}}^{j}L_{T^{\prime }}\right\Vert }{\left\Vert \tilde{S}_{T^{\prime }}^{j}\hat{L}%
_{T^{\prime }}\right\Vert }-1\right) +\frac{\tilde{S}_{T^{\prime }}^{j}\left( \hat{L}%
_{T^{\prime }}-L_{T^{\prime }}\right) }{\left\Vert \tilde{S}_{T^{\prime }}^{j}\hat{L}%
_{T^{\prime }}\right\Vert } \\
=I+II,\text{ say.}
\]
We begin with the following bound:
\begin{eqnarray*}
\frac{\left\Vert \tilde{S}_{T^{\prime }}^{j}L_{T^{\prime }}\right\Vert }{\left\Vert
\tilde{S}_{T^{\prime }}^{j}\hat{L}_{T^{\prime }}\right\Vert }-1 &=&\frac{\left\Vert
\tilde{S}_{T^{\prime }}^{j}L_{T^{\prime }}\right\Vert -\left\Vert \tilde{S}_{T^{\prime }}^{j}%
\hat{L}_{T^{\prime }}\right\Vert }{\left\Vert \tilde{S}_{T^{\prime }}^{j}\hat{L}%
_{T^{\prime }}\right\Vert } \\
&= &\frac{\left\Vert \tilde{S}_{T^{\prime }}^{j}\hat{L}_{T^{\prime
}}-\tilde{S}_{T^{\prime }}^{j}\left( \hat{L}_{T^{\prime }}-L_{T^{\prime }}\right)
\right\Vert -\left\Vert \tilde{S}_{T^{\prime }}^{j}\hat{L}_{T^{\prime }}\right\Vert
}{\left\Vert \tilde{S}_{T^{\prime }}^{j}\hat{L}%
_{T^{\prime }}\right\Vert} \\
&\leq &\frac{\left\Vert \tilde{S}_{T^{\prime }}^{j}\left( \hat{L}_{T^{\prime
}}-L_{T^{\prime }}\right) \right\Vert }{\left\Vert \tilde{S}_{T^{\prime }}^{j}\hat{L}%
_{T^{\prime }}\right\Vert } \\
&\leq &\frac{\left\Vert \tilde{S}_{T^{\prime }}^{j}\right\Vert \left\Vert \hat{L}%
_{T^{\prime }}-L_{T^{\prime }}\right\Vert }{\left\Vert \tilde{S}_{T^{\prime
}}^{j}\hat{L}_{T^{\prime }}\right\Vert} \\
&= &\frac{ \left\Vert \hat{L}%
_{T^{\prime }}-L_{T^{\prime }}\right\Vert }{\left\Vert \tilde{S}_{T^{\prime
}}^{j}\hat{L}_{T^{\prime }}\right\Vert / \left\Vert \tilde{S}_{T^{\prime }}^{j}\right\Vert}.
\end{eqnarray*}
The last equality holds because $\|\tilde{S}^j_{T'} \| > 0$ for all $T'$.

According to Assumption~\ref{a_all}(vi). $\hat{L}_{T^{\prime }} \stackrel{p}{\longrightarrow} L$.
Moreover, we deduce from (\ref{eq.ap.lemmaB5.limit}) and $A^{j}L^{-1}\neq 0$
that
\begin{eqnarray*}
0 &<&\frac{\left\Vert \tilde{S}_{T^{\prime }}^{j}\hat{L}_{T^{\prime }}\right\Vert }{%
\left\Vert \tilde{S}_{T^{\prime }}^{j}\right\Vert }\leq \frac{\left\Vert
\tilde{S}_{T^{\prime }}^{j}L_{T^{\prime }}\right\Vert +\left\Vert \tilde{S}_{T^{\prime
}}^{j}\left( \hat{L}_{T^{\prime }}-L_{T^{\prime }}\right) \right\Vert }{%
\left\Vert \tilde{S}_{T^{\prime }}^{j}\right\Vert } \\
&\leq &\frac{\left\Vert \tilde{S}_{T^{\prime }}^{j}L_{T^{\prime }}\right\Vert }{%
\left\Vert \tilde{S}_{T^{\prime }}^{j}\right\Vert }+\left\Vert \hat{L}_{T^{\prime
}}-L_{T^{\prime }}\right\Vert  \\
&\stackrel{p}{\longrightarrow} & \frac{1}{\left\Vert A^{j}L^{-1}\right\Vert }>0.
\end{eqnarray*}%
Therefore,
\begin{equation*}
0\leq \frac{\left\Vert \hat{L}_{T^{\prime }}-L_{T^{\prime }}\right\Vert }{%
\left\Vert \tilde{S}_{T^{\prime }}^{j}\hat{L}_{T^{\prime }}\right\Vert /
\left\Vert \tilde{S}_{T^{\prime }}^{j}\right\Vert } \leq o_{p}\left( 1\right)
\left\Vert A^{j}L^{-1}\right\Vert =o_{p}\left( 1\right) .
\end{equation*}
Similarly, we obtain the bound
\begin{eqnarray*}
1-\frac{\left\Vert \tilde{S}_{T^{\prime }}^{j}L_{T^{\prime }}\right\Vert }{%
\left\Vert \tilde{S}_{T^{\prime }}^{j}\hat{L}_{T^{\prime }}\right\Vert } &=&\frac{%
\left\Vert \tilde{S}_{T^{\prime }}^{j}\hat{L}_{T^{\prime }}\right\Vert -\left\Vert
\tilde{S}_{T^{\prime }}^{j}L_{T^{\prime }}\right\Vert }{\left\Vert \tilde{S}_{T^{\prime
}}^{j}\hat{L}_{T^{\prime }}\right\Vert } \\
&=&\frac{\left\Vert \tilde{S}_{T^{\prime }}^{j}L_{T^{\prime }}+\tilde{S}_{T^{\prime
}}^{j}\left( \hat{L}_{T^{\prime }}-L_{T^{\prime }}\right) \right\Vert
-\left\Vert \tilde{S}_{T^{\prime }}^{j}L_{T^{\prime }}\right\Vert }{\left\Vert
\tilde{S}_{T^{\prime }}^{j}\hat{L}_{T^{\prime }}\right\Vert } \\
&\leq &\frac{\left\Vert \tilde{S}_{T^{\prime }}^{j}\left( \hat{L}_{T^{\prime
}}-L_{T^{\prime }}\right) \right\Vert }{\left\Vert \tilde{S}_{T^{\prime }}^{j}\hat{L}%
_{T^{\prime }}\right\Vert } \stackrel{p}{\longrightarrow} 0.
\end{eqnarray*}%
Since $\tilde{S}_{T^{\prime }}^{j}L_{T^{\prime }}/\Vert
\tilde{S}_{T^{\prime }}^{j}L_{T^{\prime }}\vert=O\left( 1\right)$, we have
established that term $I$ vanishes asymptotically:
\begin{equation*}
I=o_{p}\left( 1\right) .
\end{equation*}%
Term $II$ can be bounded as follows:
\[
\left\Vert II\right\Vert  =\frac{\left\Vert \tilde{S}_{T^{\prime }}^{j}\left( \hat{%
L}_{T^{\prime }}-L_{T^{\prime }}\right) \right\Vert }{\left\Vert
\tilde{S}_{T^{\prime }}^{j}\hat{L}_{T^{\prime }}\right\Vert }\leq \frac{\left\Vert
\tilde{S}_{T^{\prime }}^{j}\right\Vert \left\Vert \hat{L}_{T^{\prime }}-L_{T^{\prime
}}\right\Vert }{\left\Vert \tilde{S}_{T^{\prime }}^{j}\hat{L}_{T^{\prime
}}\right\Vert }
\leq \frac{\left\Vert \hat{L}_{T^{\prime }}-L_{T^{\prime }}\right\Vert }{%
\left\Vert \tilde{S}_{T^{\prime }}^{j}\hat{L}_{T^{\prime }}\right\Vert /
\left\Vert \tilde{S}_{T^{\prime }}^{j}\right\Vert }\stackrel{p}{\longrightarrow} 0,
\]
and so
\begin{equation*}
II=o_{p}\left( 1\right) .
\end{equation*}%
Combining the two $o_p(1)$ results completes the proof of Part (ii). $\Box$

%
%

\begin{applemma}
\label{l_lemmaB9}
Suppose Assumption~\ref{a_all} is satisfied. Consider Case~(i) in Step~3 of
the proof of Theorem~\ref{t_asymptotics_q}(ii). Along the $\{T\}$ sequence
defined in Step~1 of the proof of Theorem~\ref{t_asymptotics_q}(ii),
\begin{equation*}
c_{T}^{\alpha }\left( \hat{A}\left( q_{T}\right) ,\hat{B}\left( q_{T}\right)
,\hat{\varphi}_{T}^{\ast }(q_{T})\right) \rightarrow _{p}c_{T}^{\alpha
}\left( A,B,\pi ^{\ast }\right) ,
\end{equation*}%
where the critical value function $c_T^\alpha(\cdot)$ is defined in~(\ref{eq_cTalphafcn1}) and~(\ref{eq_cTalphafcn2}).
\end{applemma}

\noindent {\bf Proof of Lemma~B~\ref{l_lemmaB9}.}
The proof is very similar to that of Lemma 2(a) in Andrews and Soares
(2010b) and we provide a sketch. The proof proceeds in three steps. First,
show
\begin{equation*}
\left( \hat{\xi}_{T},\hat{A}(q_{T}),\hat{B}\left( q_{T}\right) \right)
\overset{p}{\longrightarrow }(\pi ,A,B)\quad \mbox{and}\quad \hat{\varphi}%
_{T}^{\ast }(q_{T})\overset{p}{\longrightarrow }\pi ^{\ast }.
\end{equation*}
Second, show
\begin{equation*}
\mathbb{P}\left \{ \min_{v\geq -\hat{\varphi}_{T}^{\ast }(q_{T})}\left \Vert (\hat{A}%
(q_{T})^{\prime }Z_{m}-v\right \Vert _{\hat{B}(q_{T})}^{2}\leq x\right
\} \overset{p}{\longrightarrow } \mathbb{P}\left \{ \min_{v\geq -\pi ^{\ast }}\left \{
\left \Vert A^{\prime }Z_{m}-v\right \Vert _{B}^{2}\right \} \leq
x\right \} .
\end{equation*}
Third, deduce $c_{T}^{\alpha }\left( \hat{A}\left( q_{T}\right) ,\hat{B}%
\left( q_{T}\right) ,\hat{\varphi}_{T}^{\ast }(q_{T})\right) \stackrel{p}{\longrightarrow}
c_{T}^{\alpha }\left( \hat{A}\left( q_{T}\right) ,\hat{B}\left(
q_{T}\right) ,\pi ^{\ast }\right) ,$ as required for the lemma.

\noindent \emph{Proof of Step 1:} By the choice of the sequence $\{T\}$ and
the limit result in Step 1 of the proof of Theorem~\ref{t_asymptotics_q}(ii)
\begin{equation*}
\left( \hat{\xi}_{T},\hat{A}(q_{T}),\hat{B}\left( q_{T}\right) \right)
\overset{p}{\longrightarrow }(\pi ,A,B).
\end{equation*}%
Notice that if $\pi _{j}=0$, then $\xi _{T}(q_{T})<\kappa _{T}$ as $%
T\longrightarrow \infty $ and by using an argument similar to the one used in the proof of Lemma~B~\ref{l_lemmaB5}(ii),
we have $\hat{\xi}_{T}(q_{T})< \kappa _{T}$ in
probability as $T \longrightarrow \infty $. Therefore, $\mbox{plim} \; \hat{%
\varphi}_{j,T}^{\ast }(q_{T})=\mbox{plim} \; \hat{\varphi}_{j,T}(q_{T})=0=\pi
_{j}^{\ast }$ with probability one. On the other hand, if $\pi _{j}>0$, then
$\hat{\varphi}_{j,T}^{\ast }(q_{T})=\infty =\pi _{j}^{\ast }$. Therefore, $%
\hat{\varphi}_{T}^{\ast }(q_{T})\overset{p}{\longrightarrow }\pi ^{\ast }.$

\noindent \emph{Proof of Step 2}: The desired result can be obtained by the
same argument used in the proof of (S1.17) of Andrews and Soares (2010).

\noindent \emph{Proof of Step 3}: It is immediate from Step 2 and the fact
that the distribution of
\begin{equation*}
\min_{v\geq -\pi ^{\ast }} \left \Vert A^{\prime }Z_{m}-v\right
\Vert _{B}^{2}
\end{equation*}%
is continuous if $k\geq 1$, and continuous near the $(1-\alpha )^{\prime }s$
quantile, where $\alpha <1/2$, if $k=0$. $\Box$

\section{Description of Monte Carlo Experiments}
\label{appsec_derivationmc}

\renewcommand{\theequation}{C.\arabic{equation}}
\setcounter{equation}{0}

\subsection{Experiment 1: Bivariate VAR(0)}

Computations for the Monte Carlo experiment with the VAR(0) model,
e.g., Design~1 in Table~1 of the main article: $y_{t}=u_{t},$ $%
u_{t}\sim N\left( 0,\Sigma \right)$. The population identified set is given by
$F^{\theta }\left( \phi \right) =%
\left[ 0,\max \left\{ \mathcal{I}\left\{ \phi _{2}\geq 0\right\} ,\sqrt{%
\frac{\phi _{3}^{2}}{\phi _{2}^{2}+\phi _{3}^{2}}}\right\} \right] $ where $%
\phi =\left[ \phi _{1},\phi _{2},\phi _{3}\right] ^{\prime }=\left[ \Sigma
_{11}^{tr},\Sigma _{21}^{tr},\Sigma _{22}^{tr}\right] ^{\prime }$ and $%
\Sigma _{ij}^{tr}$ are the elements of $\Sigma _{tr},$ the lower triangular
matrix from the Cholesky decomposition of $\Sigma .$

It is convenient to reparameterize $q$ in spherical coordinates:
$q = q(\varphi) = \big[ \cos(\varphi) \; \sin ( \varphi) \big]'$.
However, for brevity we typically write $q$, omitting the $\varphi$ argument.
We generate a grid ${\cal Q}$ for $q$ by dividing the domain of $\varphi$, $\left[ -\frac{\pi }{2},\frac{\pi }{2}\right]$,
into equally sized partitions of length $\delta_{\varphi}$.
As discussed in the main text, since $\phi_1=\Sigma_{11}^{tr}>0$
the inequality restriction $\theta =q_{1}\phi_{1}\geq 0$
implies that $q_{1}\geq 0$. Thus, it suffices to conduct the grid search
with respect to $\varphi$ over the interval $\left[ -\frac{\pi }{2},\frac{\pi }{2}\right]$.

The following steps are repeated $N_{sim}$ times. The results reported
in the main text are averages across these repetitions. We report the
average length of the confidence intervals and compute the coverage
probability as the fraction of times for which the upper bound of $F^{\theta }\left( \phi \right)$
is contained in the confidence interval. The upper bound of the identified set
determines the lower bound of the coverage probability.

\noindent {\bf Generating Data:}
Generate a sample of length T of data from the VAR(0) using the parameters
reported in Table~1.

\noindent {\bf Estimating the Reduced-Form Parameters}

\begin{itemize}
\item Compute the sample covariance $\hat{\Sigma}=\frac{1}{T}%
\sum\limits_{t=1}^{T}( y_{t}-\bar{y}) ^{\prime }( y_{t}-%
\bar{y}) $ where $\bar{y}=\frac{1}{T}\sum\limits_{t=1}^{T}y_{t}$.
Denote by $\hat{\Sigma}_{tr}$ the lower triangular matrix from the Cholesky
decomposition of $\hat{\Sigma}.$ Then $\hat{\phi}=[ \hat{\phi}_{1},\hat{%
\phi}_{2},\hat{\phi}_{3}] ^{\prime }=[ \hat{\Sigma}_{11}^{tr},%
\hat{\Sigma}_{21}^{tr},\hat{\Sigma}_{22}^{tr}] ^{\prime }$, where $\hat{%
\Sigma}_{ij}^{tr}$ are the elements of $\hat{\Sigma}_{tr}.$

\item Estimate $\Lambda ,$ the asymptotic variance covariance matrix of $%
\hat{\phi}$, using a parametric bootstrap:
\begin{itemize}
\item Generate bootstrap samples $b=1,\ldots,B$ of
length $T$ from $y_{t}^{(b)}=u_{t}^{(b)}$ where $u_{t}^{(b)}\sim N( 0,\hat{\Sigma}%
).$
\item For each bootstrap sample, estimate $\hat{\Sigma}^{(b)}$
and compute $\hat{\phi}^{(b)}=[ \hat{\phi}_{1}^{(b)},\hat{\phi}%
_{2}^{(b)},\hat{\phi}_{3}^{(b)}] ^{\prime }$.
\item Let $\hat{%
\Lambda}=\frac{1}{B}\sum_{b=1}^{B}[ \sqrt{T}( \hat{\phi}%
^{(b)}-\hat{\phi}) ][ \sqrt{T}(
\hat{\phi}^{(b)}-\hat{\phi}) ]' $ with factorization $\hat{\Lambda} = \hat{L} \hat{L}'$.
\end{itemize}
\end{itemize}

\noindent \textbf{Computing the Confidence Intervals}

\begin{itemize}

\item \textit{Step 1: Construct a }$\left( 1-\alpha _{1}\right) $ \textit{confidence
set for q}. The following computations are executed for each $q \in {\cal Q}$.
As before, it is convenient to express $q$ in terms of the angle $\varphi$ and
generate ${\cal Q}$ by equally spaced grid points on the interval $[-\pi,\pi]$.
Recall the definition of $\xi_{1,T}\left( q\right)$ and $\xi _{2,T}\left( q\right)$ in~(\ref{eq_xijT}).

\begin{itemize}
\item If $\varphi \in \left\{ -\frac{\pi }{2},\frac{\pi }{2} \right\}$, the objective function is given by
\begin{equation*}
G^{q}\left( q;\hat{\phi},\hat{W}\right) =\min\limits_{\mu \geq 0} \;
\frac{T}{\hat{\Sigma} _{22}\left( q\right) }\left( q_{1}\hat{\phi}_{2}+q_{2}\hat{%
\phi}_{3}-\mu\right) ^{2}.
\end{equation*}

\begin{itemize}
\item If $\xi _{2,T}\left( q\right) <\kappa _{T}$, the inequality condition is
considered binding and the critical value $c^{\alpha _{1}}\left( q\right) $ is the $%
\left( 1-\alpha _{1}\right)$ quantile of a squared truncated normal $Z^{2}%
\mathcal{I}\left\{ Z\geq 0\right\}$.

\item If $\xi _{2,T}\left( q\right) \geq \kappa _{T}$, the inequality condition is considered
non-binding and $c^{\alpha _{1}}\left( q\right) =0.$

\end{itemize}

\item If $\varphi \not\in \left\{ -\frac{\pi }{2},\frac{\pi }{2} \right\}$, the objective
function is given by
\begin{equation*}
G^{q}\left( q;\hat{\phi},\hat{W}\right) =\min\limits_{\mu_{1}\geq 0,%
\mu_{2}\geq 0}T\left\Vert \hat{D}^{-1/2}\left( q\right) \left[
\begin{array}{c}
q_{1}\hat{\phi}_{1}- \mu_{1} \\
q_{1}\hat{\phi}_{2}+q_{2}\hat{\phi}_{3}- \mu_{2}%
\end{array}%
\right] \right\Vert _{\hat{B}\left( q\right) }^{2}.
\end{equation*}

\begin{itemize}
\item If $\xi _{1,T}\left( q\right) < \kappa _{T}$ and $\xi _{2,T}\left(
q\right) < \kappa _{T}$, both inequality conditions are considered binding.  For
$j=1,\ldots,n_Z$ draw $Z_{3}^{(j)}$ from $N\left( 0,I_{3}\right) $. The
critical value is the $\left( 1-\alpha _{1}\right)$ quantile of%
\begin{equation*}
\mathcal{\bar{G}}^{\left( j\right) }\left( q;\hat{B}\left( q\right) \right)
=\min \limits_{\nu \geq 0} \;  \big\| \hat{D}^{-1/2}\left( q\right) S\left(
q\right) \hat{L}Z_{3}^{(j)}-\nu \big\| _{\hat{B}\left( q\right) }^{2}.
\end{equation*}
The minimization can be executed with a numerical routine that solves
quadratic programming problems.

\item If $\xi_{1,T}\left( q\right) <\kappa _{T}$ and $\xi _{2,T}\left( q\right)
\geq \kappa _{T}$ or if $\xi _{1,T}\left( q\right) \geq \kappa _{T}$ and $%
\xi _{2,T}\left( q\right) <\kappa _{T}$, i.e., only one inequality
condition is considered binding, then $c^{\alpha _{1}}\left( q\right) $ is the $%
\left( 1-\alpha _{1}\right) th$ quantile of a squared truncated normal $Z^{2}%
\mathcal{I}\left\{ Z\geq 0\right\}$.

\item if $\xi _{1,T}\left( q\right) \geq \kappa _{T}$ and $\xi _{2,T}\left(
q\right) \geq \kappa _{T}$, then no inequality condition is considered binding and $%
c^{\alpha _{1}}\left( q\right) =0.$

\end{itemize}

\item Let $CS^{q}=\left\{ q \in {\cal Q} \mid \left( G^{q}\left( q;\hat{\phi},\hat{W}%
\right) -c^{\alpha _{1}}\left( q\right) \right) \leq 0\right\}$.

\end{itemize}

\item \textit{Step 2: Construct a }$\left( 1-\alpha _{2}\right) $ \textit{confidence
set for }$\theta $ \textit{conditional on q}:
\[
CS_{q}^{\theta }=\left[
\max \left\{ 0,q_{1}\hat{\phi}_{1}-z_{\alpha_2/2} \sqrt{ q_{1}^2 \hat{\Lambda}_{11}/ T} \right\}
,\text{ }q_{1}\hat{\phi}_{1}+ z_{\alpha_2/2} \sqrt{ q_{1}^2 \hat{\Lambda}_{11}/ T} \right],
\]
where $z_{\alpha_2/2}$ is the $(1-\alpha_2/2)$ quantile of a $N(0,1)$ distribution
and $\hat{\Lambda}_{11}$ is the $(1,1)$ element
of the matrix $\hat{\Lambda}$.

\item \textit{Step 3: Construct the }$1-\alpha $ \textit{Bonferroni set for }$%
\theta $:
Compute the minimum of the lower bounds of $CS_{q}^{\theta}$
and the maximum of the upper bounds of $CS_{q}^{\theta }$ for $q \in CS^q$.

\end{itemize}

\subsection{Experiment 2: Bivariate VAR(1)}

The computations are very similar to the computations for
the VAR(0) experiment. Thus, we focus on highlighting the differences.
The model takes the form (Designs 2 to 4 in Table 1 of the main article):
$y_{t}=A y_{t-1}+u_{t}$, where $u_t \sim N(0,\Sigma)$.
Let $\Sigma_{tr}$ denote the lower-triangular Cholesky factor of $\Sigma$.
The reduced-form parameters are given by
\[
\phi = \mbox{vec} \big( \big( A \Sigma_{tr} \big)^{\prime} \big)
     = \big[ \phi_1,\phi_2,\phi_3,\phi_4 \big]'
     = \big[ A_{11} \Sigma_{11}^{tr}+A_{12}\Sigma_{21}^{tr},
             A_{12} \Sigma_{22}^{tr}, A_{21} \Sigma_{11}^{tr}
              +A_{22} \Sigma_{21}^{tr}, A_{22} \Sigma_{22}^{tr} \big]^{\prime},
\]
where $\Sigma_{ij}^{tr}$ are the elements of $\Sigma_{tr}$.
Under our three Monte Carlo designs the identified set $F^q(\phi)$ has
a geometry similar to that of the identified set for the VAR(0) design, depicted
in Figure~1 of the main article. Roughly speaking, it is an arc located in
the Northeast section of the unit circle.
Under the parameterization of the data-generating processes (DGPs), the top-left endpoint of $F^q(\phi)$
is given by the solution of
\[
    q^2_{1,l} = \frac{1}{1+ (\phi_1/\phi_2)^2},
\]
whereas the bottom-right endpoint of $F^q(\phi)$ is given by the solution of
\[
    q^2_{1,r} = \frac{1}{1+ (\phi_3/\phi_4)^2}.
\]
The structural parameter of interest is $\theta = q_1 \phi_1 + q_2 \phi_2$.
For our Monte Carlo designs the lower bound of the identified set $F^\theta(\phi)$ is determined
by $\theta_l = q_{1,l} \phi_1 + q_{2,l} \phi_2$. The upper bound is
$\theta_u = q_{1,r} \phi_1 + q_{2,r} \phi_2$ if $q_{2,r} > 0$; or is
$\theta_u = q_{1,r} \phi_1 + q_{2,r} \phi_2$ otherwise.

As for the VAR(0) experiment, minimizations with respect to $q$
are carried out using a grid $q \in {\cal Q}$, where
$q = \big[ \cos(\varphi) \; \sin (\varphi) \big]'$ and $\varphi$ takes values
on an equally spaced grid over $[-\pi,\pi]$ with spacing $\delta_\varphi$.

\noindent {\bf Generating Data:} The DGP is now given by $y_{t}=A y_{t-1}+u_{t}$.

\noindent {\bf Estimating the Reduced-Form Parameters:} Follow the same steps as in the VAR(0) experiment.

\noindent \textbf{Bonferroni Approach}

\begin{itemize}

\item {\em Step 1: Construct a $1-\alpha_1$ confidence set for $q$}.

\begin{itemize}
\item The objective function is%
\begin{equation*}
G^{q}\left( q;\hat{\phi},\hat{W}\right) =\min\limits_{\mu_{1}\geq 0,%
\mu_{2}\geq 0}T\left\Vert \hat{D}^{-1/2}\left( q\right) \left[
\begin{array}{c}
q_{1}\hat{\phi}_{1}+q_{2}\hat{\phi}_{2}-\mu_{1} \\
q_{1}\hat{\phi}_{2}+q_{2}\hat{\phi}_{3}-\mu_{2}%
\end{array}%
\right] \right\Vert _{\hat{B}\left( q\right) }^{2}
\end{equation*}

\item If $\xi _{1,T}\left( q\right) <\kappa _{T}$ and $\xi _{2,T}\left( q\right)
<\kappa _{T}$, both inequality conditions are considered binding. For $j=1,\ldots,n_Z$ draw
$Z_{4}^{(j)}$ from $N\left( 0,I_{4}\right) $. The critical
value is the $\left( 1-\alpha _{1}\right)$ quantile of%
\begin{equation*}
\mathcal{\bar{G}}^{\left( j\right) }\left( q;\hat{B}\left( q\right) \right)
=\min\limits_{\nu \geq 0} \; T \big\| \hat{D}^{-1/2}\left( q\right) S\left(
q\right) \hat{L}Z_{4}^{(j)}-\nu \big\| _{\hat{B}\left( q\right) }^{2}
\end{equation*}

\item If $\xi _{1,T}\left( q\right) <\kappa _{T}$ and $\xi _{2,T}\left( q\right)
\geq \kappa _{T}$ or if $\xi _{1,T}\left( q\right) \geq \kappa _{T}$ and $%
\xi _{2,T}\left( q\right) <\kappa _{T}$, i.e., only one inequality
condition is considered binding, then $c^{\alpha _{1}}\left( q\right) $ is the $%
\left( 1-\alpha _{1}\right)$ quantile of a squared truncated normal $Z^{2}%
\mathcal{I}\left\{ Z\geq 0\right\}$.

\item If $\xi _{1,T}\left( q\right) \geq \kappa _{T}$ and $\xi _{2,T}\left(
q\right) \geq \kappa _{T}$, then no inequality condition is considered binding and $%
c^{\alpha _{1}}\left( q\right) =0.$

\end{itemize}

\item \textit{Step 2: Construct the $(1-\alpha_{2})$ confidence
set for $\theta $ conditional on q}. Follow the same steps as in Experiment~1.

\item \textit{Step 3: Construct the $1-\alpha$ Bonferroni set for $
\theta$}. Follow the same steps as in the Experiment~1.

\end{itemize}

\subsection{Experiment 3: Four-Variable VAR(2)}

\noindent {\bf Design.} The coefficient matrices for the DGP are given by
\begin{eqnarray*}
	A_{1}^{\prime } &=&\left[ 
	\begin{array}{cccc}
		1.001 & -0.100 & 0.302 & -0.085 \\ 
		0.065 & 0.585 & 0.089 & -0.055 \\ 
		0.126 & 0.284 & 1.072 & -0.073 \\ 
		0.233 & 0.141 & 0.056 & 1.522%
	\end{array}%
	\right], \quad
	A_{2}^{\prime } =\left[ 
	\begin{array}{cccc}
		-0.080 & 0.119 & -0.269 & 0.078 \\ 
		-0.056 & 0.262 & 0.065 & 0.013 \\ 
		-0.223 & -0.222 & -0.178 & 0.070 \\ 
		-0.230 & -0.097 & -0.069 & -0.538%
	\end{array}%
	\right] \\
	c &=&\left[ 
	\begin{array}{c}
		0.626 \\ 
		0.175 \\ 
		0.064 \\ 
		0.204%
	\end{array}%
	\right] \text{ \ }\Sigma =\left[ 
	\begin{array}{cccc}
		0.542 & -0.124 & 0.199 & 0.095 \\ 
		-0.124 & 1.164 & 0.129 & -0.369 \\ 
		0.199 & 0.129 & 0.912 & -0.263 \\ 
		0.095 & -0.369 & -0.263 & 0.549%
	\end{array}%
	\right].
\end{eqnarray*}

\section{Further Details on the Empirical Analysis}
\label{appsec_empirical}

\renewcommand{\theequation}{D.\arabic{equation}}
\setcounter{equation}{0}


The construction of the data set follows \cite{AruobaSchorfheide2011}.
Unless otherwise
noted, the data are obtained from the FRED2 database maintained by
the Federal Reserve Bank of St. Louis. Per capita  output is
defined as real GDP (GDPC96) divided by the civilian
non-institutionalized population (CNP16OV). The population
series is provided at a monthly frequency and converted to quarterly
frequency by simple averaging. We take the natural log of per capita
output and extract a deterministic trend by OLS regression over
the period 1959:I to 2006:IV. The deviations from the linear trend
are scaled by 100 to convert them into percentages.
Inflation is defined as the log difference of the GDP deflator (GDPDEF), scaled
by 400 to obtain annualized percentage rates.
Our measure of nominal interest rates corresponds to the federal
funds rate (FEDFUNDS), which is provided at monthly frequency and
converted to quarterly frequency by simple averaging.
We use the sweep-adjusted M2S series provided by Cynamon, Dutkowsky and Jones (2006).
This series is recorded at monthly frequency without seasonal adjustments.
The EVIEWS default version of the X12 filter is applied to remove seasonal
variation. The M2S series is divided by quarterly nominal GDP to obtain inverse
velocity. We then remove a linear trend from log inverse velocity and scale the
deviations from trend by 100. Since our VAR is expressed in terms of real money
balances, we take the sum of log inverse velocity and real GDP.
Finally, we restrict our quarterly observations to the period from 1965:I to 2005:I.

\section*{Additional References}

\begin{description}
	\item Aliprantis, Charalambos D. and Kim C. Border (2006): {\em Infinite-Dimensional Analysis}, 3rd Edition,
	Springer Verlag, New York.
\item Andrews, Donald and Patrick Guggenberger (2009): ``Validity of Subsampling and
`Plug-in Asymptotics' Inference for Parameters Defined by Moment Inequalities,''
{\em Econometric Theory}, {\bf 25}, 669-709.
\item Andrews, Donald and Gustavo Soares (2010): ``Supplement to `Inference for Parameters Defined by Moment Inequalities Using Generalized Moment Selection,' ''{\em Econometrica Supplementary Material}.
\item Aruoba, Boragan and Frank Schorfheide (2011): ``Sticky Prices versus Monetary Frictions:
      An Estimation of Policy Trade-offs,'' {\em American Economic Journal: Macroeconomics},
      {\bf 3}, 60-90.
\item Border, Kim C. (2007): ``Alternative Linear Inequalities,'' {\em Manuscript}, California Institute of Technology.
\item Border, Kim C. (2010): ``Introduction to Correspondences,'' {\em Manuscript}, California Institute of Technology.
\item Cynamon, Barry, Donald Dutkowsky, and Barry Jones (2006):
      ``Redefining the Monetary Aggregates: A Clean Sweep,''
      {\em Eastern Economic Journal}, {\bf 32}, 661-672.
\end{description}

\end{appendix}

\end{document}